\documentclass[useAMS,usenatbib]{mn2e}

\usepackage{graphicx}
\usepackage{times}
\usepackage{natbib}
\usepackage{color}
\usepackage{todonotes}

\newcommand{\apjs}{ApJS}

\newcommand{\apjl}{ApJL}

\newcommand{\aap}{A \& A}

\newcommand{\enzo}{\it{\small ENZO}}

\begin{document}
 
\title[Properties of cosmic filaments] {A survey of the thermal and non-thermal properties of cosmic filaments}
\author[C. Gheller, F. Vazza]{C. Gheller$^1$, F. Vazza$^{4,3,2}$\thanks{E-mail: franco.vazza2@unibo.it}\\
$^{1}$ Swiss Plasma Center, EPFL, SB SPC Station 13 - 1015 Lausanne, Switzerland\\
$^{2}$Istituto di Radio Astronomia, INAF, Via Gobetti 101, 40121 Bologna, Italy\\
$^{3}$ Hamburger Sternwarte, Gojenbergsweg 112, 21029 Hamburg, Germany\\
$^{4}$ Dipartimento di Fisica e Astronomia, Universit\'{a} di Bologna, Via Gobetti 92/3, 40121, Bologna, Italy
}

\date{Received / Accepted}
\maketitle
\begin{abstract}
In this paper, we exploit a large suite of {\enzo} cosmological magneto-hydrodynamical simulations adopting uniform mesh resolution, to investigate the properties of cosmic filaments under different baryonic physics and magnetogenesis scenarios. We exploit a isovolume based algorithm to identify filaments and determine their attributes from the continuous distribution of gas mass density in the simulated volumes. The global (e.g. mass, size, mean temperature and magnetic field strength, enclosed baryon fraction) and internal (e.g. density, temperature, velocity and magnetic field profiles) properties of filaments in our volume are calculated across almost four orders of magnitude in mass. The inclusion of variations in non-gravitational physical processes (radiative cooling, star formation, feedback from star forming regions and active galactic nuclei) as well as in the seeding scenarios for magnetic fields (early magnetisation by primordial process vs later seeding by galaxies) allows us to study both the large-scale thermodynamics and the magnetic properties of the Warm-Hot Intergalactic Medium (WHIM) with an unprecedented detail. We show how the impact of non-gravitational physics on the global thermodynamical properties of filaments is modest, with the exception of the densest gas environment surrounding galaxies in filaments.  Conversely, the magnetic properties of the WHIM in filament are found to dramatically vary as different seeding scenarios are considered. We study the correlation between the properties of galaxy-sized halos and their host filaments, as well as between the halos and the local WHIM in which they lie. Significant general statistical trends are reported.
\end{abstract}

\label{firstpage} 
\begin{keywords}
galaxy: clusters, general -- methods: numerical -- intergalactic medium -- large-scale structure of Universe
\end{keywords}

\section{Introduction}
\label{sec:intro}

Galaxy surveys and numerical simulations show that the large scale structure of the Universe is organised in form of filaments, halos and voids. Numerical simulations predict also that a large fraction of the baryonic matter (around 50\%) resides in the form of plasma in such cosmic web, at densities $\sim$10-100 times the average cosmic value and $10^5$-$10^7$ K temperatures, forming the “Warm-Hot Intergalactic Medium” (WHIM, see e.g. \citealt{1996Natur.380..603B,2001ApJ...552..473D}).
Direct observations of the cosmic web are challenging, due to its extremely low mass density. The cooler phases of the intergalactic gas ($\sim 10^4 \rm K$) contain a small fraction of neutral hydrogen, producing the characteristic Ly-$\alpha$ absorption. Indications of possible detection of filamentary structures emerged from the analysis of soft X–ray \citep[e.g.][]{2003A&A...410..777F,2008A&A...482L..29W,2010ApJ...715..854N,2016xnnd.confE..27N}, or of the Sunyaev-Zeldovich effect \citep[e.g.][]{2013A&A...550A.134P,2017arXiv170905024T,2017arXiv170910378D}. 

The first X-ray imaging of the terminal part of four filaments connected to the virial radius of cluster A2744 has been reported by \citet{2015Natur.528..105E} using XMM-Newton. By using the Sardinia Radio Telescope (SRT), \citet{2018MNRAS.tmp.1093V} reported the possible detection of large-scale diffuse emission around a giant filaments connecting clusters. However, the emission could be associated to the outer regions of clusters rather than to the WHIM in filaments. Finally, the possible detection of a Faraday Rotation excess produced  by filaments overlapping the polarised emission of a giant radio galaxy at $z=0.34$ has been proposed by \citet{2018arXiv181107934O}, based on LOFAR observations. Statistical techniques via cross-correlation analysis have also attempted to detect the signature from cosmic filaments emitting in radio, but reported only upper limits \citep[e.g.][]{vern17,brown17}. Complementary to this, statistical studies of the Faraday Rotation measurement from background sources were used to limit the magnetisation of the cosmic web intervening onto high redshift polarised sources \citep[][]{Blasi.Burles..1999,2015RAA....15.1629X,2015A&A...575A.118O,2016PhRvL.116s1302P}.

In the radio domain, the interest in detecting filaments is rapidly growing, triggered by the fast development of precursors and pathfinders of the Square Kilometer Array (SKA), which will have the potential of scratching the surface of the magnetic cosmic web \citep[e.g.][]{
2011JApA...32..577B,va15ska}. Current radio observatories like LOFAR, ASKAP, MWA and MeerKAT may already lead to early detection in a non-negligible number of targets \citep[][]{va15radio}, and sophisticated techniques for the automated detection of very faint and diffuse emission on $\sim \rm ~degree$ scales  in such surveys  are already available \citep[e.g.][]{cosmodeep}. 

Filaments represent an ideal environment to investigate the past epochs that led to present cosmic structures.  In fact, their dynamics is less violent and complex than that of galaxy clusters or of galaxies, with adiabatic physics (besides gravity) dominating the gas dynamics, but with other physical processes influencing the behaviour and the properties of the gas component. Among these processes, energy injection from supernova explosion or AGN jet heating have already proved to be crucial to explain observed features of dense environments (e.g. preventing runaway cooling in dense galaxy clusters cores, or regulating the star formation efficiency (see \citet{borgani08,2018arXiv181001883M} and references therein). In filaments, thanks to their more gentle evolution, the interplay of the different physical processes can be better understood and their effects on observable signatures effectively disentangled. Furthermore, filamentary structures are expected to preserve many traces of the original environment in which the process of gravitational clustering started. This is particularly relevant when magnetic fields come into play. Although all the cosmological structures are expected to be substantially magnetised, the origin and the evolution of such magnetic fields is currently unsettled. Filaments should retain memory of the initial magnetic seed fields since they should not host strong dynamo amplification \citep[e.g.][]{ry08,donn09,va14mhd}. Furthermore, although astrophysical processes can contribute to the evolution of such primordial seeds, in close connection with the dynamical history of the cosmic structures, their signatures should remain confined to the highest density regions, influencing only partially the overall filaments' volume.

When first accreted onto filaments, the cosmic gas is shock heated by strong shocks, $\mathcal{M} \sim 10-100$  \citep[e.g.][]{ry03,pf06}. Downstream of these shocks, supersonic turbulence is injected  \citep[e.g.][]{ka07,ry08,wi17}. When such turbulence gets dissipated, a tiny fraction of the resulting energy is expected to feed the  amplification of magnetic fields  \citep[][]{ry08,2011PhRvL.107k4504F,jones11,2013NJPh...15b3017S,po15}. Most of cosmological MHD  simulations targeting filaments have found little presence of volume-filling dynamo amplification of seed fields in filaments \citep[][]{br05,va14mhd,2015MNRAS.453.3999M,va17cqg}, which is well explained by the little time available for amplification in such environment, magnetic eddies being rapidly advected onto the surrounding cluster environment before being significantly amplified \citep[e.g.][]{va14mhd}. 

In this scenario,  most of filaments' volume must be filled with magnetic fields whose evolution mostly come from  compression/rarefaction of primordial magnetic field lines, following $B  \approx B_{0} \cdot ({n_e} /{\langle n \rangle)}^{\alpha_B}$, where $B_0$ is the seed field, $\langle n \rangle$ is the cosmic mean (gas) density and $\alpha_B \approx 2/3$ for isotropic gas compression. If this picture is qualitatively correct, then from the observation of the present day magnetic field level of filaments, it would be possible to infer the amplitude of primordial magnetic fields to a good approximation.

In  a different model, the  predominant  mechanism  for  the origin  of  cosmic  magnetic  fields  is  the  ”magnetic  pollution” by active galactic nuclei and galactic activities \citep[e.g.][]{donn09}. Relatively low magnetisation is expected in the vast majority of filaments, due to a combination of the adiabatic decrease of the field, and of the scarcity of  magnetic field sources there \citep[e.g.][]{xu09,2015MNRAS.453.3999M,va17cqg}. Furthermore, magnetic fields would be originated following the distribution of galaxies and AGN, filling in an intermittent and discontinuous way the filament's volume. This would make their characterisation even more challenging.

Assessing the magnetisation level of filaments is of great interest also for other studies, from the origin and composition of ultra high-energy cosmic rays (UHECRs) propagating across the Universe \citep[e.g.][]{Sigl:2003ay,Dolag:2003ra,hack16,hack18} to the study of rotation measure for cosmological sources, like quasars and Fast Radio Bursts \citep[e.g.][]{2016ApJ...824..105A,va18frb}. For example, it has been recently argued that the excess of $\sim 6 \cdot 10^{19} \rm eV$ events reported by the Telescope Array may be due to the trapping of UHECRs by filaments local to the Virgo cluster \citep[][]{2019arXiv190100627K}.
\\

\begin{figure*}
\includegraphics[width=0.95\textwidth]{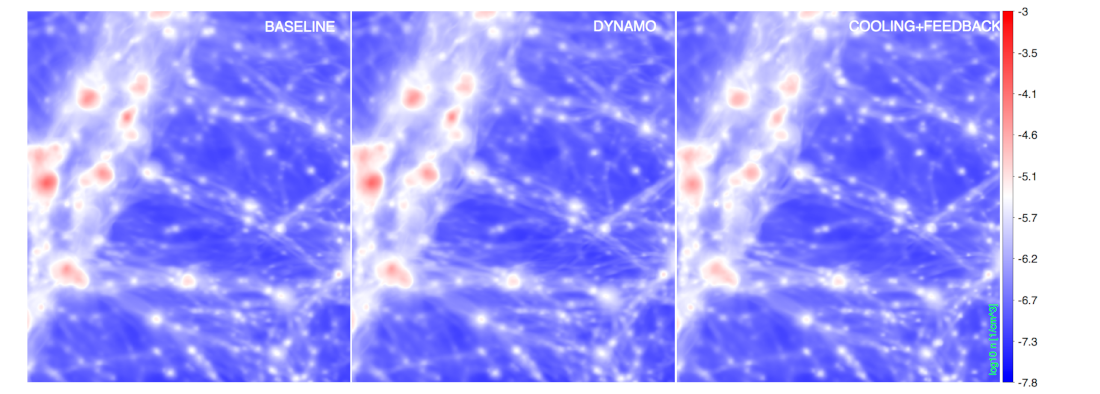}
\includegraphics[width=0.95\textwidth]{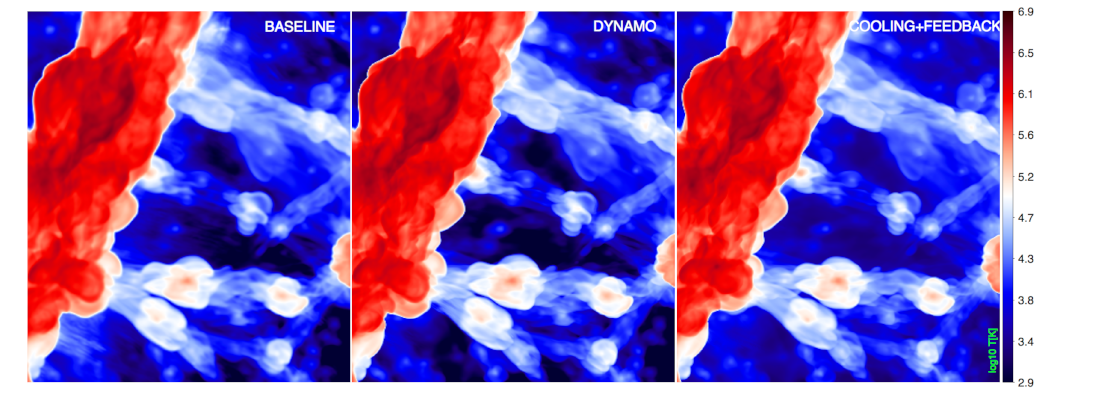}
\includegraphics[width=0.95\textwidth]{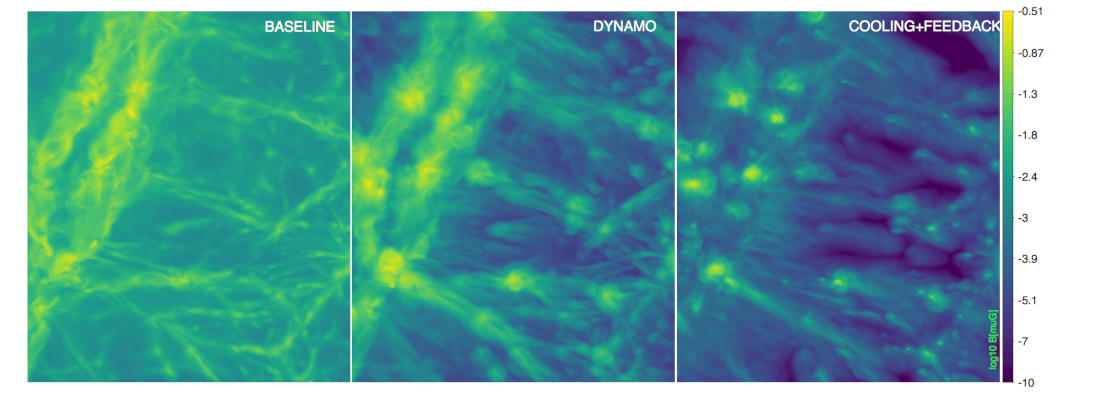}
\caption{Projected gas density (top), gas temperature (centre) and magnetic field (bottom) for a  $10^3 ~\rm Mpc^3$ cubic selection of the simulated volume at z=0; the first column shows the Baseline primordial model, the second the sub-grid dynamo (DYN5) model and the third column the astrophysical model including cooling and energy feedback from star formation and supermassive black holes (CSFBH3 model).}
\label{fig:mapT}
\end{figure*}

In this work, we address several of the above topics by studying in detail the thermal and non-thermal (magnetic) properties of the gas within cosmic filaments my means of cosmological magneto-hydrodynamics (MHD) simulations. Different physical models have been considered, selecting a number of simulations from the recent ”Chronos++ suite” \citep[][]{va17cqg}, a collection of runs aimed at exploring a number of scenarios for the emergence of large-scale magnetic fields, including different physical and astrophysical processes, encompassing primordial, dynamo and astrophysical magnetogenesis scenarios. 

Filaments are identified within the simulated datasets using the methodology already adopted in \citet{gh15} and \citet{gh16}. Such methodology is based on the calculation of baryonic matter density isovolumes to identify filamentary structures. In this, it is different from other approaches used for N-body simulations, that can be broadly grouped in geometric and tessellation methods, based on the topological analysis of the density field by means of sophisticated mathematical approaches (e.g. \cite{2005A&A...434..423S}; \cite{2008MNRAS.383.1655S}; 
\cite{2010arXiv1006.4178A};
\cite{2010MNRAS.407.1449G}; 
\cite{2015MNRAS.454.3341C})
and morphological methods, that classify the 3D distribution of matter based on the density Hessian and/or the tidal or velocity shear fields (e.g. \cite{2007MNRAS.375..489H}; \cite{2007A&A...474..315A}; \cite{2014MNRAS.441.2923C}). Exploiting the same methodology, it was also possible to extract both the galaxy halos, in order to correlate their properties with those of the hosting filaments, and the local environment where halos reside, in order to investigate its characteristics compared to those of the resident halos. In particular, the cosmological missing baryons problem is addressed, by analysing the baryon fraction of the halos, of the local environments and of the overall filament.

The details of the simulations, the code used for the runs, the models selected from the Chronos++ suite are presented in Section 2, together with a summary of the adopted filament extraction methodology. Section 3 is dedicated to the discussion of the statistical and internal properties of the filaments extracted from the different runs, while Section 4 focus on the statistical properties of the halos and their correlations with the hosting filament and local environment. In Section 5 we discuss the various results, while Section 6 draws the conclusions.

\begin{table*}
\begin{center}
\caption{Main parameters of the runs in the Chronos++ suite used in the present work. From left to right, columns defines: the presence of radiative cooling or star forming particles, the critical gas number density $n_*$ to trigger star formation in the \citet{2003ApJ...590L...1K} model, the time-scale for star formation $t_*$, the thermal feedback efficiency and the magnetic feedback efficiency ($\epsilon_{\rm SF}$ and $\epsilon_{\rm SF,b}$) from star forming regions; the efficiency of Bondi accretion $\alpha_{\rm Bondi}$ in the \citet{2011ApJ...738...54K} model for SMBH; the thermal feedback efficiency and the magnetic feedback efficiency ($\epsilon_{\rm BH}$ and $\epsilon_{\rm BH,b}$) from SMBH; the intensity of the initial magnetic field, $B_0$; the presence of sub-grid dynamo amplification at run time; the ID of the run and some additional descriptive notes.  All simulations evolved a $85^3 \rm Mpc^3$ volume using $1024^3$ cells and dark matter particles, starting at redshift $z=38$. The name convention of all runs is consistent with \citet{va17cqg}.}
\footnotesize
\centering \tabcolsep 2pt
\begin{tabular}{c|c|c|c|c|c|c|c|c|c|c|c|c|c}
  cooling & star form. & $n_*$ & $t_*$ & $\epsilon_{\rm SF}$ & $\epsilon_{\rm SF,b}$  & $\alpha_{\rm Bondi}$ & $\epsilon_{\rm BH}$ &$\epsilon_{\rm BH,b}$  & $B_0$   & dynamo & ID & description \\ 
  & & [$1/\rm cm^3$]  & [$\rm Gyr$]  &  & & &  &  &  [G] & &  & & \\  \hline \hline
   n &n& - & -& -& -& -  &-& -& $10^{-9}$ & no &  P(Baseline) & primordial,unifor, \\   
   n &n& -  & -& -& -& -  & -& -&$10^{-9}$ & no &  Z2 & primordial, tangled\\
   n &n& -  & -& -& -& - &-&  -&$10^{-18}$  & $10 \cdot \epsilon_{\rm dyn}(\mathcal{M})$  & DYN5 & low primordial, efficient dynamo\\
    n &n& -  &- & -& -& -  &-&  -&$10^{-18}$  & $\epsilon_{\rm dyn}=0.04$  & DYN7 & low primordial, inefficient dynamo \\
    n &n& - &- & -& -&  -  & -& -&$10^{-11}$  &  $\epsilon_{\rm dyn}(\mathcal{M})$  & DYN8 & high primordial, dynamo \\ \hline
  
    y & y &0.001 & 1.5& $10^{-9}$ &  $0.01$&  -  &-& -& $10^{-18}$  & -  & CSF2 & star formation, weak feedback  \\
    y & y& 0.001& 1.5 & $10^{-8}$ & 0.01 &  - & -& -&$10^{-11}$  & - & CSFB11 & star formation, high primordial field   \\
    y & y& 0.0002& 1.0 & $10^{-6}$ &  0.1  & - & -& -&$10^{-18}$  & -  & CSF5 & star formation, strong  feedback  \\
   y &y&0.0001 &1.5 & $10^{-8}$ & 0.01 & $10^3$ fix. &0.05 & 0.01 &   $10^{-18}$  & -  & CSFBH2 & star formation, BH, constant $(\frac{0.01 M_{\odot}} {\rm yr} )$ \\
   y &y&0.001 & 1.0 & $10^{-7}$ &  0.1 & $10^3$& 0.05 & 0.1 & $10^{-18}$  & - & CSFBH3 & star formation, BH, variable accr. rate \\
    y &y&0.0002 & 1.0 & $10^{-7}$ & 0.1 & $10^2$ & 0.05 & 0.1 &  $10^{-18}$  & -  & CSFBH5 & star formation, BH, strong  feedback  \\

  \end{tabular}
  \end{center}
\label{table:tab1}
\end{table*}

\section{The Numerical Approach}
\label{sec:numerics}

The simulations presented in this paper have been performed with the cosmological Eulerian code  {\enzo} \citep{enzo14},
with  a fixed mesh resolution. The code has been customised by our group mainly with the purpose of including different mechanisms for the seeding of magnetic fields in cosmology, as explained in detail in \citet{va17cqg}.

The magneto-hydrodynamics (MHD) solver adopted in our simulations implements the conservative Dedner formulation \citep[][]{ded02}, which uses hyperbolic divergence cleaning to enforce the $\nabla \cdot \vec{B} = 0$ condition. The Dedner cleaning method has been already tested by 
several works in the literature, showing that despite the relatively large rate of dissipation introduced by its ``cleaning waves'', it always converges to the correct solution as resolution is increased,  at variance with other possible ``divergence cleaning'' methods \citep[][]{2013MNRAS.428...13S,2016MNRAS.455...51H,2016MNRAS.461.1260T}.  We refer the reader to more recent reviews for a broader discussion of
the resolution and accuracy of different MHD schemes  in properly resolving the dynamo in cosmological simulations \citep{review_dynamo}.

The MHD solver adopts the Piecewise Linear Method reconstruction technique; fluxes at cell interfaces are calculated using the Harten-Lax-Van Leer  (HLL) approximate Riemann solver. Time integration is performed using the total variation diminishing (TVD) second-order Runge-Kutta (RK) scheme \citep[][]{1988JCoPh..77..439S}.  We used the GPU-accelerated MHD version of {\enzo} by \citet[][]{wang10}, which overall ensures a $\sim \times 4$ speedup compared to the more strandard CPU version of the code when applied to the large, $1024^3$ uniform grids used here.  Constant spatial resolution has the advantage of providing the best resolved description of magnetic fields even in low-density regions, which would typically not be refined by AMR schemes.

A known limitation in our numerical model is the diffusivity of the HLL scheme combined with the Dedner cleaning, which reduce the actual dynamical range. Compared to other MHD methods such as as the Constrained Transport, the Dedner scheme is more affected by small-scale dissipation of magnetic fields,  due to the $\nabla \cdot \vec{B}$ cleaning waves it generates to keep the numerical divergence under control. 
Nevertheless, several works in the literature have shown that the 
method is robust and accurate for most tests in MHD, also including magnetic turbulence in idealised setups \citep[e.g.][]{wa09,wang10,kri11, enzo14}.
Various astrophysical applications showed that this method converges to the right solution in  control  tests, unlike other common cleaning or $\nabla\cdot \vec{B}$ preserving  techniques \citep[][]{2013MNRAS.428...13S,2016MNRAS.455...51H,2016MNRAS.461.1260T}.
Moreover, resolution tests presented by our group in \citet{va14mhd} and \citet{va18mhd} have convincingly showed how at sufficiently high resolution, the scheme can properly model small-scale dynamo in a large Reynolds number regime. 
Given the fairly limited spatial resolution of our the simulations  here  ($83.3$ kpc/cell), the negligible impact of the small-scale dynamo here may be  
exacerbated by the limited resolution. Based on a dedicated suite of higher resolution simulations, in  \citet{va14mhd} we concluded that the development of the small-scale dynamo in filaments is hampered by little dynamical time available for amplification in this fast moving flow, hence the lack of dynamo amplification here looks physically motivated. However, the simplistic hydro-MHD view cannot exclude that in real plasmas "microscopic" small-scale instabilities arising from kinetic plasma effects \citep[e.g.][for a review]{2014MNRAS.440.3226M} can further promote a small-scale dynamo starting from sub-kpc scales. For this reason, a subset of our models (``subgrid dynamo models", see below) has been explicitly designed to bracket the maximal amount of magnetic field amplification that would be possible in our filaments, which is basically set by the available kinetic energy.

Our study uses the suite of cosmological MHD simulations at a fixed mesh resolution, known as the ``Chronos++ suite'' {\footnote{http://cosmosimfrazza.myfreesites.net/the\_magnetic\_cosmic\_web}}. It encompasses a large number of models which explore various plausible scenarios for the evolution of large-scale magnetic fields, with the goal of bracketing the distribution of magnetic fields outside of virialized halos.  In this work we focus on one "baseline" non-radiative simulation with a simple primordial seeding of magnetic fields, and we contrast it with 10 additional resimulations out of the full suite of $24$ models presented in \citet{va17cqg}, which overall give a good match to the observed  cosmic average star formation rate as well as to observed galaxy cluster scaling relations,  as discussed there. 

The Chronos++ suite includes three basic classes of models: {\it primordial}, {\it dynamo} and {\it astrophysical} magnetogenesis scenarios (or a mix of the them).  In detail: 

\begin{itemize}
\item{\it primordial} models:  we assumed the existence of weak and volume-filling magnetic fields at the beginning of the simulation. These are the only seed of magnetisation in our simulated universe. We set either a spatially uniform seed field value for every component of the initial magnetic field $B_0$ (Baseline model in Table 1), or a primordial field with orientation orthogonal to
the initial 3-dimensional velocity field from the ``Zeldovich approximation'', used by the standard cosmological initial condition generators (Z2 model in Table 1) which ensures a $\nabla \cdot \vec{B}=0$ field, being the Zeldovich approximation irrotational by construction.  We ensured by construction that  $(\langle B^2\rangle)^{1/2}=B_0$, which allows us to closely compare the above two models.  We consider here $B_0=1 ~\rm nG$ (comoving) at  $z_{\rm in}=38$, consistent with the most recent upper limits derived from the analysis of the Cosmic Microwave Background by  \citet[][]{PLANCK2015}, of order $B_0 \leq 5 ~\rm nG$ for fields with a coherence scale of $\sim ~\rm Mpc$, although even lower limits have been suggested \citep[e.g.][]{2014PhRvD..89d3523T}. 
\item {\it dynamo} models: we estimated at run-time via sub-grid modelling the small-scale dynamo amplification of seed field of primordial origin  ($B_0 = 10^{-9} \rm nG$ in runs DYN5 {\footnote{We notice that $B_0 = 10^{-9} \rm nG$ is lower than the $\geq 10^{-7}-10^{-6} \rm ~nG$ upper limits derived from the non-detection of the Inverse Compton Cascade around  high-$z$ blazar sources \citep[e.g.][]{2010Sci...328...73N,2014ApJ...796...18A,2015PhRvD..91l3514C} (see however \citealt{2012ApJ...752...22B} for a different interpretation of these results). However, the dynamical role of such tiny magnetic fields is negligible anyway in our runs, and our arbitrary choice is just meant to allow us clearly locating the magnetisation bubbles associated with astrophysical sources in our simulated volume.}} and DYN7 or $B_0=0.01 \rm ~nG$ in run DYN8). The dissipation of solenoidal turbulence is estimated following \citet{fed14}, by extrapolating the information resolved at our fixed $83.3 ~\rm kpc$ cell resolution on smaller scales. These models attempt to bracket the possible residual amount of dynamo amplification which can be attained, on energy grounds, in our cosmic structures, but may be lost because of a lack of resolution and/or for the onset of small-scale plasma instabilities that cannot be captured from first principles in our simple hydro-MHD model. 
In this work we consider models which give a reasonable match to the magnetic field strength in galaxy clusters.  More details on this procedure can be found in \citet{va17cqg}, as well as in Appendix B.
\item {\it astrophysical} models: magnetic fields are seeded by a) stellar activity and/or b) feedback by supermassive black holes (SMBH), simulated at run time using prescriptions available in {\enzo}  \citep[e.g.][]{2003ApJ...590L...1K,2011ApJ...738...54K,enzo14}, with tuneable star formation efficiency, timescales and feedback parameters. Our assumed accretion model for SMBH follows from the  spherical Bondi-Hoyle formula, without taking into account the Eddington accretion limit (run CSFBH3 and CSFBH5) or assuming a fixed $0.01 ~\rm M_{\odot}/yr$ accretion rate (CSFBH2) which are all options available in the {\enzo} public version, that we have tuned to produce realistic accretion and feedback results at the spatial and mass resolution covered here. In all cases we consider a "boost" factor to the mass growth rate of SMBH ($\alpha_{\rm Bondi} =100$ in CSFBH5  or $=1000$ otherwise), again to account for the effect of coarse resolution in properly resolving the mass accretion rate onto our simulated SMBH particles. We have extended {\enzo} coupling thermal feedback to the direct injection of additional magnetic energy via bipolar jets, with an efficiency with respect to the feedback energy computed at run-time, $\epsilon_{\rm SF,b}$ and $\epsilon_{\rm BH,b}$ for the stellar and supermassive black hole feedback, respectively. For the subset of runs analysed in this work we used $\epsilon_{\rm SF,b}=10\%$ and $\epsilon_{\rm BH,b}=1\div 10\%$ for the magnetic feedback, while the different values for the feedback efficiency (referred to the $\epsilon_{\rm SF, BH} = \dot{M} c^2$ energy accreted by star forming or black hole particles) are listed in Table 1. Further details can be found in  \citet{va17cqg}. The initial seed field in all the runs in set to $B_0=10^{-9} \rm ~nG$, with the exception of run CSFB11 which started from $B_0 = 0.01 ~\rm nG$. A more detailed overview of the above methods is given in Appendix A.
Clearly, the fixed spatial resolution of our runs is in principle too coarse to properly model galaxy formation processes with sufficient detail (even massive galaxies are resolved within a few cells at most). However, in \citet{va17cqg} we showed how our star-formation recipes (based on \citealt{2003ApJ...590L...1K}) are specifically calibrated to properly reproduce the cosmic star formation history in the volume (see Appendix of 
\citealt{va17cqg}). Moreover, the combination of cooling and stellar/AGN feedback has been tuned so that the scaling relations of galaxy clusters and groups are well reproduced (see Appendix of \citealt{va17cqg}). In \cite{va13feedback} and \cite{va16scienzo} we showed that these models can also fairly describe the innermost density, temperature and entropy profiles of galaxy clusters when going to a higher resolution. In summary, our previous works suggest that the feedback recipes implemented in our simulations can effectively mimic the large-scale effect and energetic of galaxy formation recipes at a higher resolution, hence they can be effectively used to model the large-scale effects on thermal and magnetic feedback on the surrounding distribution of filaments.

\end{itemize}

All the runs adopted the $\Lambda$CDM cosmology, with density parameters $\Omega_{\rm BM} = 0.0478$ (BM representing the Baryonic Matter), $\Omega_{\rm DM} = 0.2602$ (DM being the Dark Matter),  $\Omega_{\Lambda} = 0.692$ ($\Lambda$, being the cosmological constant) and a Hubble constant $H_0 = 67.8$ km/sec/Mpc \citep[][]{2016A&A...594A..13P}.  The initial redshift is $z=38$, the spatial  resolution is $83.3 ~\rm kpc/cell$ (comoving) and the constant mass resolution of  $m_{\rm DM}=6.19 \cdot 10^{7}M_{\odot}$ for dark matter particles. 

\begin{figure*}
\includegraphics[width=0.995\textwidth]{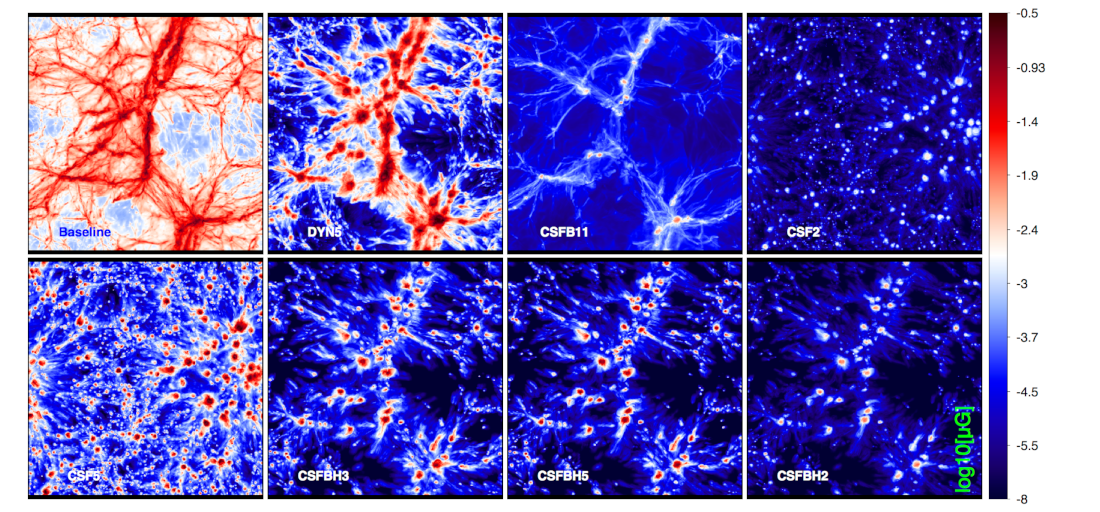}
\caption{Projected (mass-weighted) magnetic field strength across the entire $85^3~\rm Mpc^3$ volume at $z=0$ for  11 of ours models (see labels).}
\label{fig:mapB}
\end{figure*}

\subsection{Filaments Identification}
\label{sec:visit}

Filaments are identified in the data produced by the different simulations according to the procedure presented in \cite{gh15},
based on the VisIt data visualisation and analysis framework \citep{Childs11visit:an}. Here we summarise the main steps of such procedure.

The mass density of the baryonic component is used in order to separate over- from under-dense regions. The {\it Isovolume}-based approach, described in \citet{Meredith2004}, has been adopted in order to accurately trace the boundaries between the warm/hot phase which characterises collapsed cosmological structures and the cold under-dense phase typical of voids. Isovolumes are then segmented into distinct objects applying a {\it Connected Components} filter. Identified overdense isovolumes undergo a further cleaning procedure in order to eliminate objects that either cannot be classified as filaments or that can be affected by large numerical errors, due, in particular to the limited spatial resolution. The main steps of the identification procedure are:

\begin{enumerate}
\item {\it Identification and removal of large clumps.}
Galaxy clusters are identified using the Isovolume algorithm out of the highest peaks of the
matter distribution. They are discarded by removing from the data all cells identified as
part of the cluster.

\item {\it Filament identification.}
The Isovolume algorithm is used once more on the residual cells to identify the volumes
with:
\begin{equation}
\delta_{\rm BM} = {\varrho_{\rm BM}\over\varrho_0} \ge a_{\rm fil}, \label{eq:BM}
\end{equation}
where $\delta_{\rm BM}$ is the baryonic matter (BM) over-density,
$\varrho_{\rm BM}$ is the cell's baryonic mass density, $a_{\rm fil}$ is a proper threshold, whose value has been set to 1 after extensive experimentation \citep{gh15}, and $\varrho_0$ is the critical density.
The Connected Components algorithm is then used to combine cells belonging to distinct (i.e. non intersecting) filaments, assigning to each filament an Id (an integer number) and tagging each cell belonging to a filament with that Id. The same identification criterion, but using the total (baryonic matter + dark matter) density, is used to extract the galactic halos (see Section \ref{sec:galaxies}).

\item {\it Shape selection.}
We use the following two-stage cleaning procedure to remove possible round-shaped, isolated
structures that could be still present in the identified objects.
First, we retain only elongated objects, with:
\begin{equation}
{\rm MAX}(r_{xy}, r_{xz}, r_{yz}) > \alpha ,
\end{equation}
where $r_{ab}$ is the ratio between axes $a$ and $b$ of the bounding box and $\alpha$ was set to 2.
Second, among the remaining objects, we accept as filaments all those whose volume is smaller than a given fraction of the corresponding bounding box volume.

\end{enumerate}

The resulting methodology depends on six parameters whose influence and optimal setting are extensively discussed in \citep{gh15}.

The whole procedure is implemented within the VisIt framework, which provides all the required
numerical methods and supports big data processing through a combination of optimised algorithms, the
exploitation of high-performance computing architectures, in particular
through parallel processing, and the support of client-server capabilities.

\begin{table}
\caption{Number of filaments identified in the different simulation at all masses (column 2) and with $M_{\rm BM} \geq 5 \cdot 10^{12} M_{\odot}$ (column 3).}
\centering \tabcolsep 5pt
\begin{tabular}{l|c|c|}
ID    & $N_{fil}$ & $N_{fil,12}$\\  
\hline
P(Baseline)&740 & 169\\
Z2&         689 & 174\\
DYN5&       876 & 176\\
DYN7&       851 & 173\\
DYN8&       927 & 176\\
CSF2&       862 & 174\\
CSFB11&     891 & 172\\
CSF5&       822 & 155\\
CSFBH3&     862 & 175\\
CSFBH5&     834 & 162\\
CSFBH2&     819 & 155\\

\end{tabular}
\label{tab:filnumber}
\end{table}

\begin{table}
\caption{Best-fit parameters of the mass-length relation,  $log_{\rm 10}(L)=\alpha + \beta ~ log_{\rm 10}(M_{\rm BM})$, for our samples of simulated filaments. The fit is calculated considering only filaments with $M_{\rm BM} \geq 5 \cdot 10^{12} M_{\odot}$.}
\centering \tabcolsep 5pt
\begin{tabular}{|c|c|c}
   \hline
   run &
   \multicolumn{2}{c}{Mass-Length}   \\  \hline
   ID & $\alpha$ & $\beta$ \\ \hline
  P(Baseline)&     -5.657$\pm$      0.703 &     0.521$\pm$    0.054\\
Z2&     -5.671$\pm$      0.720 &     0.521$\pm$    0.055\\
DYN5&     -5.145$\pm$      0.701 &     0.485$\pm$    0.054\\
DYN7&     -5.250$\pm$      0.711 &     0.492$\pm$    0.054\\
DYN8&     -6.070$\pm$      0.689 &     0.559$\pm$    0.053\\
CSF2&     -4.786$\pm$      0.669 &     0.459$\pm$    0.051\\
CSFB11&     -4.887$\pm$      0.642 &     0.467$\pm$    0.049\\
CSF5&     -5.372$\pm$      0.606 &     0.504$\pm$    0.046\\
CSFBH3&     -5.416$\pm$      0.604 &     0.508$\pm$    0.046\\
CSFBH5&     -4.735$\pm$      0.567 &     0.455$\pm$    0.043\\
CSFBH2&     -4.815$\pm$      0.709 &     0.460$\pm$    0.054\\
\end{tabular}
\label{tab:fitML}
\end{table}

\begin{table}
\caption{Best-fit parameters of the mass-temperature relation, $log_{\rm 10}(T)=\alpha + \beta ~ log_{\rm 10}(M_{\rm BM})$, for the same objects of Table \ref{tab:fitML}.}
\centering \tabcolsep 5pt
\begin{tabular}{|c|c|c}
   \hline
   run &
   \multicolumn{2}{c}{Mass-Temperature}   \\  \hline
   ID & $\alpha$ & $\beta$ \\ \hline
P(Baseline)&      2.309$\pm$      0.691 &     0.315$\pm$    0.053\\
Z2&      2.521$\pm$      0.726 &     0.301$\pm$    0.056\\
DYN5&      2.397$\pm$      0.814 &     0.309$\pm$    0.062\\
DYN7&      2.478$\pm$      0.800 &     0.304$\pm$    0.061\\
DYN8&      3.123$\pm$      0.799 &     0.254$\pm$    0.061\\
CSF2&      2.219$\pm$      0.799 &     0.322$\pm$    0.061\\
CSFB11&      1.857$\pm$      0.807 &     0.349$\pm$    0.062\\
CSF5&      1.744$\pm$      0.903 &     0.363$\pm$    0.069\\
CSFBH3&      1.929$\pm$      0.883 &     0.349$\pm$    0.068\\
CSFBH5&      1.695$\pm$      0.876 &     0.369$\pm$    0.067\\
CSFBH2&      1.787$\pm$      0.767 &     0.355$\pm$    0.059\\
\end{tabular}
\label{tab:fitMT}
\end{table}

\begin{table}
\caption{Best-fit parameters of the mass-cosmic baryon fraction relation,  $log_{\rm 10}(f_{\rm b}/f_{\rm b,cosm})=\alpha + \beta ~ log_{\rm 10}(M_{\rm BM})$, for the same objects of Table \ref{tab:fitML}.}
\centering \tabcolsep 5pt
\begin{tabular}{|c|c|c}
   \hline
   run &
   \multicolumn{2}{c}{Mass-Baryon Fraction}   \\  \hline
   ID & $\alpha$ & $\beta$ \\ \hline
P(Baseline)&    -0.322$\pm$      0.143 &    0.026$\pm$    0.011\\
Z2&    -0.314$\pm$      0.118 &    0.025$\pm$   0.009\\
DYN5&    -0.329$\pm$      0.223 &    0.0269$\pm$    0.017\\
DYN7&    -0.328$\pm$      0.234 &    0.026$\pm$    0.018\\
DYN8&    -0.305$\pm$      0.133 &    0.024$\pm$    0.010\\
CSF2&    -0.178$\pm$      0.173 &    0.013$\pm$    0.013\\
CSFB11&    -0.170$\pm$      0.163 &    0.013$\pm$    0.012\\
CSF5&  -0.003$\pm$      0.165 &  -0.004$\pm$    0.012\\
CSFBH3&   -0.046$\pm$      0.199 &  -0.002$\pm$    0.015\\
CSFBH5&    -0.153$\pm$      0.262 &   0.005$\pm$    0.020\\
CSFBH2&    -0.208$\pm$      0.144 &    0.016$\pm$    0.011\\
\end{tabular}
\label{tab:fitMFb}
\end{table}

\begin{table}
\caption{Best-fit parameters of the mass-mean magnetic field relation,  $log_{\rm 10}\langle |B|\rangle =\alpha+\beta ~log_{\rm 10}(M_{\rm BM})$, for the same objects of Table \ref{tab:fitML}.}
\centering \tabcolsep 5pt
\begin{tabular}{|c|c|c}
   \hline
   run &
   \multicolumn{2}{c}{Mass-$|\langle B \rangle|$}   \\  \hline
   ID & $\alpha$ & $\beta$ \\ \hline
P(Baseline)&     -1.822$\pm$      0.734 &    0.031$\pm$    0.0564\\
Z2&     -1.881$\pm$      0.642 &    0.044$\pm$    0.049\\
DYN5&     -4.348$\pm$      0.705 &     0.210$\pm$    0.054\\
DYN7&     -4.187$\pm$      0.694 &     0.203$\pm$    0.053\\
DYN8&     -4.616$\pm$      0.827 &     0.188$\pm$    0.063\\
CSF2&     -3.638$\pm$       1.516 &   -0.016$\pm$     0.116\\
CSFB11&     -2.690$\pm$      0.795 &   -0.028$\pm$    0.061\\
CSF5&     -4.004$\pm$       1.158 &     0.169$\pm$    0.089\\
CSFBH3&     -3.110$\pm$       1.199 &    0.073$\pm$    0.092\\
CSFBH5&     -2.278$\pm$       1.535 &  -0.003$\pm$     0.118\\
CSFBH2&     -3.654$\pm$       1.571 &    0.026$\pm$     0.121\\
\end{tabular}
\label{tab:fitM_Bmean}
\end{table}

\begin{table}
\caption{Best-fit parameters of the mass-maximum magnetic field relation,  $log_{\rm 10}(B_{\rm max})=\alpha+\beta ~log_{\rm 10}( M_{\rm BM})$, for the same objects of Table \ref{tab:fitML}.}
\centering \tabcolsep 5pt
\begin{tabular}{|c|c|c}
   \hline
   run &
   \multicolumn{2}{c}{Mass-$B_{\rm max}$}   \\  \hline
   ID & $\alpha$ & $\beta$ \\ \hline
P(Baseline)&     -3.177$\pm$      0.667 &     0.201$\pm$    0.051\\
Z2&     -3.730$\pm$      0.709 &     0.249$\pm$    0.054\\
DYN5&     -5.868$\pm$      0.938 &     0.395$\pm$    0.072\\
DYN7&     -5.835$\pm$      0.926 &     0.397$\pm$    0.071\\
DYN8&     -7.690$\pm$      0.985 &     0.497$\pm$    0.076\\
CSF2&     -4.825$\pm$       1.036 &     0.185$\pm$    0.079\\
CSFB11&     -3.723$\pm$      0.978 &     0.142$\pm$    0.075\\
CSF5&     -3.725$\pm$      0.676 &     0.256$\pm$    0.052\\
CSFBH3&     -3.932$\pm$      0.865 &     0.245$\pm$    0.066\\
CSFBH5&     -3.108$\pm$      0.720 &     0.175$\pm$    0.055\\
CSFBH2&     -4.593$\pm$       1.040 &     0.207$\pm$    0.079\\
\end{tabular}
\label{tab:fitM_Bmax}
\end{table}

\section{Properties of the Filaments}
\label{sec:results}

In Figure \ref{fig:mapT} we give the visual impression of the 3-dimensional distribution of gas density, temperature and magnetic field magnitude in a  $10^3 ~\rm  Mpc^3$ sub-volume for three reference runs in our sample: the baseline non-radiative primordial run, the sub-grid dynamo model (DYN5) and the cooling+feedback run with astrophysical seeding of magnetic fields (CSFBH3).
The overall morphology of thermal gas properties on $\geq \rm Mpc$ scales is very similar in all runs, even if the gas distribution in the radiative run is significantly clumpier. As generally found also in the outskirts of galaxy clusters \citep[e.g.][]{nala11,va13clump}, also in this case the combination of radiative cooling and feedback heating enhances the level of gas clumpiness ($(\langle \rho^2 \rangle)^{0.5}/\langle \rho \rangle$), at least sufficiently away from the active sources of feedback.
The gas density in galaxy groups in the field is generally smoother in the radiative run while their core is hotter, as an effect of ongoing thermal feedback from the SMBH active in it. On the other hand, the distribution of magnetic fields in the volume is dramatically different at all scales, becoming progressively less diffuse and more concentrated onto halos going from the primordial scenario to the dynamo to the astrophysical one. While in the first case filaments have a floor magnetisation of $\sim 1- 10 ~\rm nG $ resulting from the compression of the primordial magnetic field lines, in the dynamo case they get significantly magnetised only where the flow is turbulent enough to boost a dynamo, which happens only in the proximity of substructures within filaments. Finally, in the astrophysical scenario the activity of star-formation and SMBH feedback could only magnetise "bubbles" around halos, leaving most of the filaments volume un-magnetized (down to the $\sim 10^{-18} ~\rm G$ floor value assumed here).  
The variety of possible distributions of magnetic fields across 8 different simulations is shown in Figure  \ref{fig:mapB}.  In galaxy clusters, most of the models generate magnetic field whose magnitude is of the order observed values ($\sim 0.1-1 ~\rm \mu G$). In the vast majority of the cosmic web, instead, the differences in magnitude and volume filling factors of magnetic fields can be large. We will analyse such differences in detail in the following sections.

Table \ref{tab:filnumber} shows the number of filaments identified in the different models using the procedure described in Section \ref{sec:visit}. Considering all masses, the number of detected filaments is highly variable, due to the influence of small objects strongly affected by the local physics and the resolution effects. Selecting objects with $M_{\rm BM} \geq 5 \cdot 10^{12} M_{\odot}$ the number counts become homogeneous, with the exception of the models with strong feedback. The feedback, contrasting the infall of matter into the forming structures, leads to a decrease of the total number of filaments.

\begin{figure*}
\includegraphics[width=0.49\textwidth]{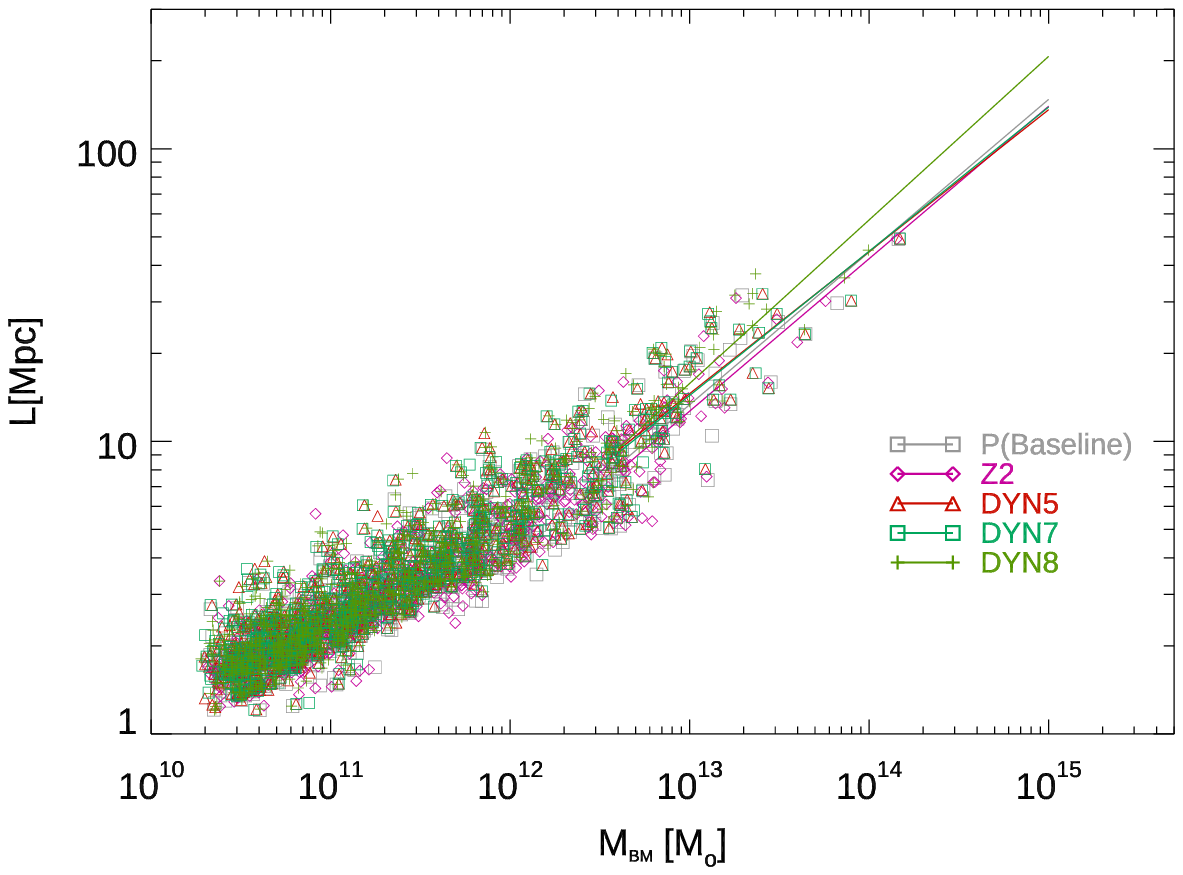}
\includegraphics[width=0.49\textwidth]{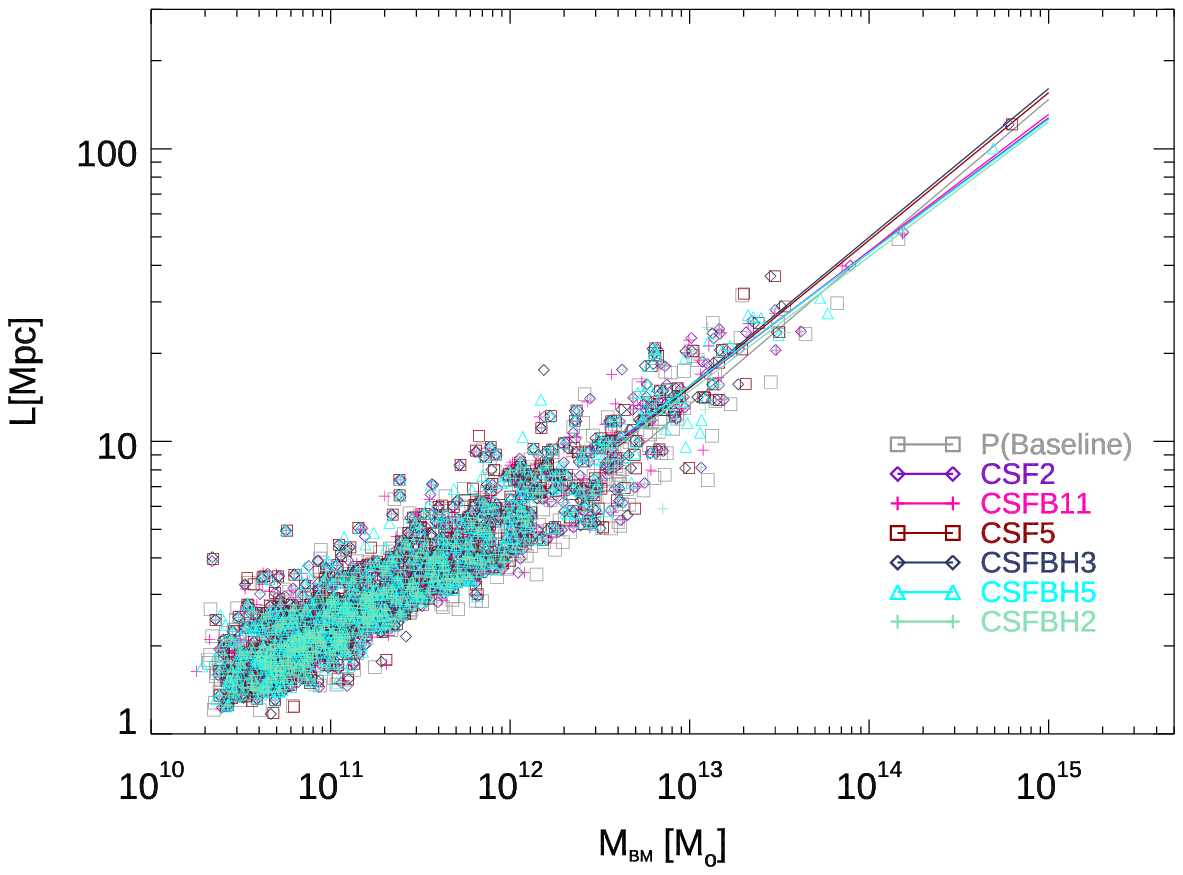}
\includegraphics[width=0.49\textwidth]{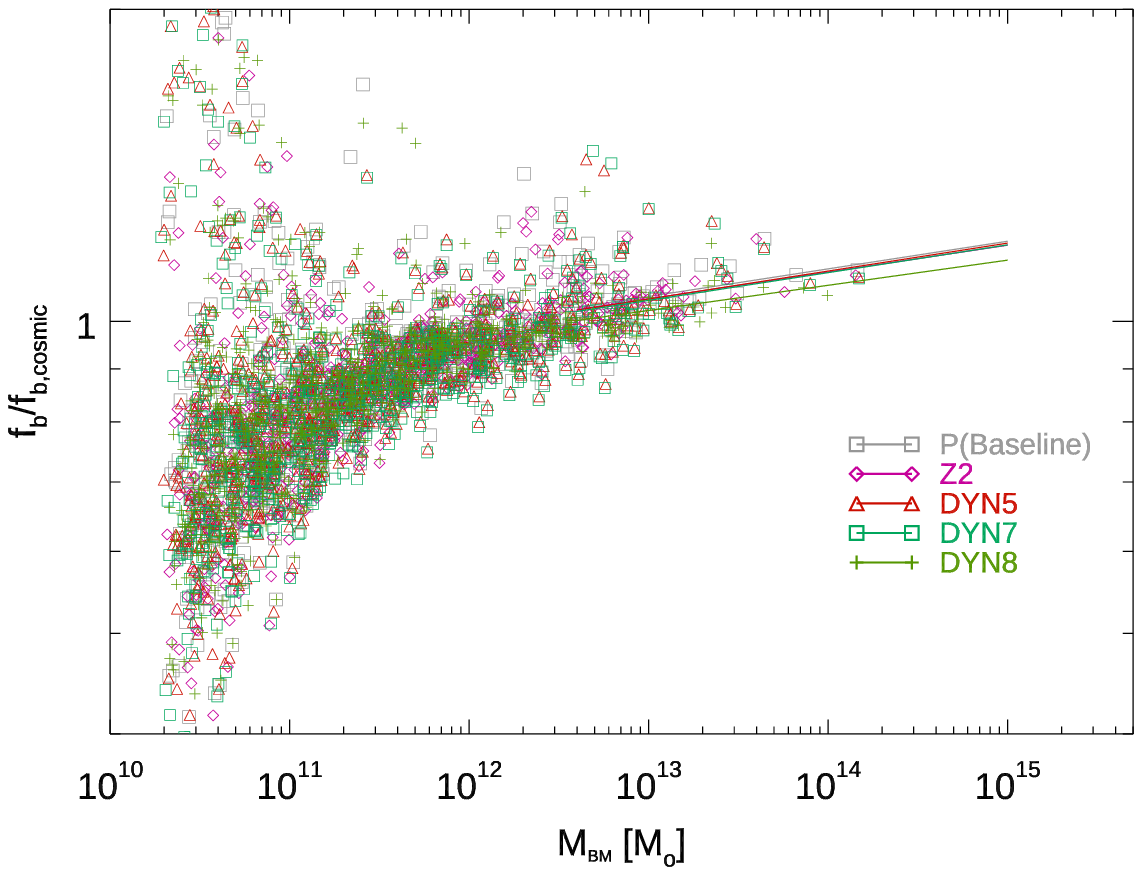}
\includegraphics[width=0.49\textwidth]{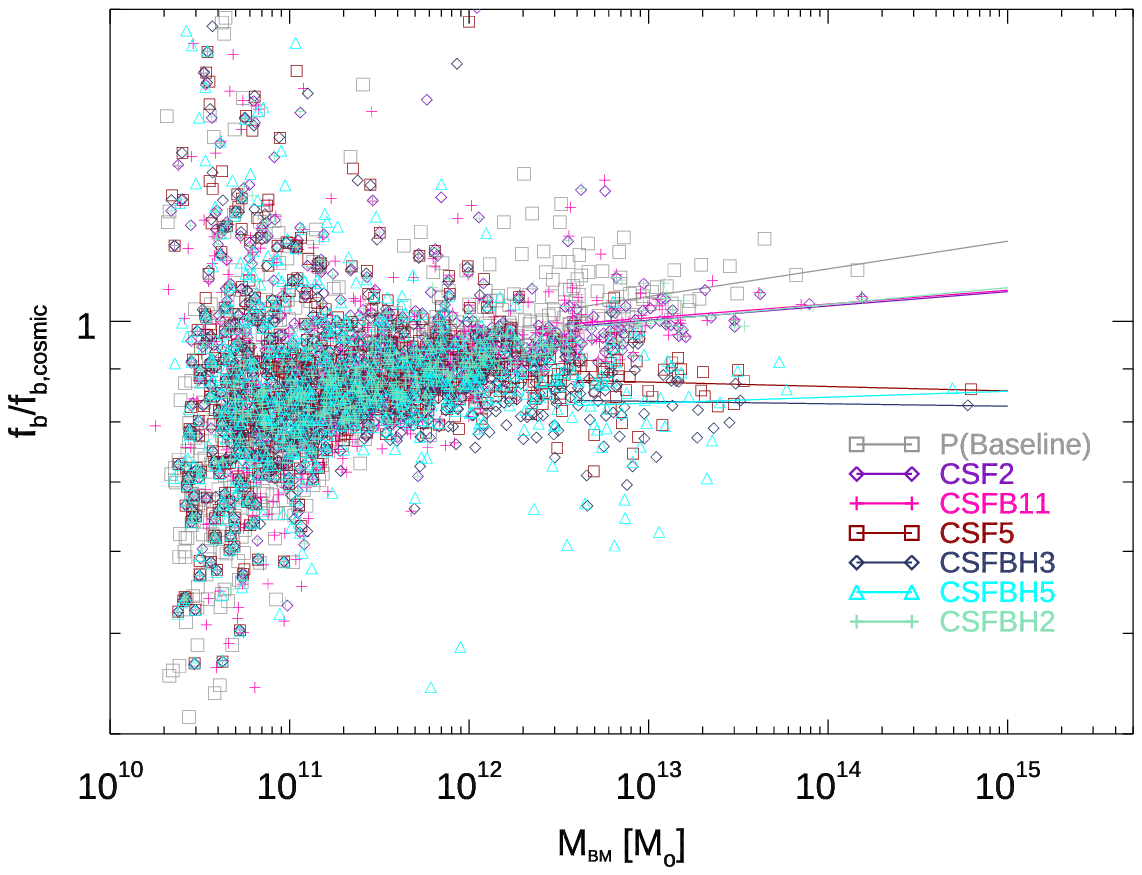}
\includegraphics[width=0.49\textwidth]{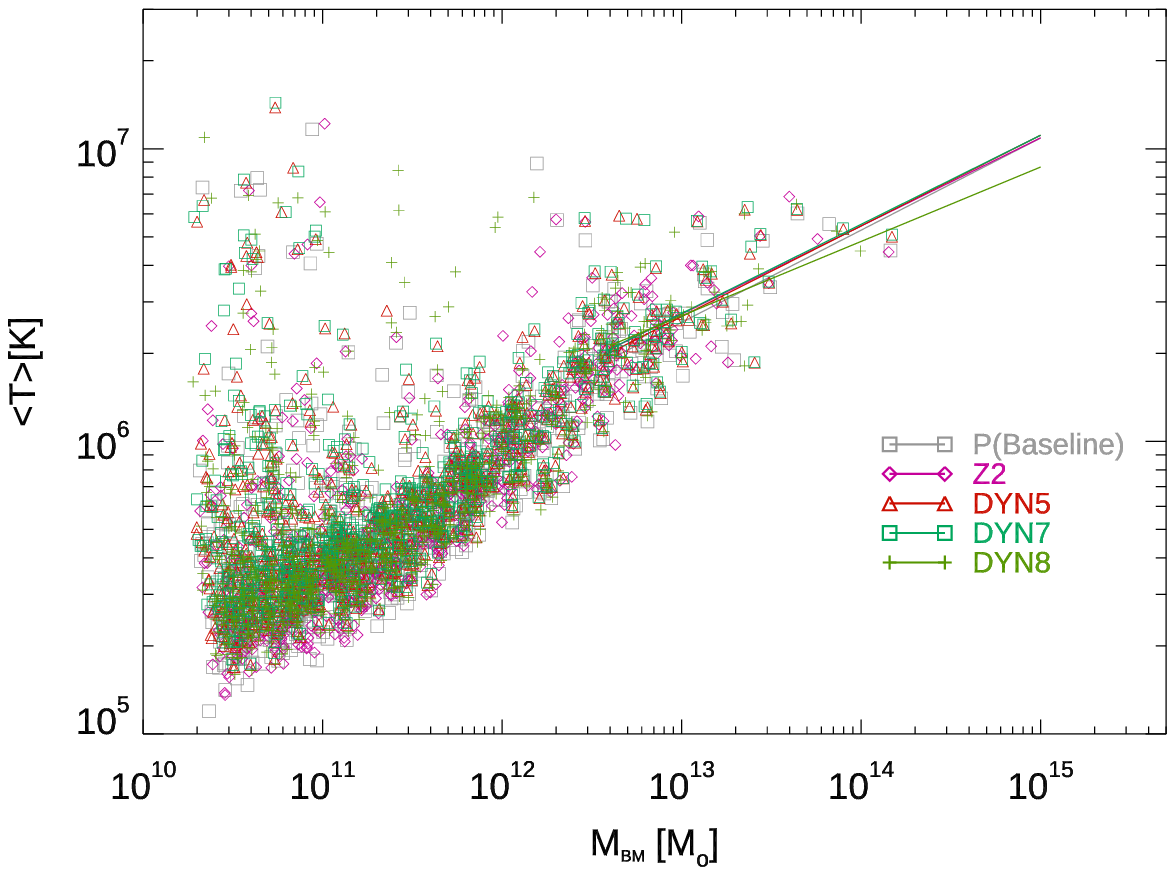}
\includegraphics[width=0.49\textwidth]{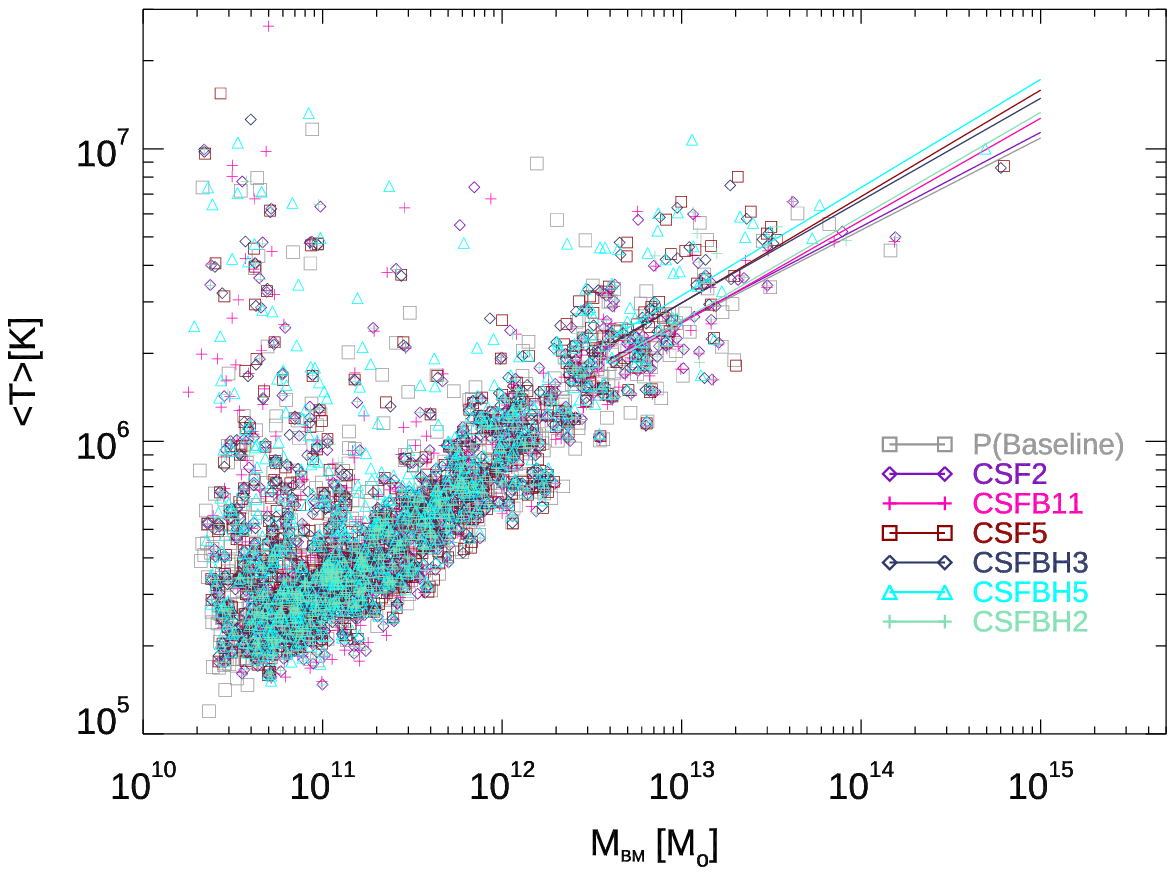}
\caption{Scaling relations between the total gas mass in filaments and: a) the filament length (top); the baryon fraction (centre) and the mass-weighted gas temperature (bottom) for all our identified objects at $z=0$.}
\label{fig:ml}
\end{figure*}   

\begin{figure}
\includegraphics[width=0.45\textwidth]{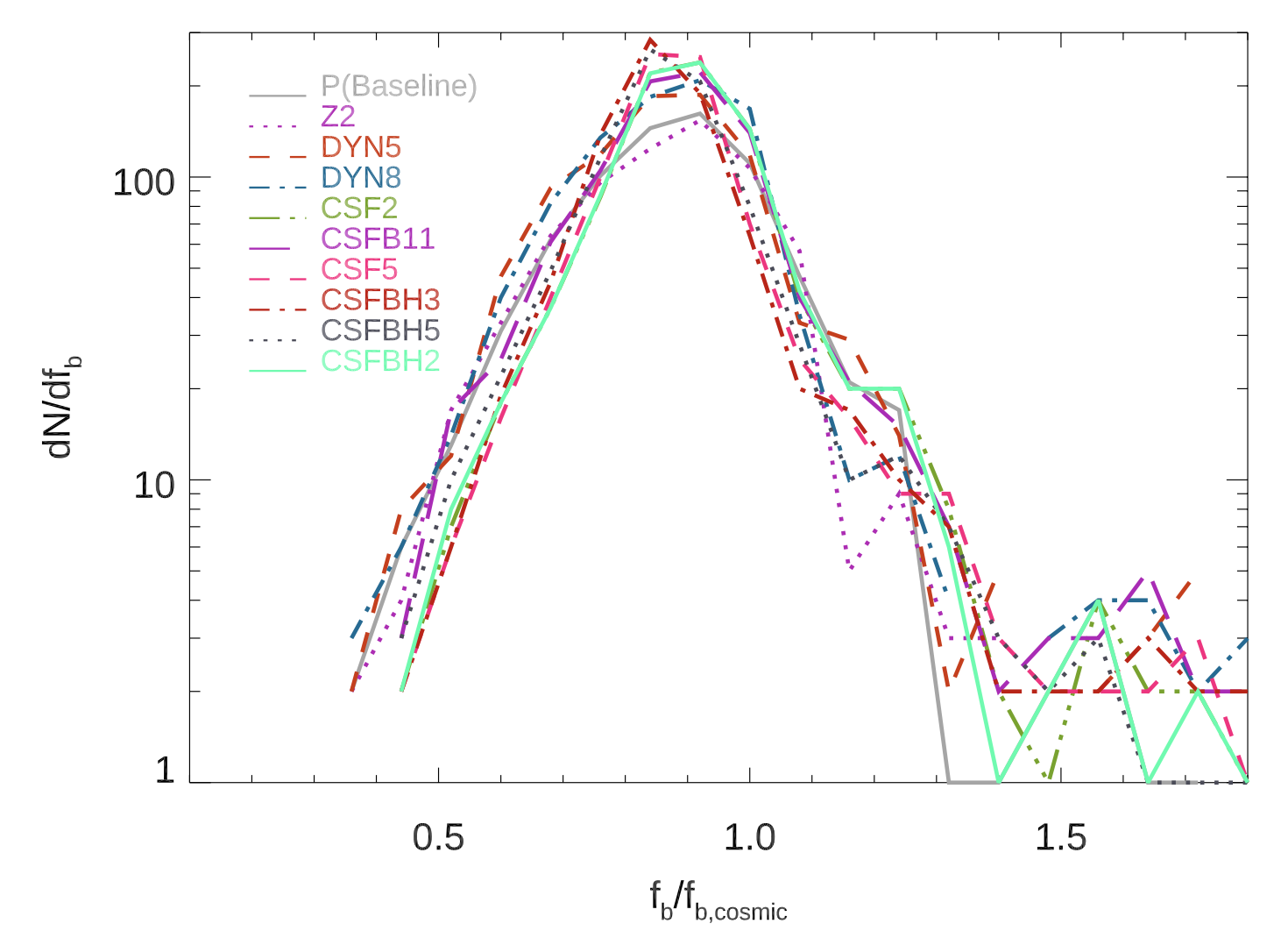}
\caption{Number distribution function of filaments as a function of their enclosed baryon fraction at $z=0$.}
\label{fig:dist_fb}
\end{figure}  

\begin{figure*}
\includegraphics[width=0.49\textwidth]{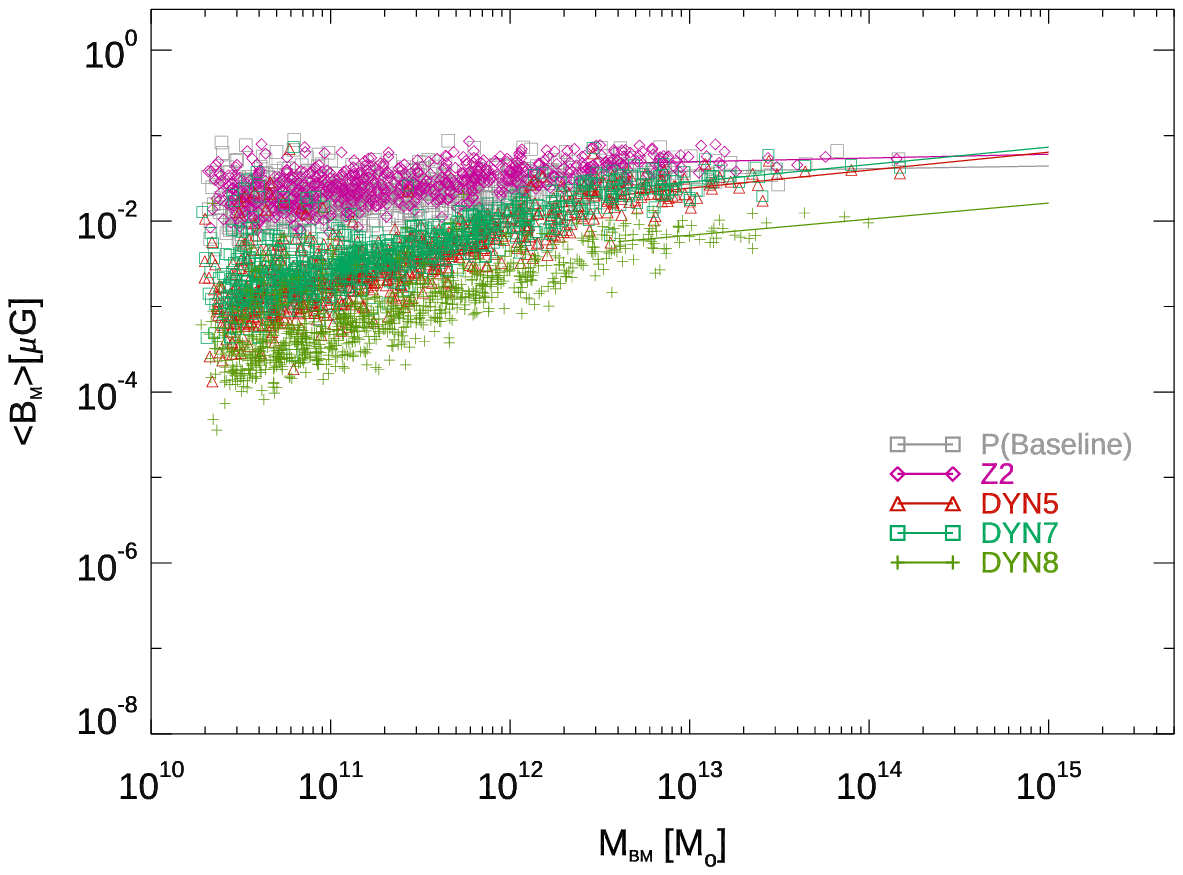}
\includegraphics[width=0.49\textwidth]{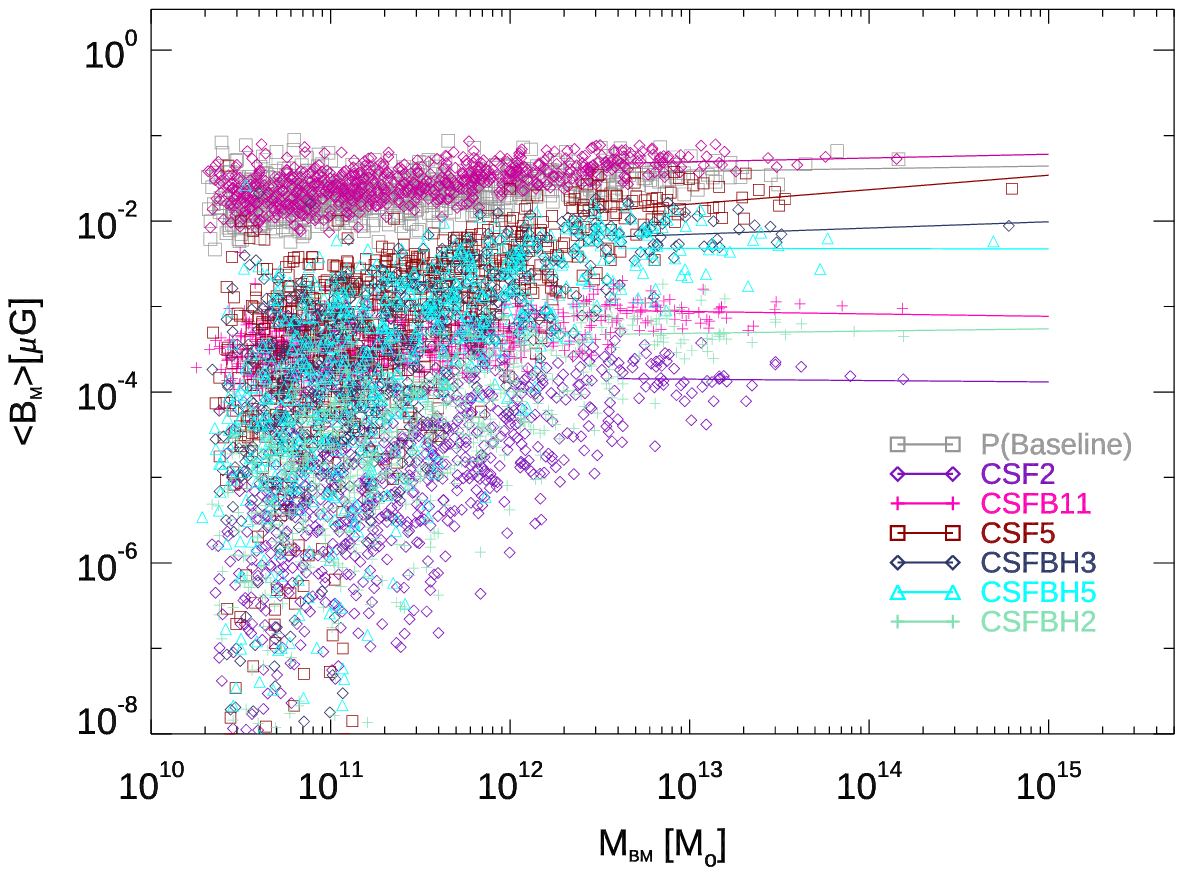}
\includegraphics[width=0.49\textwidth]{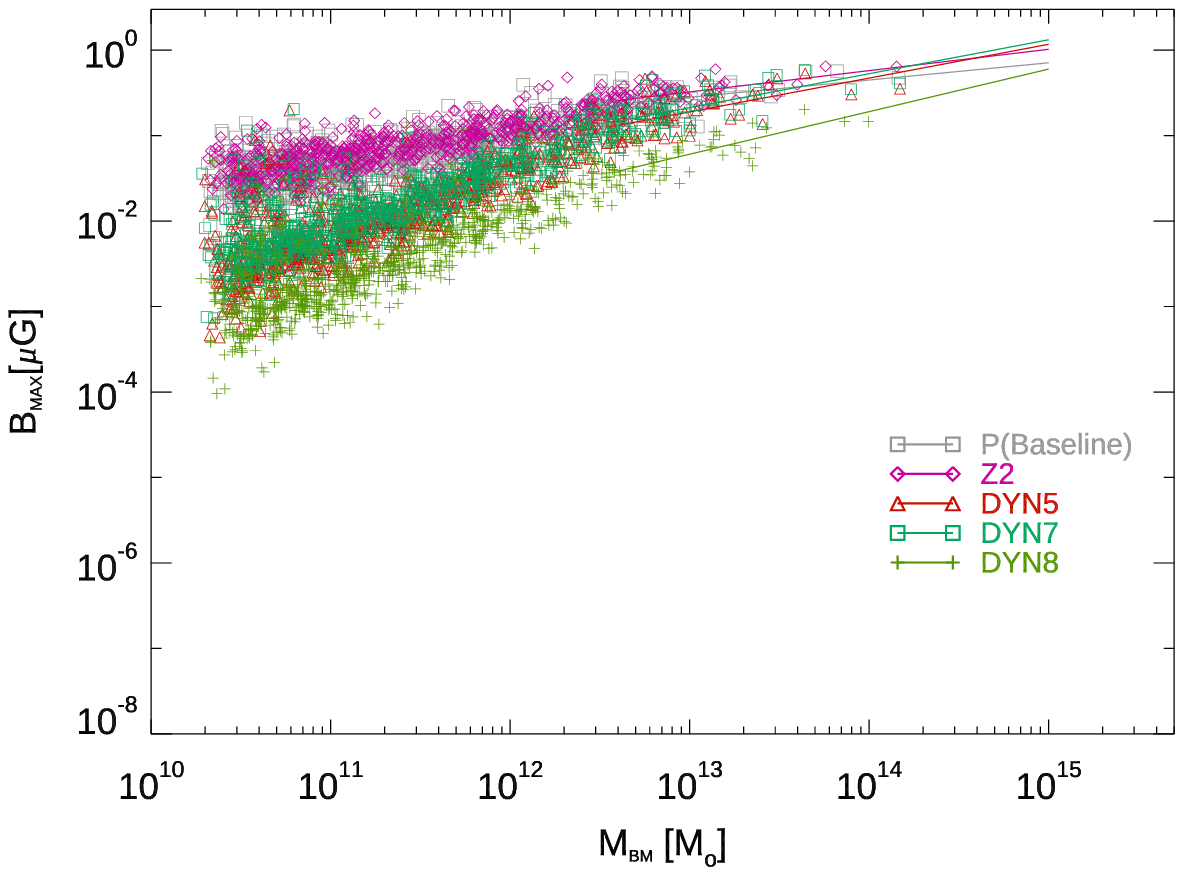}
\includegraphics[width=0.49\textwidth]{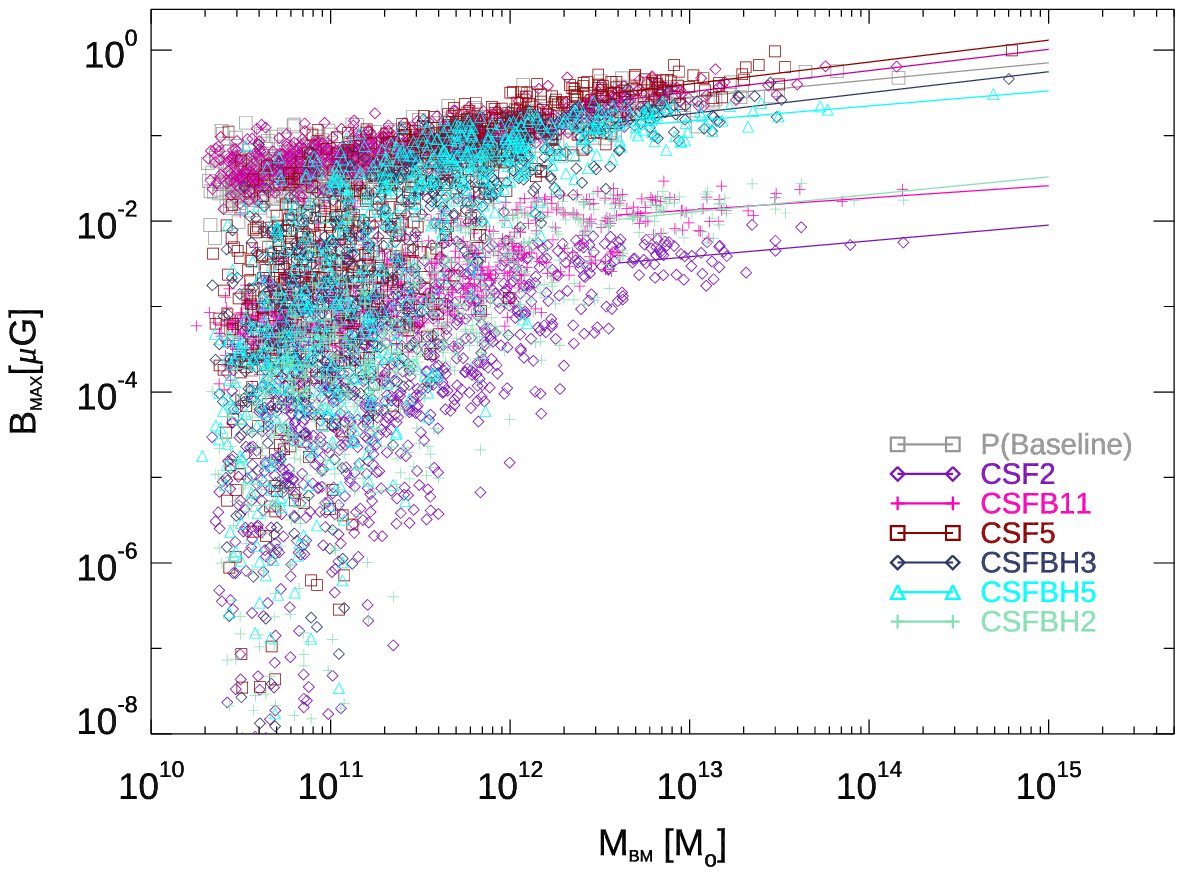}
\caption{Scaling relations between the total gas mass in filaments and the mean mass weighted magnetic field strength (top) or the maximum magnetic field strength (bottom) for all our identified objects at $z=0$. }
\label{fig:mb}
\end{figure*}   

\begin{figure}
\includegraphics[width=0.45\textwidth]{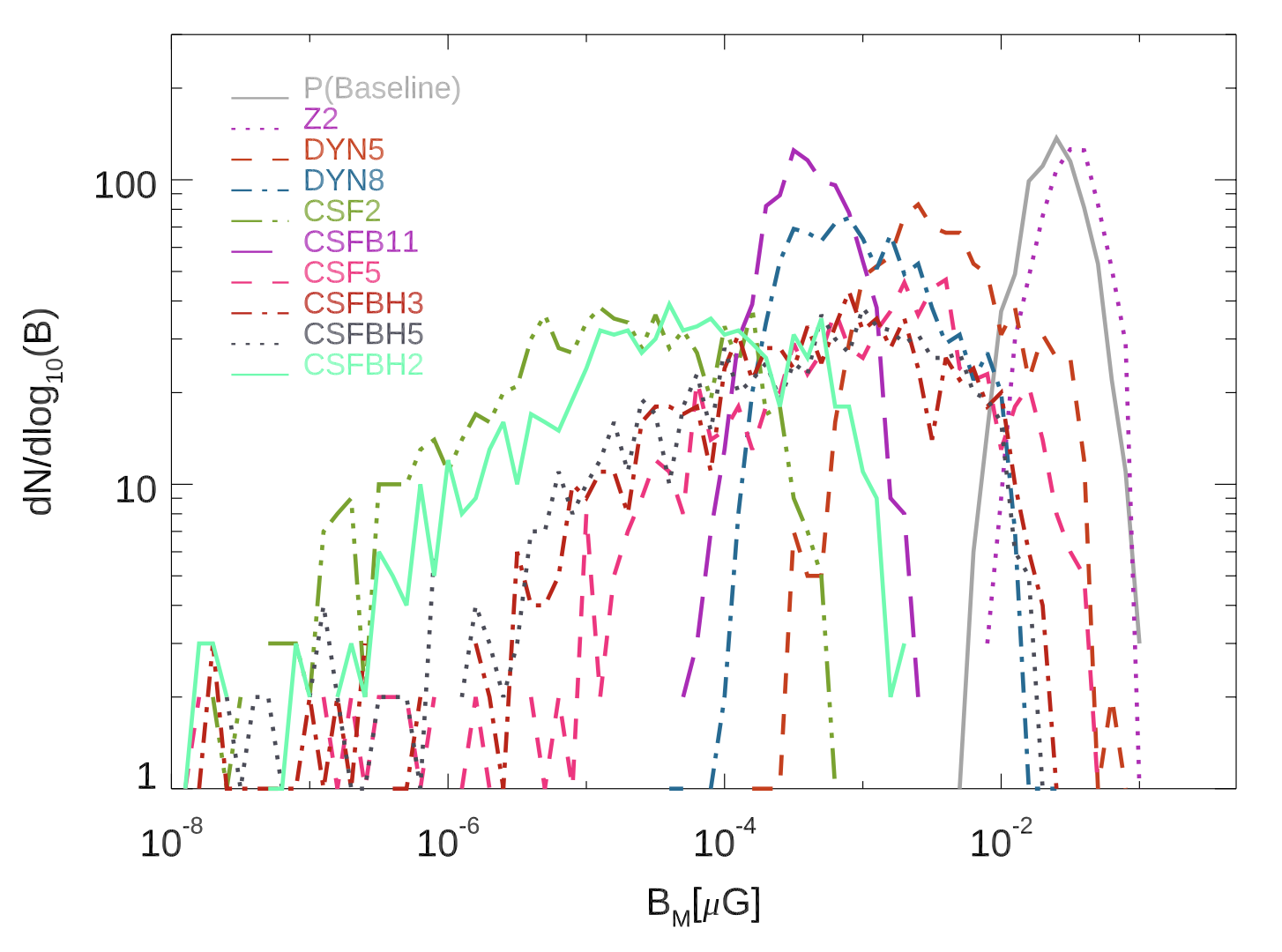}
\caption{Number distribution function of filaments as a function of their average magnetic field at $z=0$.}
\label{fig:dist_B}
\end{figure} 
\subsection{Statistical Properties}

Similar to our previous works \citep[][]{gh15,gh16},
we studied the scaling relations of global quantities of the filaments in our sample as a function of physical variations. For each filament we have calculated the total gas mass, the filament length, approximated as the diagonal of the bounding box enclosing the filament as determined by our filament finder, the average baryon fraction within each filament (normalised to the cosmic baryon fraction) and the mass-weighted mean gas temperature and magnetic field for all our identified objects at $z=0$. All quantities are calculated considering the cells selected by our finder as part of the filament. Cells at the boundaries are described as variable number of faces polyhedra, accurately adapting to the complex geometry of the filament.

Figure \ref{fig:ml} gives the scaling relations between the total gas mass in filaments, the filament length,  the baryon fraction within each filament (normalised to the cosmic baryon fraction) and the mass-weighted mean gas temperature for all our identified objects at $z=0$. We also give in Tables \ref{tab:fitML}-\ref{tab:fitMFb} the best fit relations between the baryonic mass \footnote{As in our previous papers \citep[][]{gh15,gh16}, for the filaments we use the {\it baryonic} rather than the {\it dark matter} (or total) mass for our scaling relations, the latter being affected by non-negligible numerical errors in the lowest density regions, due to its particle-based representation.} of filaments and length, temperature and baryon fraction, using a simple $log_{\rm 10}(Y) = \alpha + \beta~ log_{\rm 10}(X)$ relation, where $X=M_{\rm BM}$. The same fitting relation, for different quantities $X$ and $Y$, will be adopted throughout the paper, fitting parameters being always referred as $\alpha$ and $\beta$.
 
The same relations have been investigated also in \citet{gh16}. We now extend the analysis investigating the role played by different thermal and non thermal physical processes on the population of cosmic filaments. 
The lowest mass filaments in our distributions  ($\leq 5 \cdot 10^{12} M_{\odot}$)  typically present a large scatter in all scaling relations, which makes our fitting procedure unstable. The origin of this scatter is twofold. On one hand, the smallest filaments can be affected by numerical effects, i.e. the limited numerical resolution makes it difficult to properly resolve the properties of the smallest filaments, their size getting comparable to the numerical resolution. For these small objects, the gravitational potential is under-resolved, numerical pressure affects the force balance and the finite DM mass resolution can introduce spurious graininess in the mass distribution (and hence in the gravitational potential). At the same time, small-size filaments are more easily affected by the random effect of cooling and feedback from halos around and within them, which introduce an additional strong scatter in their thermodynamic properties. For example, run CSFBH5, which presents overall the strongest effect of feedback from star formation, due to the low threshold density used to form stars, is found to produce the largest scatter in the temperature, baryon fraction and magnetic field strength for   $\leq 5 \cdot 10^{12} M_{\odot}$ filaments. 
This second mechanism has a physical origin and changes with the model. The related dependence of the level of scatter from the baryonic mass, different for each model, may in principle offer an (observationally challenging) way of testing magnetic field scenarios in the radio band, as we discuss in Sec.~\ref{sec:discussion}. Here, however, we want to extract robust estimates of the scaling relations of our objects, in a regime in which the random impact of single stellar and AGN activity is reduced, and therefore in the following we discuss our best-fit estimates restricted to $M_{\rm BM} \geq 5 \cdot 10^{12} M_{\odot}$ filaments.

Our most important findings, shown in Figure \ref{fig:mb} and \ref{fig:dist_B} and listed in Tables \ref{tab:fitML} to \ref{tab:fitM_Bmax}, can be summarised as follows. 

First, the impact of  model variations on the geometric and baryonic properties of filaments is overall modest, and the filaments population looks remarkably similar across $4-5$ orders of magnitude in mass, despite the variety of investigated physical setups. The relation between the length of the filaments and their baryonic mass is always close to $L \propto M_{\rm BM}^{0.5}$, consistent with \citet{gh16}, while only modest differences in normalisation are found between models. As expected, the overall evolution of the large scale structure of the universe is guided by gravity and pressure forces, other local physical processes playing only a minor role. The different physical models analysed in these paper cannot be disentangled through the study of global geometric or thermodynamics properties.\\

Second, the mean gas temperature of all our objects falls within the expected range for the WHIM: $10^5 ~\rm K\le T \le 10^{7} ~\rm K$. 
In models with cooling and feedback, the temperature of filaments with mass $M_{\rm BM} \approx 10^{14}M_{\odot}$ is a factor $\sim 2$ higher than in non-radiative runs. At the low mass end of the distribution ($10^{10}-10^{11} M_{\odot}$), cooling and feedback models are a factor $\sim 2$ colder. This leads to a relation between gas temperature and baryonic mass always steeper for cooling and feedback models compared to non radiative runs (e.g. $T \propto M_{\rm BM}^{0.35}$ vs $T \propto M_{\rm BM}^{0.3}$).
Moreover, we measure a significant scatter in the temperature values towards the lowest masses as feedback events from neighbouring AGNs or star forming regions (either within filaments or next to them) affect the thermodynamics of the WHIM, contributing a significant amount of energy per particle. The long-term effect of cooling is to trigger thermal and magnetic feedback, that can indeed partially affect the WHIM thermodynamics on scales of several $\sim \rm Mpc$. The relative trend of temperature in filaments simulated with radiative and non radiative setups is qualitatively similar to our earlier analysis in \citet{gh15}. A quantitative comparison is challenging due to the different resolution and feedback scheme (which is here more sophisticated). Comparing our non-radiative runs to those with the same mass and spatial resolution presented in in \citet{gh16}, we find that the temperature-mass relation for our baseline model is slightly flatter ($T \propto M_{\rm BM}^{0.3}$ vs. $T \propto M_{\rm BM}^{0.431}$). This is connected to the more restrictive mass range we considered here, $M_{\rm BM} \geq 5 \cdot 10^{12} M_{\odot}$, which minimises the large scatter produced by low mass objects, which are potentially the most affected by numerical effects due to the fixed spatial and force resolution. We refer to \citet{gh16} for further discussion on the influence of numerical resolution on the scaling relations.

Third, the baryon fraction enclosed in filaments displays in all models a flat relation with gas mass, i.e. $f_{\rm b}/f_{b,cosm}\propto M_{\rm BM}^{(-0.004 \div 0.03)} $   in the $ \geq 5 \cdot 10^{12} M_{\odot}$ mass range. The normalisation significantly varies across models, with the tendency of filaments in non-radiative to retain slightly more  baryons in filaments than the cosmic average. In the strongest feedback models, on the other hand, the average baryon fraction drops to $\sim 80-90\%$ of the cosmic mean. This can be ascribed to the combined effect of cooling and feedback, that on one hand condenses more gas from filaments into halos and away from the WHIM phase, and on the other hand can push baryons even outside filaments, following strong feedback events \citep[][]{2006ApJ...650..560C}. Once more, at the low mass end of our distribution the scatter is large, as already found in  \citet{gh16}, again due to a combination of the more extreme effect of AGN feedback onto small mass systems\citep[e.g.][]{2006MNRAS.373.1265O}, as well as to the limited mass resolution for dark matter. The number distribution of baryon fraction across our entire sample is shown in Figure \ref{fig:dist_fb}. In general, the scatter at high baryon fraction, typically found in runs with stronger feedback effects, suggest that large variations due to the AGN-feedback cycle are an important physical mechanism to vary the baryonic (and possibly chemical) content of cosmic filaments across the cosmic volume \citep[][]{2006ApJ...650..560C,2016MNRAS.457.3024H,2018arXiv181001883M}. \\

Finally, the magnetic properties of the WHIM in filaments across our sample and for the different models as a function of the baryonic mass show the most striking differences.
Figures \ref{fig:mb} and \ref{fig:dist_B} present the scaling relations between the filament's baryonic mass and the corresponding maximum and mean (mass-weighted) magnetic field strength, as well as the number distribution of magnetic field strength in our runs.  
The mean magnetic field in filaments is roughly constant with mass in the high mass range (i.e. $\langle |B| \rangle \propto M_{\rm BM}^{-0.04 \div 0.2}$) with a normalisation dependent on the seeding model. This is because, without efficient small-scale dynamo amplification, the average distribution of magnetic fields in the volume of filaments mainly depends on the input seed field, modulo the compression factor \citep[][]{donn09,va17cqg}.  The normalisation in the primordial scenarios investigated here is $\sim 10^2-10^3$ higher than in the dynamo or astrophysical scenarios. However, the magnetisation of the largest filaments is similar, i.e. $\langle |B|\rangle\sim 50 \rm nG$ , as an effect of the $\propto M_{\rm BM}^{0.2}$ scaling relation. On the other hand, for $M_{\rm BM}\leq 10^{12} M_{\odot}$ objects, the differences are large. \\

The differences in magnetic fields grow in radiative runs. Such differences follow from the trend of mean magnetic fields and matter overdensity \citep[][]{va17cqg}:  the astrophysical seeding scales with the number of feedback sources in the environment (and the number of galaxies strongly depends on the size of filaments, e.g. \citealt{gh16}). Only for the highest efficiency scenario investigated here (CSF5, with an assumed  $\epsilon_{\rm SF,b}\sim 10\%$ magnetisation efficiency from star formation feedback) the magnetisation in $M_{\rm BM} \geq 5\cdot 10^{12} M_{\odot}$ filaments reaches the  $\langle |B|\rangle\sim 10 \rm nG$ level, while in the same mass range this is $\sim 0.1 ~\rm nG$ for the lowest efficiency models. 
We observe steeper relations and (obviously) larger differences in normalisation for the relation between the maximum magnetic field and the host baryonic mass of filaments, which is helpful to highlight the presence of rare sub-regions in filaments where small-scale dynamo amplification can be present at some degree. In this case we observe $B_{\rm max} \sim 0.5 ~\rm \mu G$ in the non-radiative scenarios, and up to $\sim  1 ~\rm \mu G$ in the strong feedback scenario CSF5. 

While the above trends apply to the statistical properties of the filaments population in the cosmic volume, in the following Section we will focus on their internal properties, in a range of scales that may in principle be probed by future X-ray or radio observations. 

\subsection{Resolved properties of filaments}
\label{subsec:profiles}

\begin{figure*}
\centering
\begin{tabular}{c|c}
\includegraphics[width=0.48\textwidth]{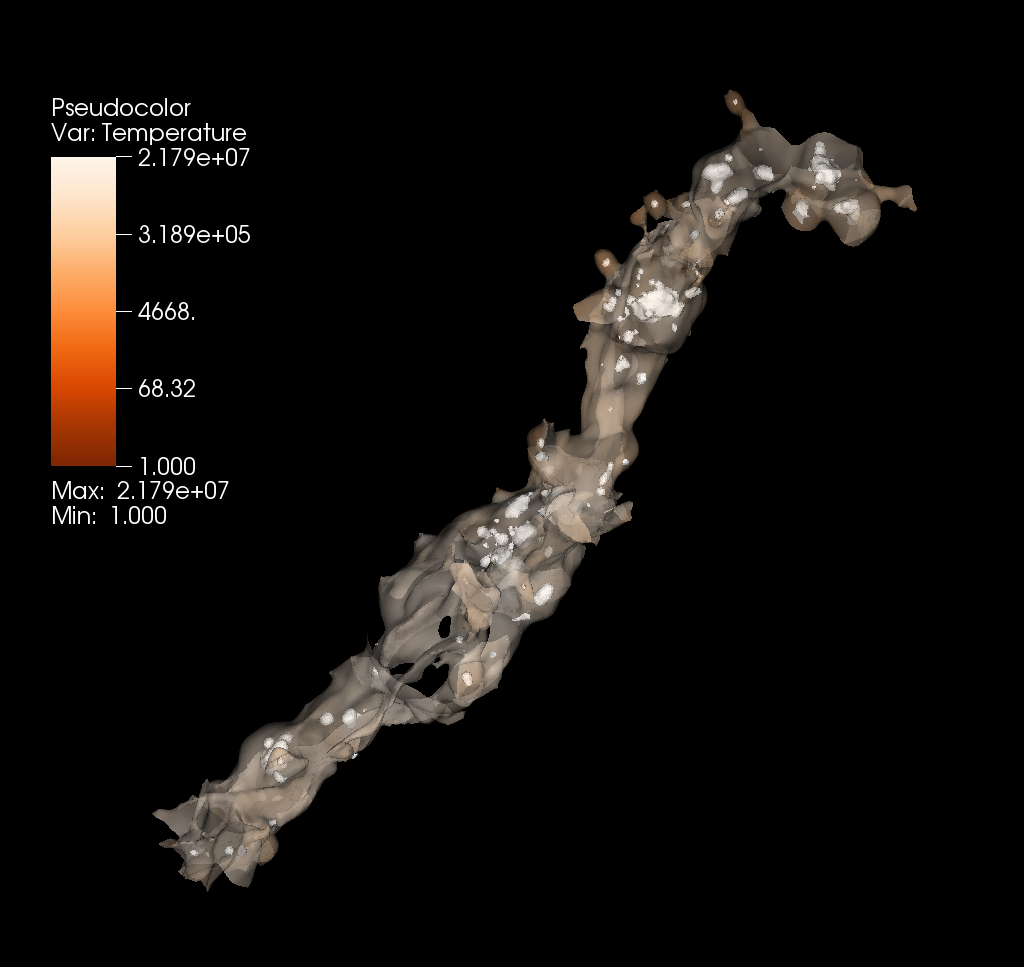} &
\includegraphics[width=0.48\textwidth]{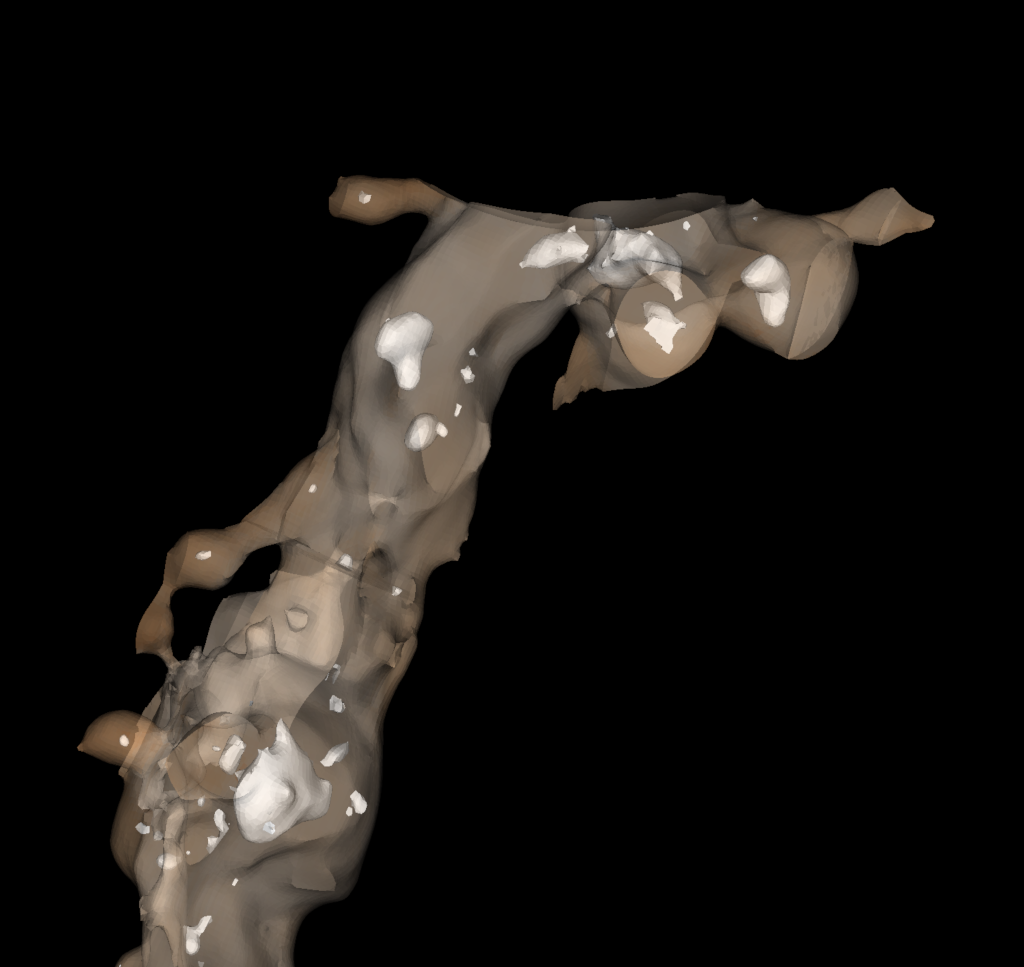}
\end{tabular}
\caption{Three dimensional rendering of the filament selected for the profiles analysis, coloured by the temperature. Galactic halos are also displayed (white structures). The filament is extracted from  Baseline simulation, but the same object is identified with a similar shape also in all the other models. The right panel zooms to the top part of the filament to show in more details the halos.}
\label{fig:filament00001}
\end{figure*}

Characterising the trends of thermodynamic and magnetic properties along the major and minor axis of filaments can yield interesting information on their internal structure, similar to what routinely done for galaxy clusters. However, filaments have complex shapes, often tilted, folded and with secondary ``branches'', that pose a number of practical problems while attempting to identify the ``spine'' of a filament, making it difficult to compute such profiles for a large number of objects \citep[e.g.][]{gh15}. 

We have initially focused on a $M_{BM} \sim 7.04 \times 10^{13} M_{\odot}$ filament that could be identified in all the different simulations, with a simple enough geometry that allows us to remove secondary branches. The resulting object, whose isodensity contours are shown in Figure  \ref{fig:filament00001}, is suitable for standard profile analysis assuming approximate cylindrical geometry. Fields of interest have been averaged within cylindrical shells at increasing distance from the filament's spine. The location of the spine is identified computing the centre of total mass of horizontal parallel planes crossing the filament, once this is rotated  along the z-axis: $X_s= [\sum x_i \cdot (m_{\rm gas}+m_{\rm DM})_i]/[(m_{\rm gas}+m_{\rm DM})_i]$ and $Y_s= [\sum y_i \cdot (m_{\rm gas}+m_{\rm DM})_i]/[(m_{\rm gas}+m_{\rm DM})_i]$ for the $X_s$ and $Y_s$ coordinates, respectively. 

\begin{figure*}
\includegraphics[width=0.45\textwidth]{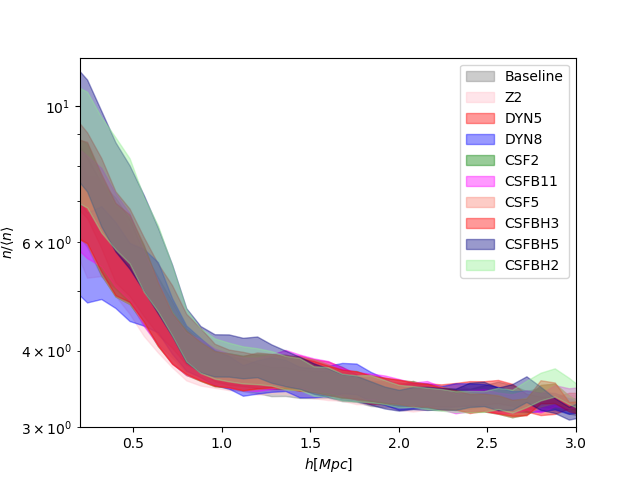}
\includegraphics[width=0.45\textwidth]{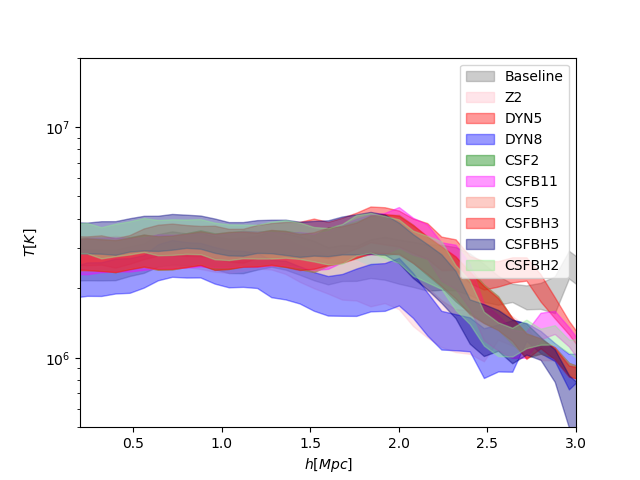}
\includegraphics[width=0.45\textwidth]{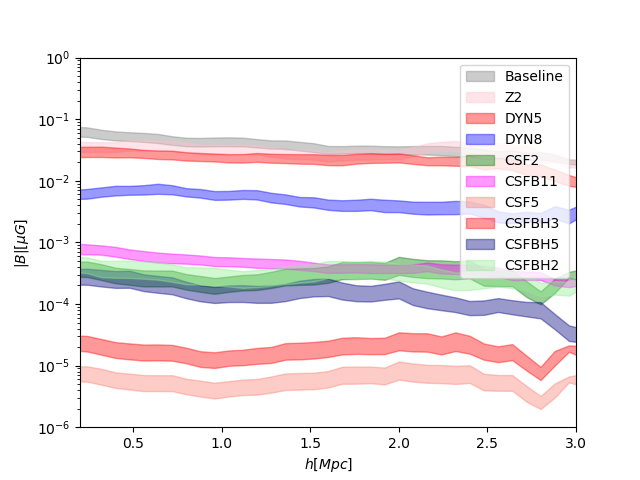}
\includegraphics[width=0.45\textwidth]{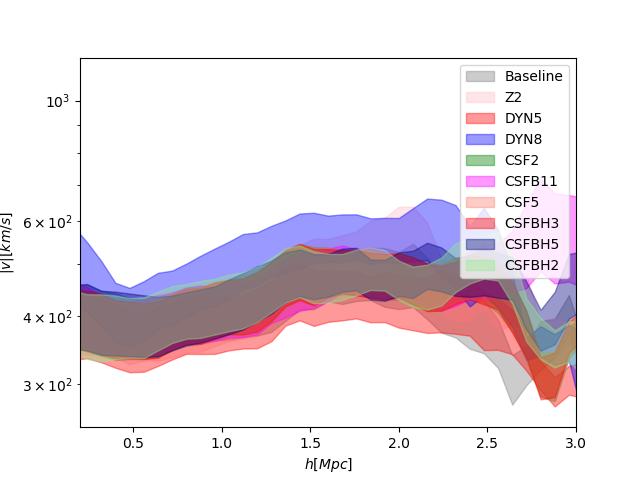}
\caption{Average trend of gas density (top left),  temperature (top right), magnetic field strength (bottom left) and velocity module (bottom right) as a function of the scale height from the filament spine, the filament in Figure \ref{fig:filament00001}, extracted from our simulated models. The shaded areas show the 33-66 \% range around the median relation.}
\label{fig:prof1}
\end{figure*}

The calculated profiles are presented in Figure \ref{fig:prof1} and are referred to the height, $h$, defined as the distance of the cylindrical shell from the filament's spine on the cutting planes. The cells associated to halos within the same volume have been removed (halos appears as white objects embedded in the filament in Figure \ref{fig:filament00001}). Small-scale differences in the profiles are likely an effect of the different time-scales of feedback processes in the different runs, and they are not considered significant.  However, we observe the tendency of the central gas density and temperature to be higher in cooling and feedback runs. This is a combined effect of the slightly enhanced compression of gas onto the filament's spine promoted by cooling with the non-gravitational heating effect by AGN and stellar winds from surrounding halos, which mostly affect the temperature profile at $h \geq 1.5 ~\rm Mpc$. The velocity profiles are remarkably similar across the different models, with a rather constant value comprised between $v_f\sim 400-600 ~\rm km/s$ at heights bigger than 1 Mpc, and a significant drop towards the filament's spine. The temperature is uniform from the spine to about 2 Mpc at a mean temperature around $10^6$K for most of the models. At this temperature, the sound speed in the WHIM is $c_s \sim 250-300 ~\rm km/s$, which means that gas motions internal to the filament are transonic, $v_f/c_s\sim 1.5-2$ . The magnetic field profile is extremely flat in all models, and ranges from $\sim 50 \rm nG$ in primordial and dynamo scenarios, to $\sim 0.01 ~\rm nG$ in the CSF2 scenario. \\

\begin{figure*}
\includegraphics[width=0.45\textwidth]{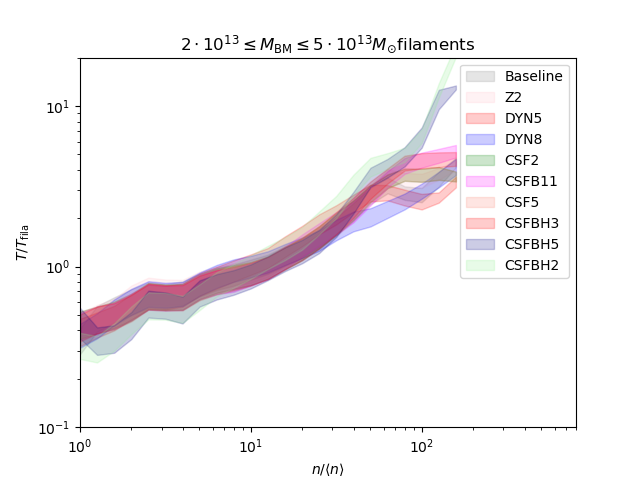}
\includegraphics[width=0.45\textwidth]{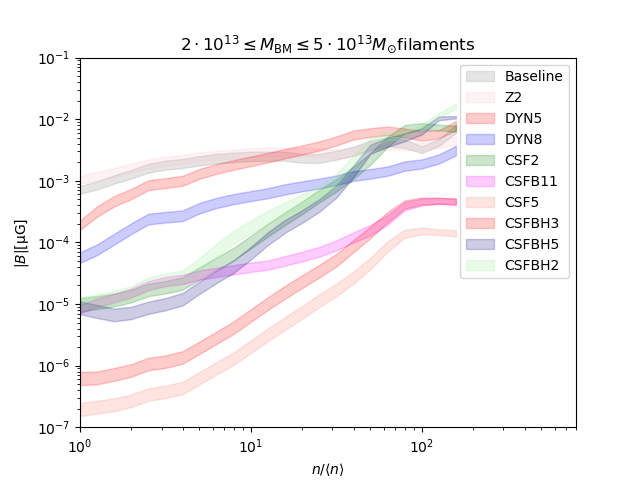}
\includegraphics[width=0.45\textwidth]{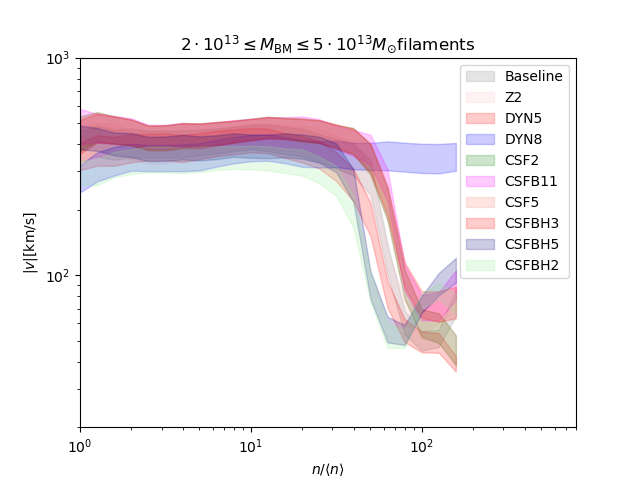}
\includegraphics[width=0.45\textwidth]{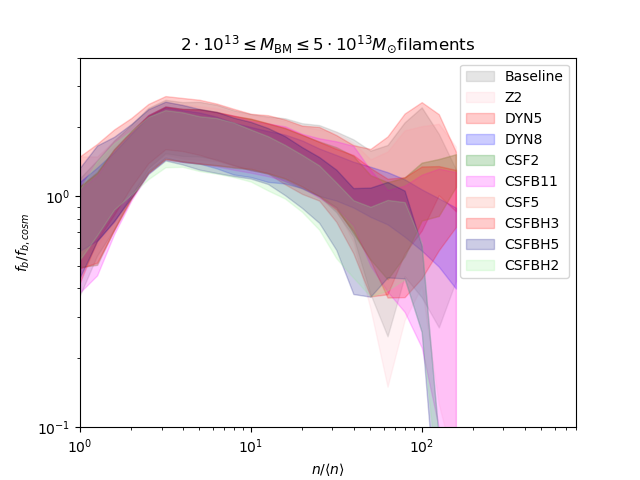}
\caption{Average temperature (top left), magnetic field strength (top right), velocity field (lower left) and baryon fraction (lower right) for  a sample of $10$ filaments in the gas mass range $M_{\rm BM}=2-5 \cdot 10^{13} M_{\odot}$, for all models in the paper. The shaded areas show the 33-66 \% range around the median relation.}
\label{fig:prof2}
\end{figure*}

Given the monotonic dependency of the gas density from the height $h$, following \citet{gh16}, we compute the median value of the quantities of interest as a function of this variable. In this way, we can bypass the limitations related to the complex geometry of the filaments and extend the analysis to a larger number of objects. Figure \ref{fig:prof2} gives the 33-66 \% range around the median relation of gas temperature, magnetic field strength, velocity module and baryon fraction for 10 filaments identified in all runs and in the $M_{\rm BM}=2-5 \cdot 10^{13} M_{\odot}$ mass range. 
In the case of gas temperature, we used the $(\langle T \rangle ,M_{\rm BM})$ scaling relations measured in the previous section which are basically identical in this mass range to normalise temperatures for different masses, while we leave the normalisation free to change for all other quantities, considering that the average magnetic field and the baryon fraction do not show a clear trend with mass, while the gas overdensity at which filaments form is always the same and shall not be further renormalized.

We find that the differences between models are overall mild, and get more pronounced at high enough gas particle number density ($n/\langle n \rangle > 50$), which mark region close to the filament spine but also close to overdense substructures in the filaments'volume. These substructures are located where the sources of feedback are, hence the differences mostly reflects the different impact of feedback processes to the WHIM surrounding galaxies in filaments (see next section). All runs consistently show that in this mass range the baryon fraction is $\sim 1-3$ times the cosmic average for ($2 \leq n/\langle n \rangle \leq 50$), regardless of the details of gas physics, while at higher densities the impact of different physical prescriptions is large. 
The $(B,n/\langle n \rangle)$ relation shows instead larger differences at all overdensities, with a $10^2$ difference at high overdensities, and a $\sim 10^4$ difference in the low density range. This trend is consistent with the global $(B,n/\langle n \rangle)$ reported in \citet{va17cqg}, although there, they were referred to the entire distribution of baryons in the cosmic volume, and not specific to filaments. 

\section{Galactic Halos in Filaments}
\label{sec:galaxies}

\begin{figure}
\includegraphics[width=0.45\textwidth]{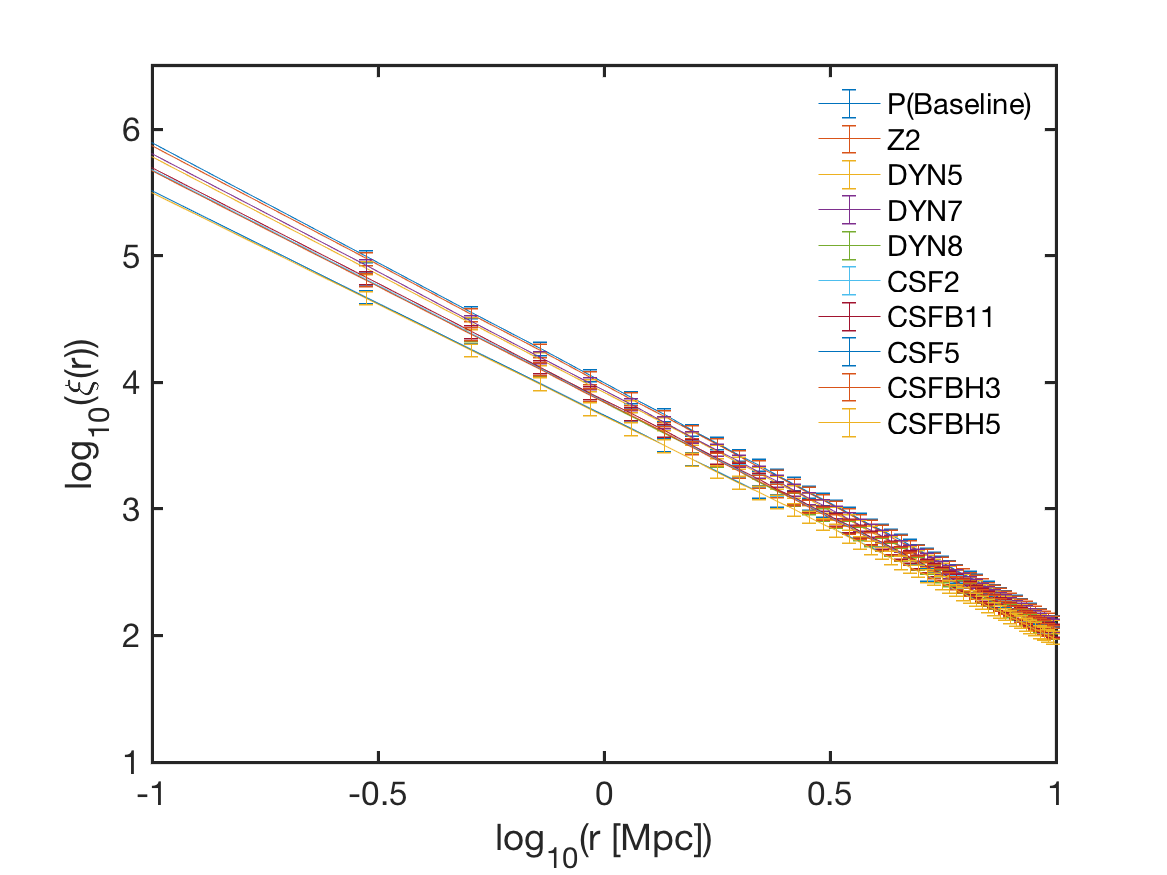}
\caption{Three dimensional two-points correlation function in the range 0.1 to 10 Mpc, in arbitrary units, of the halos extracted from the different runs. 1-$\sigma$ error bars are shown. The correlation function follows a power law whose exponent averaged over the different runs is $\beta = -1.839 \pm 0.002$, consistent with that expected for galaxies in the standard $\Lambda$CDM model.}
\label{fig:correlation}
\end{figure}

Adopting the selection procedure described in Sec. \ref{sec:visit}, it is possible to extract over-dense objects tracing the population of halos within filaments. The adopted overdensity selection criterion uses the total (dark+gas matter) density. The overdensity threshold to excise halos is set to the value of $a_{gal} = 100$, which defines halos undergoing virialization, in line with our previous results in \citet{gh16}. The simulations presented in this paper cannot follow the internal properties of forming galaxies in a robust way, in particular at the lowest masses of our halo distribution. Galaxy formation physics, in fact, is missing in the non-radiative runs and, in general, the spatial resolution in all the simulations is not sufficient to trustworthy resolve the innermost structure of galaxies. However, the statistical and physical properties of the hosting halos are well resolved and described by our approach, as shown in \citet{gh16}, where the same procedure has been adopted to compare the resulting halo catalogues with the GAMA data \citep[e.g.][]{alp14a}, reporting a significant degree of similarity between the properties of the halos identified in our simulations and real galaxies in the survey. We have also calculated the 3D two-points correlation function of the halos, presented in Figure \ref{fig:correlation}, that resulted to follow for all models the expected power law with an exponent averaged over the different runs $\beta = -1.839 \pm 0.002$ in the range 0.1 to 10 Mpc and slightly different normalisations (in arbitrary units), further showing that the properties of our halo catalogues are consistent with those expected for galaxies in the standard $\Lambda$CDM model. The similarity of the correlation curves for the different models, indicates also that the clustering properties of the halos are only mildly influenced by the different physical setup characterising the various simulations. 

\subsection{Statistical properties of the halos in the filaments}
\label{sec:halos}

The number of halos identified in the different runs is presented in Table \ref{tab:galnumber}. Severely under-sampled objects, with volume smaller than $2^3$ computational cells, have been removed. The resulting halo masses are typically $M_{\rm h}\geq 10^{11} M_{\odot}$, although few objects at smaller masses are found. The number of halos tends to be larger in models with radiative physics,  showing that these processes can favour the process of galaxy formation.  

\begin{table}
\caption{Number of halos identified in the different simulation.}
\centering \tabcolsep 5pt
\begin{tabular}{l|c|}
ID    & $N_{h}$ \\  
\hline
P(Baseline)&1588 \\
Z2&         1507 \\
DYN5&       1778 \\
DYN7&       1757 \\
DYN8&       1881 \\
CSF2&       1864 \\
CSFB11&     1823 \\
CSF5&       1992 \\
CSFBH3&     1873 \\
CSFBH5&     2012 \\

\end{tabular}
\label{tab:galnumber}
\end{table}

\begin{figure*}
\centering
  \includegraphics[width=0.45\textwidth]{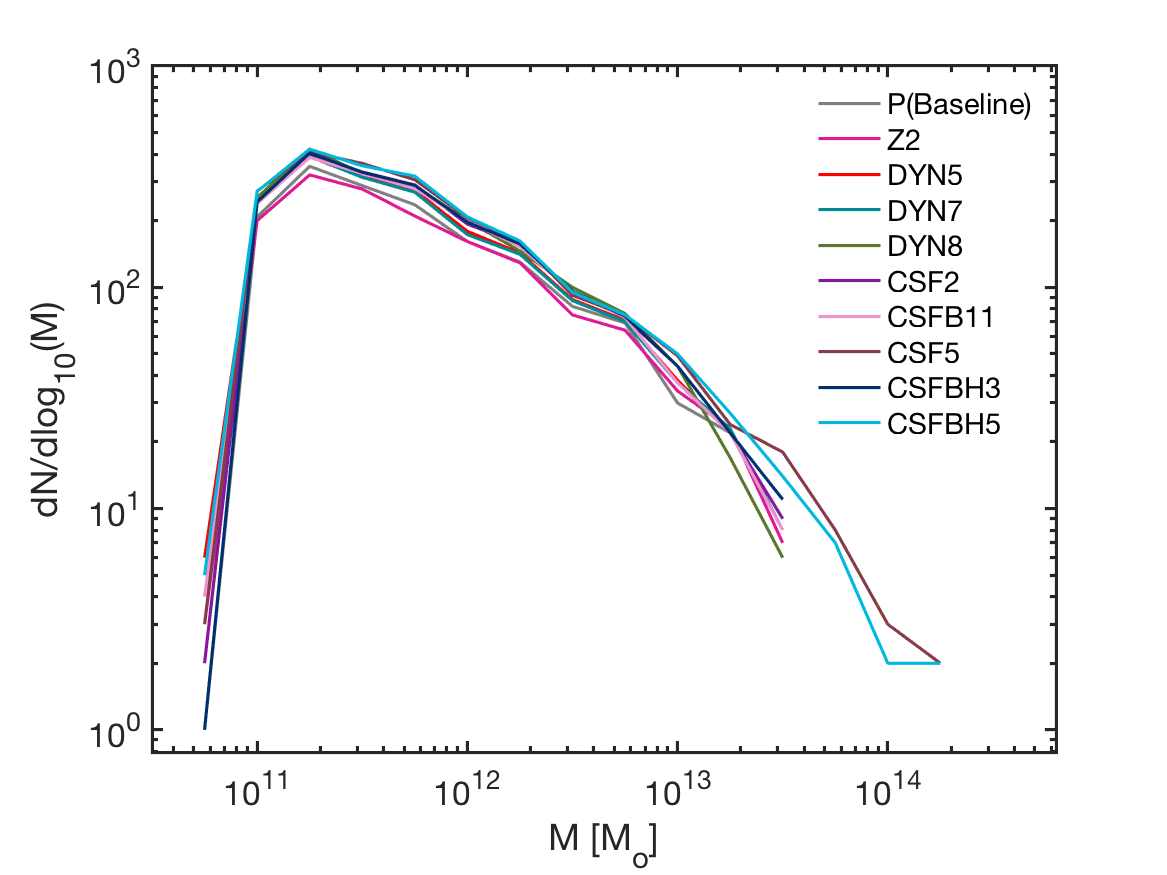} 
  \includegraphics[width=0.45\textwidth]{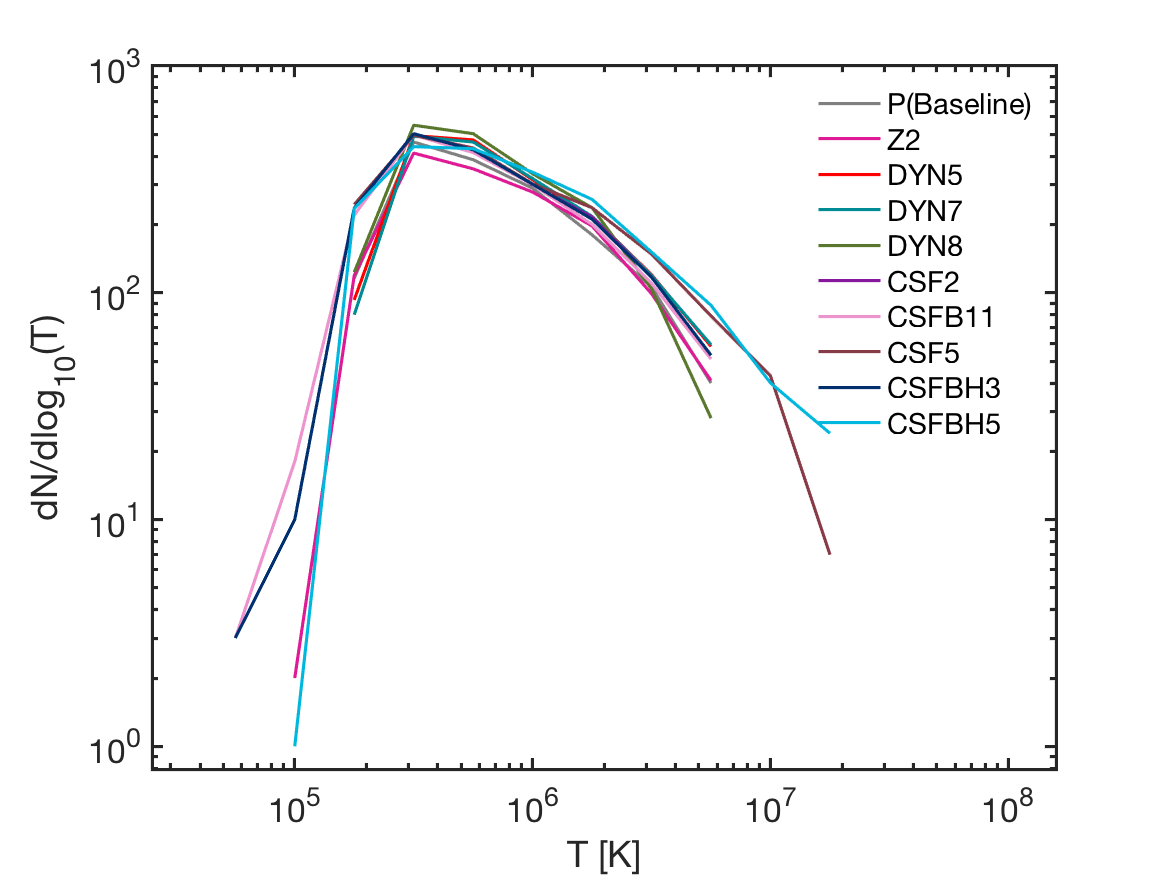}
  \includegraphics[width=0.45\textwidth]{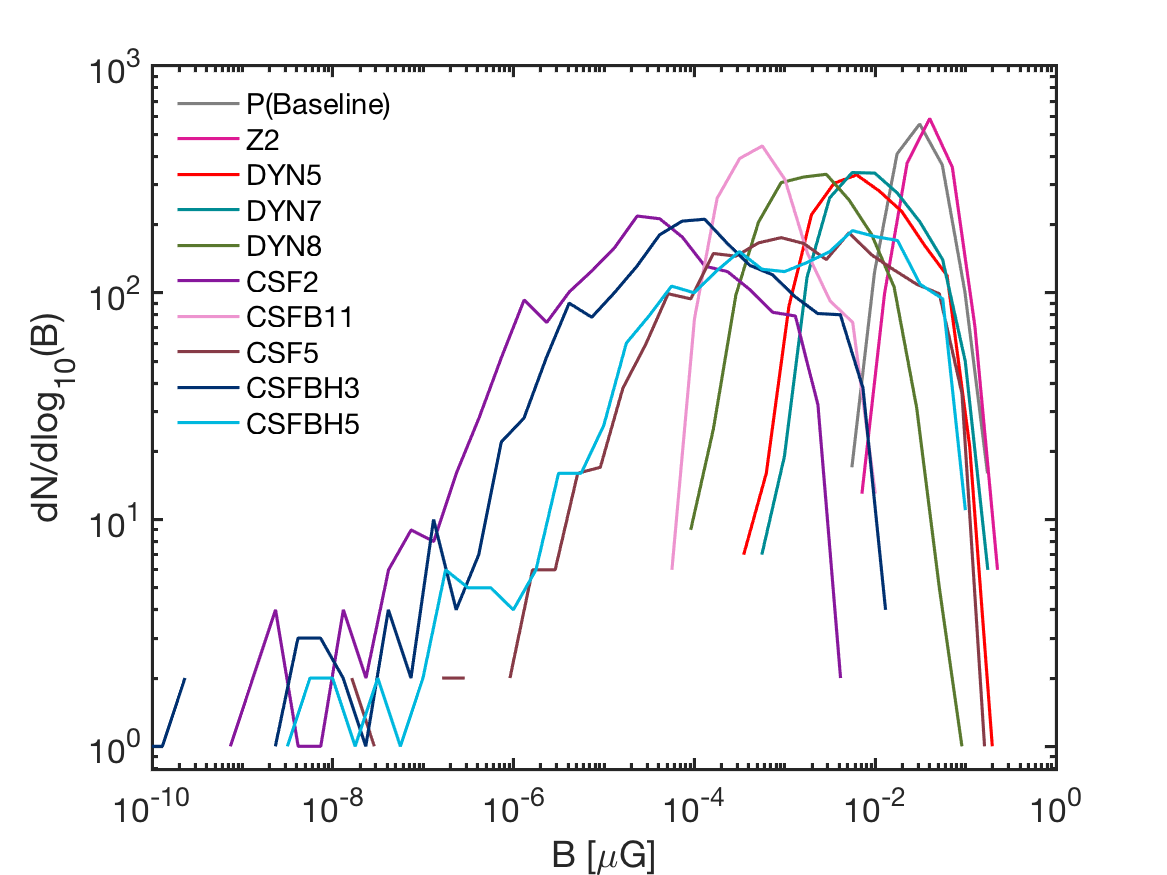} 
  \includegraphics[width=0.45\textwidth]{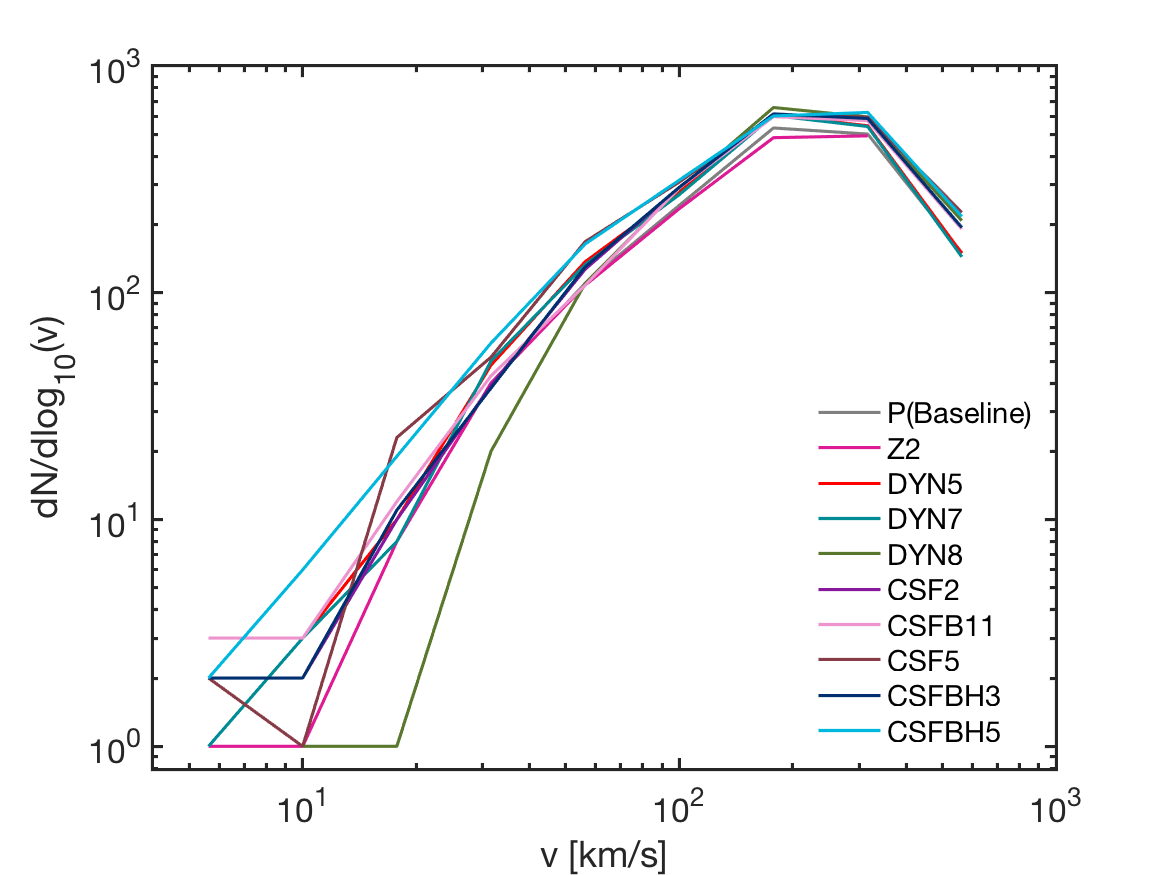}
\caption{Distribution function for total mass, mass weighted average temperature and magnetic field strength and average velocity of halos in filaments in our runs.}
\label{fig:galdist}
\end{figure*}

The distributions of the number of halos with given mass and average temperature (top row), average magnetic field strength and mean velocity (bottom row) are given in Figure \ref{fig:galdist}. 
For the mass function above $\sim 2\cdot 10^{11} M_{\odot}$, the number of objects decreases with increasing mass. The distributions are rather similar for the different models. However, models CSF5 and CSFBH5 present a tail with masses around $10^{14} M_{\odot}$. In these runs, the density threshold for star formation is 5 times smaller than in the other runs. This leads to a more efficient star formation, removing gas pressure support from within the halos, enhancing the cooling of further gas onto the halo volume and, finally, increasing the halos gas content.  The sharp drop in the number counts at masses below $10^{11} M_{\odot}$ is instead due to the adopted selection criterion, which removes most of the least massive objects. Also for the temperature, the distributions follow similar patterns in all the models, with average temperatures between $10^{4.5}$ and $10^{7.5}$K. Even when cooling and feedback are included in the simulation, they tend to self-regulate, leading to average temperature whose values are close to those of runs without any cooling or heating active. Models CSF5 and CSFBH5 show their peculiar behaviour also in the temperature distribution, in particular at the highest temperatures. Once more, the highly efficient feedback processes influence the halos properties, leading to an increase in the energy input in the gas, hence to objects with temperature higher than in all the other models.
Differences among the runs can be found also at the lowest temperatures. The primordial and dynamo models have halos in a smaller temperature range compared to the astrophysical models, their thermodynamics being controlled exclusively by adiabatic compression and shock waves, that heat all the gas at temperatures between $10^5$ and $10^7$ K. The models CSF2 and CSFB11 are among the least efficient star feedback models and they do not include feedback from supermassive black holes, hence the associated thermal feedback is smaller than for the other astrophysical models. At the same time, cooling is active, decreasing the temperature of the shocked gas. This results in overall colder objects compared to all the other runs.

The mean magnetic field distributions show remarkable differences between the simulations. The two primordial models have similar trends, with the highest values of the magnetic field peaked between $10^{-2}$ and $10^{-1}\mu$G and with a narrow distribution. The three dynamo models have a similar shape, with peak between $10^{-3}$ and $10^{-2}\mu$G and a broader distribution, encompassing magnetic fields from $10^{-4}$ to slightly more than $10^{-1}\mu$G. When cooling and feedback are active, the distributions result to be even broader, with a tail of few objects with magnetic field B$< 10^{-7}\mu$G. The number of halos grows with B until reaching a maximum which is different for the various models and then it declines, sharply for the models CSF5 and CSFBH5 and more gently in the other models. The CSFB11 model is an exception: due to the high primordial magnetic seed and the low star formation efficiency, its behaviour is close to that of the dynamo models at a slightly lower range of $|B|$. We remark that, while these trends are perfectly in line with those presented for the filaments, for halos the lack of resolution and physical details in the description of such overdense environment hampers any significant small-scale dynamo amplification. Our flows within halos in fact, have a magnetic Reynolds number  $R_m \ll 100$ (defined as the ratio between magnetic induction terms and magnetic diffusivity, i.e. $U \cdot L/\eta_m$, where $U$ is the characteristic flow velocity on the scale $L$ and $\eta_m$ is the magnetic diffusivity), which severely prevents the formation of the small-scale dynamo, at variance with what is commonly accepted to happen in real galaxies \citep[e.g.][]{beck13,2013NJPh...15b3017S,schober13,2016MNRAS.457.1722R,ma17}.

Finally, the velocity distribution is found to behave similarly in all models for most of the halos. As expected, the different physics in the various models does not affect the overall kinematic properties of the halos. At low velocity values large differences are present, but the low number statistics does not allow reaching firm conclusions. 

\begin{figure*}
\centering
  \includegraphics[width=0.45\textwidth]{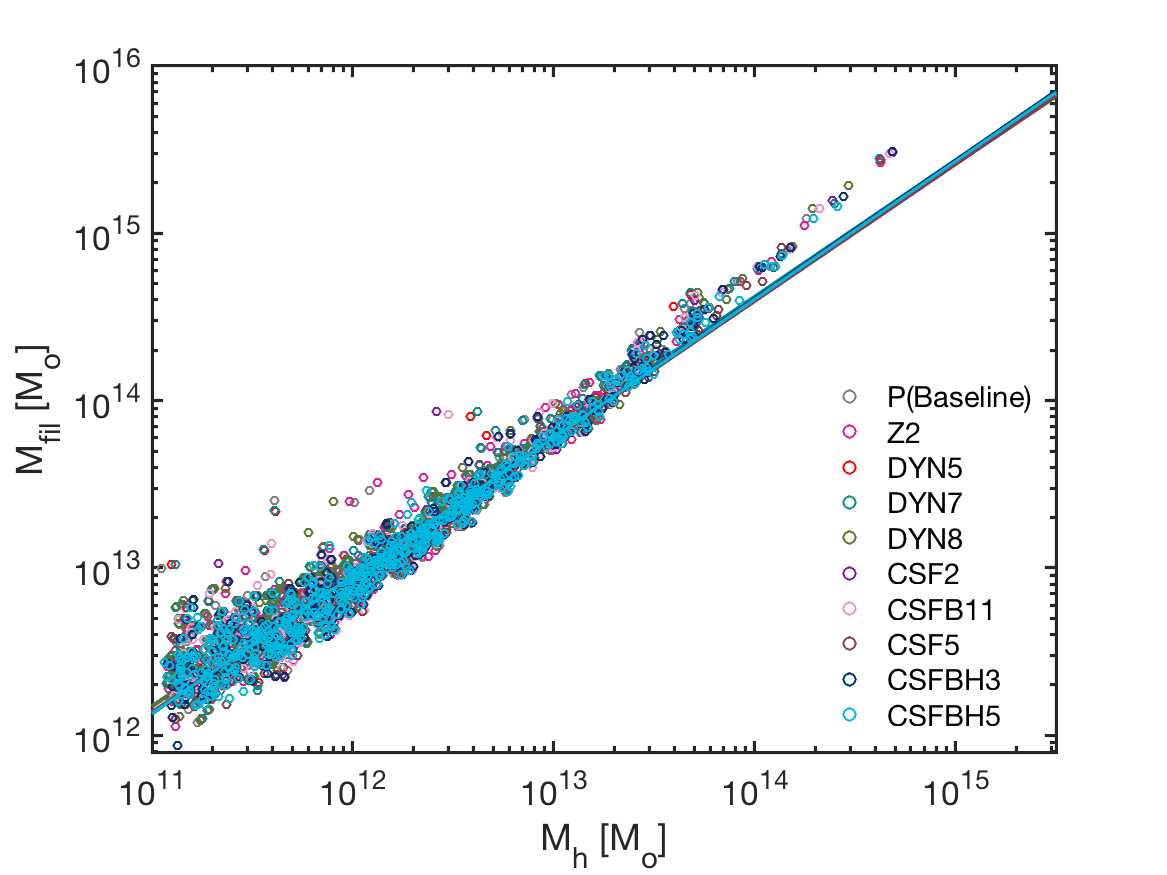} 
  \includegraphics[width=0.45\textwidth]{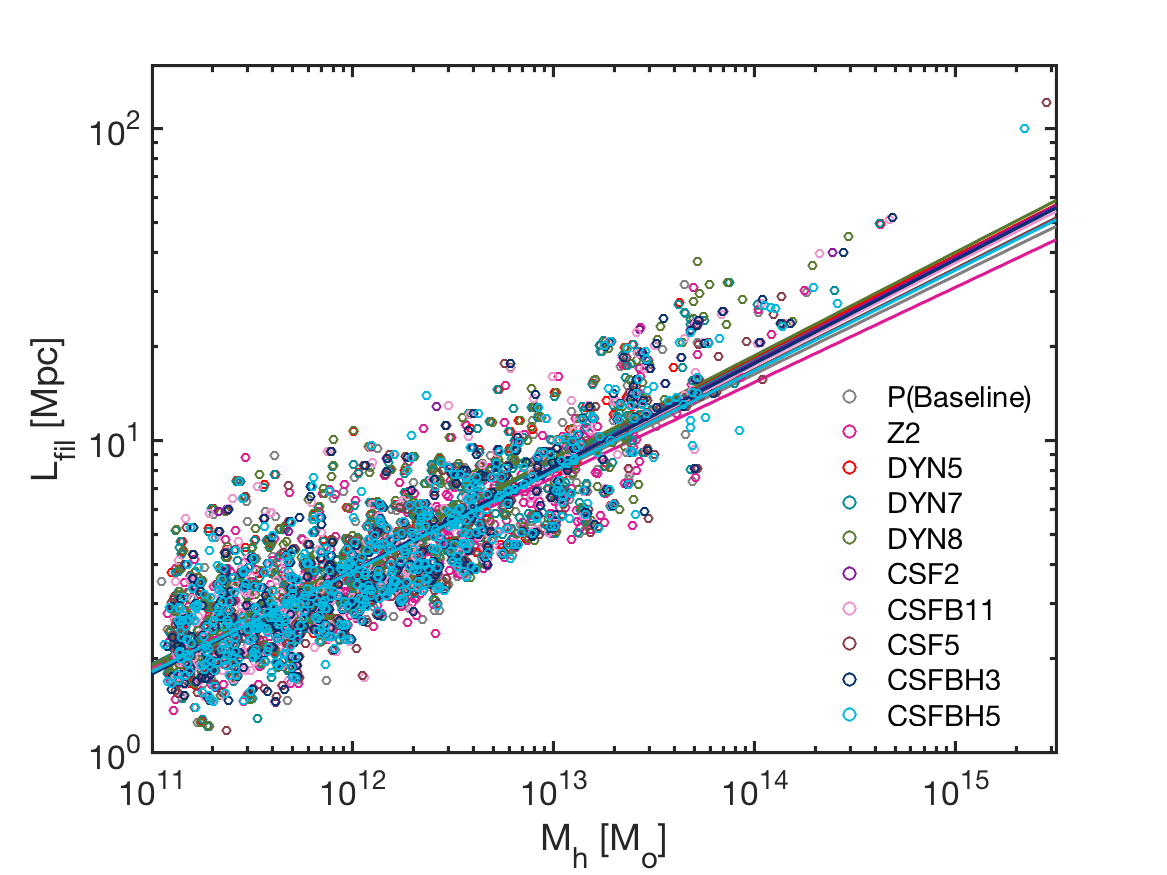} 
  \includegraphics[width=0.45\textwidth]{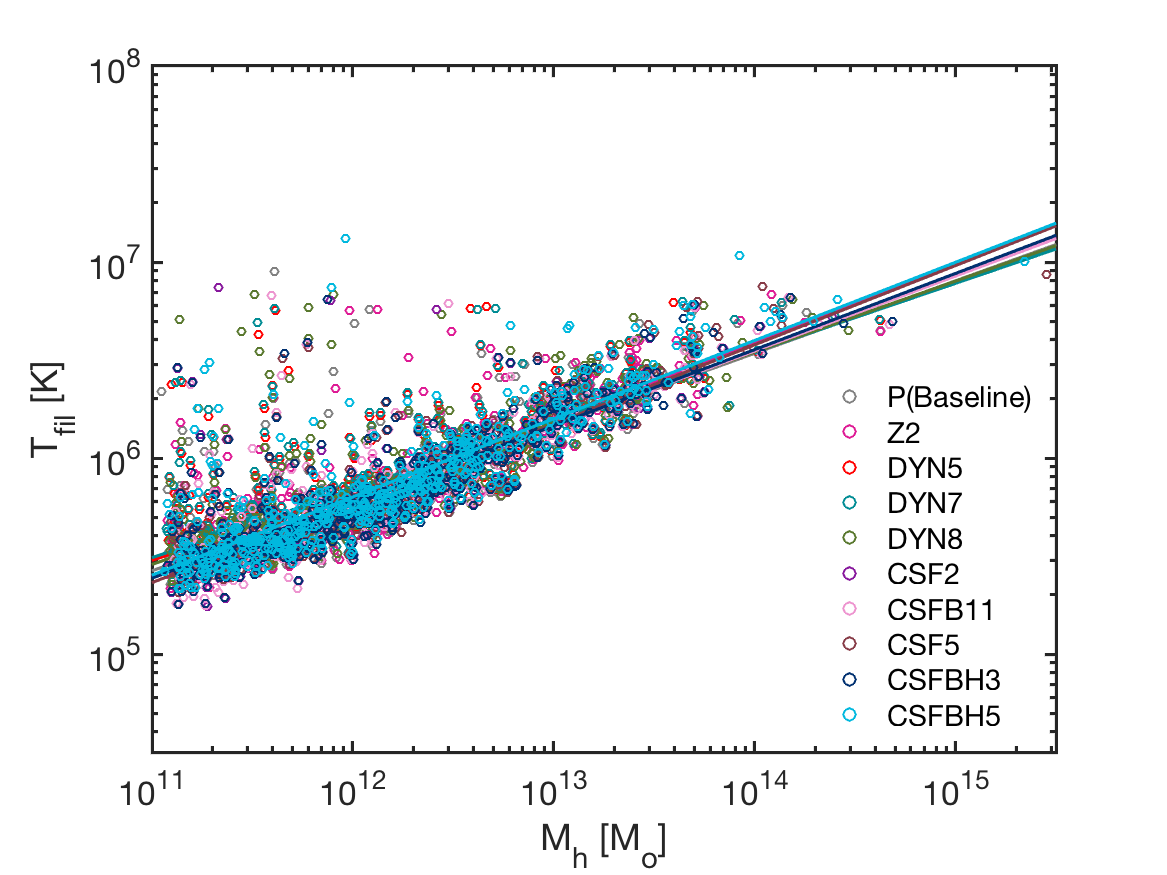} 
  \includegraphics[width=0.45\textwidth]{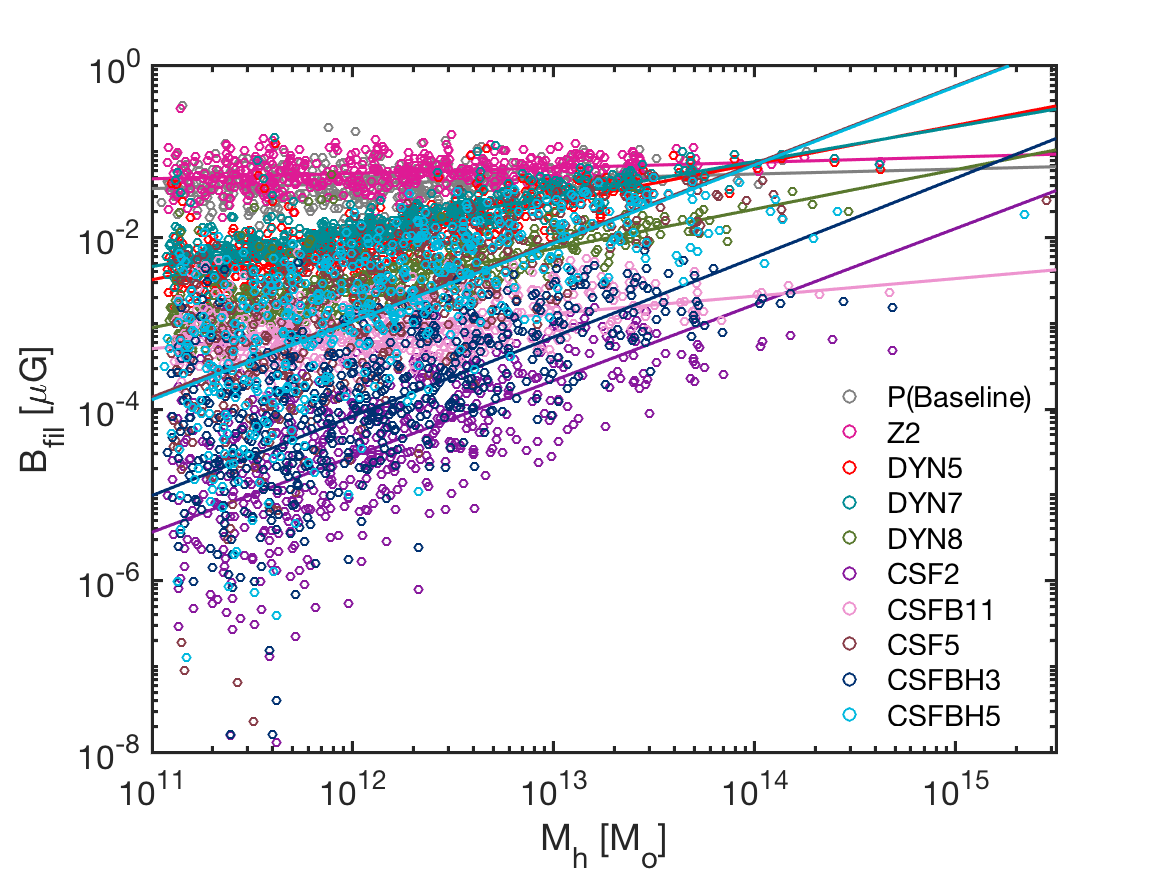}
\caption{Correlation between the total mass of halos in a filament and filament related quantities in the different models: total mass (top left), length (top right), mean temperature (bottom left), mean magnetic field (bottom right). All averages are mass weighted. The displayed fitting curves are calculated considering all the halo masses.}
\label{fig:crossfil}
\end{figure*}

Halo properties, in particular masses, can be correlated to those of the hosting filaments.
The dependence of the total mass, the length, the average temperature and magnetic field of the filaments from the total mass of embedded halos is shown in Figure \ref{fig:crossfil}. All distributions have been fitted with the same power law introduced above. The mass and length distributions become steeper at masses above $M^*_{TOT,h} = 10^{13}M_{\odot}$, hence fitting curves have been calculated considering both all the masses and only objects with mass above $M_{TOT,h}^*$. Since the trends are very similar in all models, average fitting parameters could be calculated. The results are presented in Table \ref{tab:galfilfit}. 

The mass of the filaments tends to grow almost linearly with that of the resident halos. The trend is even closer to linearity when only high total halo masses are considered. The length of the filaments scales approximately as $L_{\rm fil} \propto M_{TOT,h}^{1/3}$ including all objects, while it tends to scale as $\propto M_{TOT,h}^{1/2}$ restricting to higher masses, due to the tendency of bigger filaments to be thinner and more elongated than the smaller ones. For the mean temperature, no changes in $\beta$ with the halo mass are found, with a trend given by the power law $T_{\rm fil} \propto M_{TOT,h}^{1/3}$ at all scales. For the magnetic field results change significantly with the class of models, with the same trends we already observed in Figure \ref{fig:mb}. 

The above scaling are consistent with what previously found with non-radiative simulations in \citet{gh16}, confirming that the total halo mass is trustworthy and effective proxy of the host filament's properties. 


\begin{table*}
\caption{Fitting parameters averaged among the different models of the relations presented in Figure \ref{fig:crossfil}. The parameter $\beta$ is the exponent of the fitting power law, while $\alpha$ is the normalisation. The second and third columns consider all the data, the fourth and fifth only filaments with total mass in halos $>10^{13} M_{\odot}$.}
\centering \tabcolsep 5pt
\begin{tabular}{c|c|c|c|c|}
Quantity & \multicolumn{2}{l}{$M_{TOT,h} > 10^{10} M_{\odot}$} & 
   \multicolumn{2}{l}{$M_{TOT,h} > 10^{13} M_{\odot}$}  \\  \hline
 & $\alpha$ & $\beta$ & $\alpha$ & $\beta$ \\  \hline
  $M_{fil}$ & $3.153\pm0.089$ & $0.818\pm0.006$ & $0.843\pm0.232$ & $0.995\pm0.018$ \\
  $L_{fil}$ & $-3.303\pm0.101$ & $0.324\pm0.009$ & $-5.751\pm0.470$ & $0.509\pm0.036$ \\
  $T_{fil}$ & $1.248\pm0.251$ & $0.379\pm0.019$ & $1.325\pm0.339$ & $0.376\pm0.024$ \\
\end{tabular}
\label{tab:galfilfit}
\end{table*}

\subsection{Halos and their local environment}
\label{sec:halosandfil}

The distribution of galaxies naturally provides a clue of how baryonic matter is distributed in the cosmic web \citep[e.g.][]{alp14a}. The overdense and relatively hot environment surrounding galaxies and the associated halos in filaments is a natural target for observations. The characterisation of the properties of such environment is therefore meaningful in order to study the correlation between galaxies, the local gas distribution and the overall properties of the hosting filaments.

\begin{figure*}
\centering
\begin{tabular}{c|c}
  \includegraphics[width=0.45\textwidth]{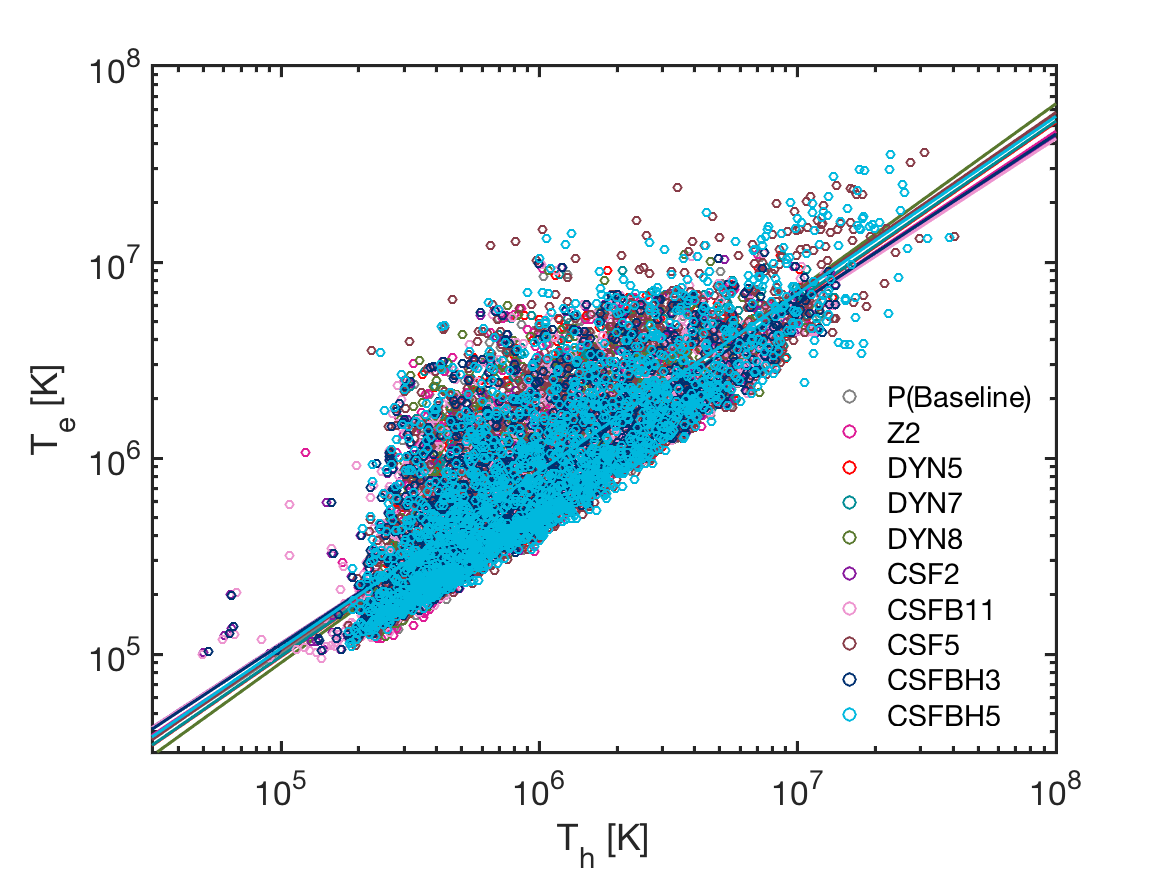} &
  \includegraphics[width=0.45\textwidth]{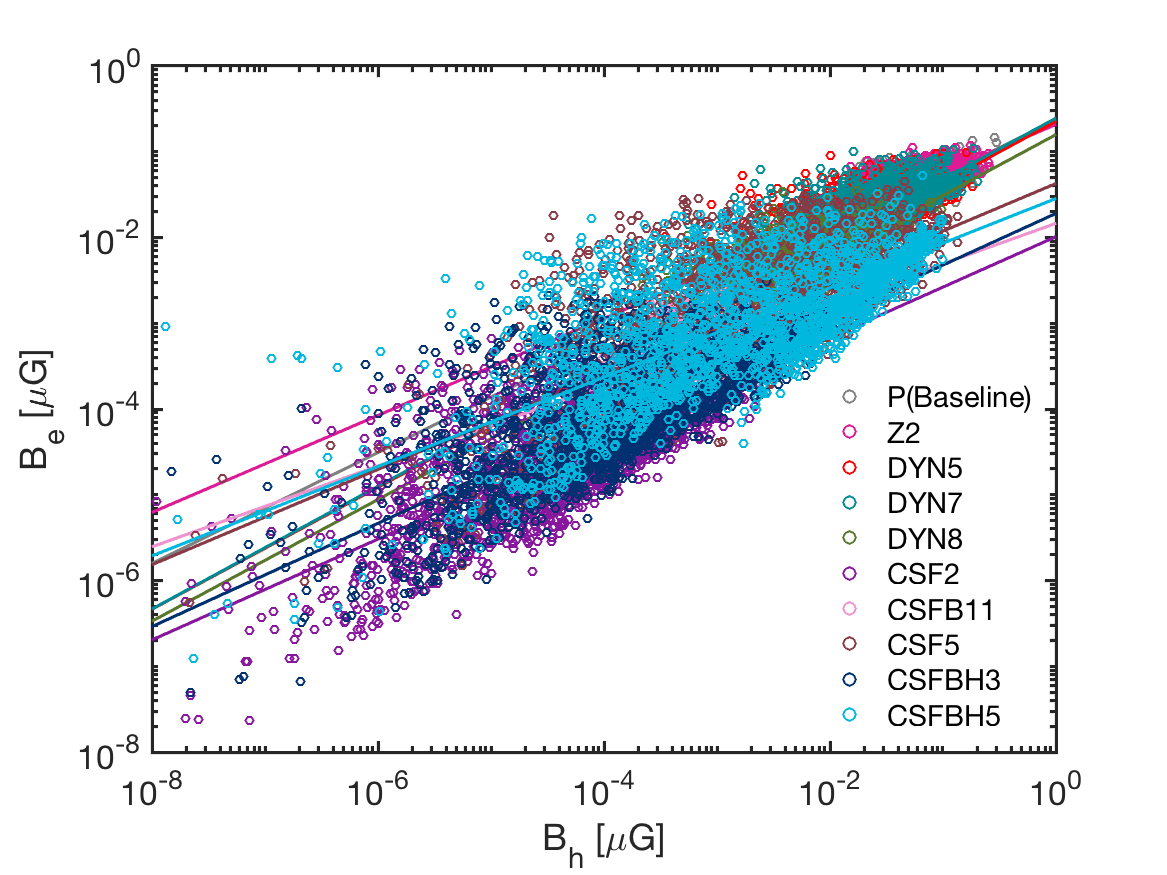} \\
  \includegraphics[width=0.45\textwidth]{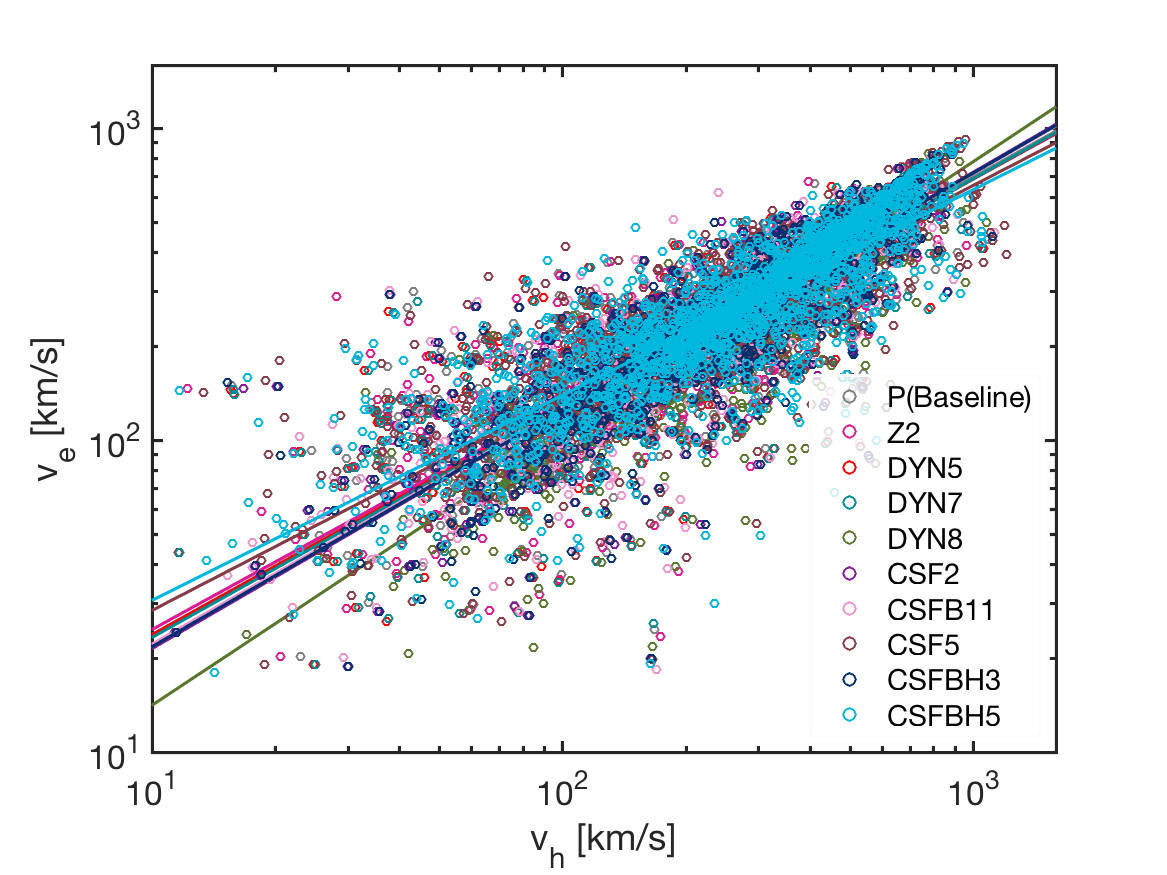} &
  \includegraphics[width=0.45\textwidth]{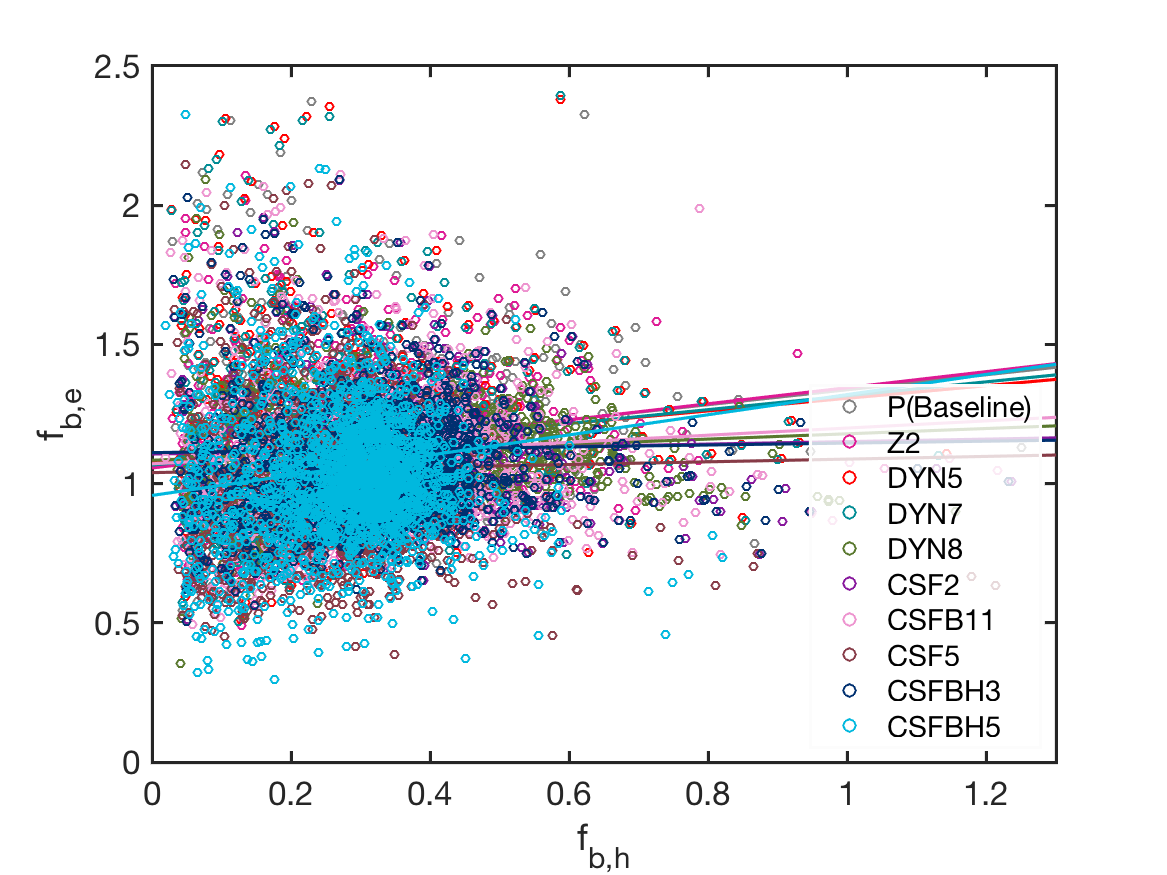}
  \end{tabular}
\caption{Correlation between halos (subscript h) quantities and their external (subscript e) environment: mass averaged temperature (top left), mass averaged magnetic field (top right), mean velocity (bottom left), baryon fraction, bottom right.}
\label{fig:cross}
\end{figure*}

In order to identify the galaxies' local environment, we have extracted spherical regions centred on each halo, with radius ten times that of the halo itself. 
Figure \ref{fig:cross} shows the relation of the properties of the resulting environment volumes and the halos lying within. In general the properties within and outside of identified halos are correlated, albeit with a large scatter, exhibiting trends that, once more, have been fit as $\rm log_{\rm 10}(X_e) = \alpha + \beta~ \rm log_{\rm 10}(X_h)$, where $X_e$ and $X_h$ are various average quantities calculated for the environment and for the corresponding halo respectively. 

In all models, the average temperature of the external WHIM (top left panel) scales with the mean halo temperature, with almost the same exponent, despite the large scatter. Its value, averaged among the different runs, is $\beta = 0.894\pm 0.029$. The average normalisation is $\alpha = 0.547\pm 0.175$. Halos result to be at slightly higher temperature that the surrounding medium, this difference slowly growing with the halo mass. 
Similar considerations can be made for the velocity (bottom left panel), with a comparable spread of the points distribution around the average best fit curve, with $\beta = 0.742\pm 0.057$ and $\alpha = 0.620\pm 0.146$. The only exception is the DYN8 model, which has a fitting curve remarkably different from all the others, showing that, if sufficiently strong, the magnetic field can affect the dynamical properties of the gas \citep[e.g.][]{2016MNRAS.456L..69M}. In general, these trends show that halos move consistently with the embedding environment. Once more, the magnetic field presents peculiar signatures associated with the different physical properties of the various models, which affects both the normalisation and the scaling. The fitting parameters $\alpha$ and $\beta$ are presented in Table \ref{tab:fitBB}. In primordial and dynamo models the magnetic field intensity of both halos and their surrounding environment is comprised between $0.0001$ and $0.2\mu G$. Halos with the highest magnetic field are embedded in environments with approximately the same value of $|B|$. For smaller values of the halos magnetic field, the environment tends to have slightly bigger values of $|B|$, as confirmed by the slope of the fitting curves, always less than one. This trend is even more evident in astrophysical models, that typically encompass a broader range of $|B|$, with both slopes and normalisation significantly lower than that of primordial and dynamo models. However, halos in astrophysical models at the lowest values of the magnetic field lies in environments with an even lower magnetisation.

\begin{table}
\caption{Best-fit parameters of the halo to environment magnetic field relation,  $log_{\rm 10}(B_e)=\alpha + \beta ~ log_{\rm 10}(B_h)$.}
\centering \tabcolsep 5pt
\begin{tabular}{|c|c|c}
   \hline
   run &
   \multicolumn{2}{c}{Mass-Length}   \\  \hline
   ID & $\alpha$ & $\beta$ \\ \hline
P(Baseline)& $-0.628\pm0.020$ & $0.646\pm0.014$\\
Z2&     $-0.680\pm0.018$ & $0.565\pm0.014$\\
DYN5&   $-0.645\pm0.025$ & $0.711\pm0.012$\\
DYN7&   $-0.607\pm0.024$ & $0.715\pm0.012$\\
DYN8&   $-0.800\pm0.032$ & $0.709\pm0.012$\\
CSF2&   $-1.990\pm0.047$ & $0.587\pm0.010$\\
CSFB11& $-1.835\pm0.037$ & $0.471\pm0.011$\\
CSF5&   $-1.370\pm0.034$ & $0.555\pm0.011$\\
CSFBH3& $-1.719\pm0.042$ & $0.602\pm0.011$\\
CSFBH5& $-1.545\pm0.033$ & $0.520\pm0.011$\\
CSFBH2& $-1.182\pm0.563$ & $0.608\pm0.085$\\
\end{tabular}
\label{tab:fitBB}
\end{table}

Finally, the bottom right panel of Figure \ref{fig:cross} compares the baryonic fraction of the halos to that of their neighbourhood.  The enclosed baryonic fraction in the halos is referred to the hot/warm gas component, while we do not consider here the mass locked into stars and SMBH particles. In all the models the baryon fraction within halos is, as expected, lower than the cosmic average, with values around $\sim 10-50 \%$ of such average. These values are typically higher than those for galaxy groups \citep[e.g.][]{2009ApJ...703..982G,2010MNRAS.407..263A} and galaxies \citep[e.g.][]{0004-637X-635-1-73,2012ApJ...759..138P,2014AJ....147..134Z} reported in literature. This is the effect of the limited spatial resolution combined to our halo identification procedure, which lead to extract volume significantly larger than those typical for those kind of objects, including matter at higher baryon fraction from the local environment, which, in fact, is found to be at baryon fraction $  f_{\rm b}/f_{\rm b,cosm}\approx 1$, significantly higher than that of the halo but also of the overall filament. The gas matter expelled from the halo  enriches the surrounding environment for all models and at all masses \citep[e.g.][]{2006MNRAS.373.1265O}, with a mild trend of the baryonic fraction to increase with the mass for non-feedback models, since no cooling mechanisms are present capable of decreasing the gas pressure in the halo, allowing it to fall in the forming object. The scatter around the fitting curves is large. However, the environmental baryon fraction almost never gets smaller than $f_{\rm b} \approx 0.5 ~f_{\rm b,cosm}$. 

\begin{figure*}
\centering
\begin{tabular}{c|c}
  \includegraphics[width=0.45\textwidth]{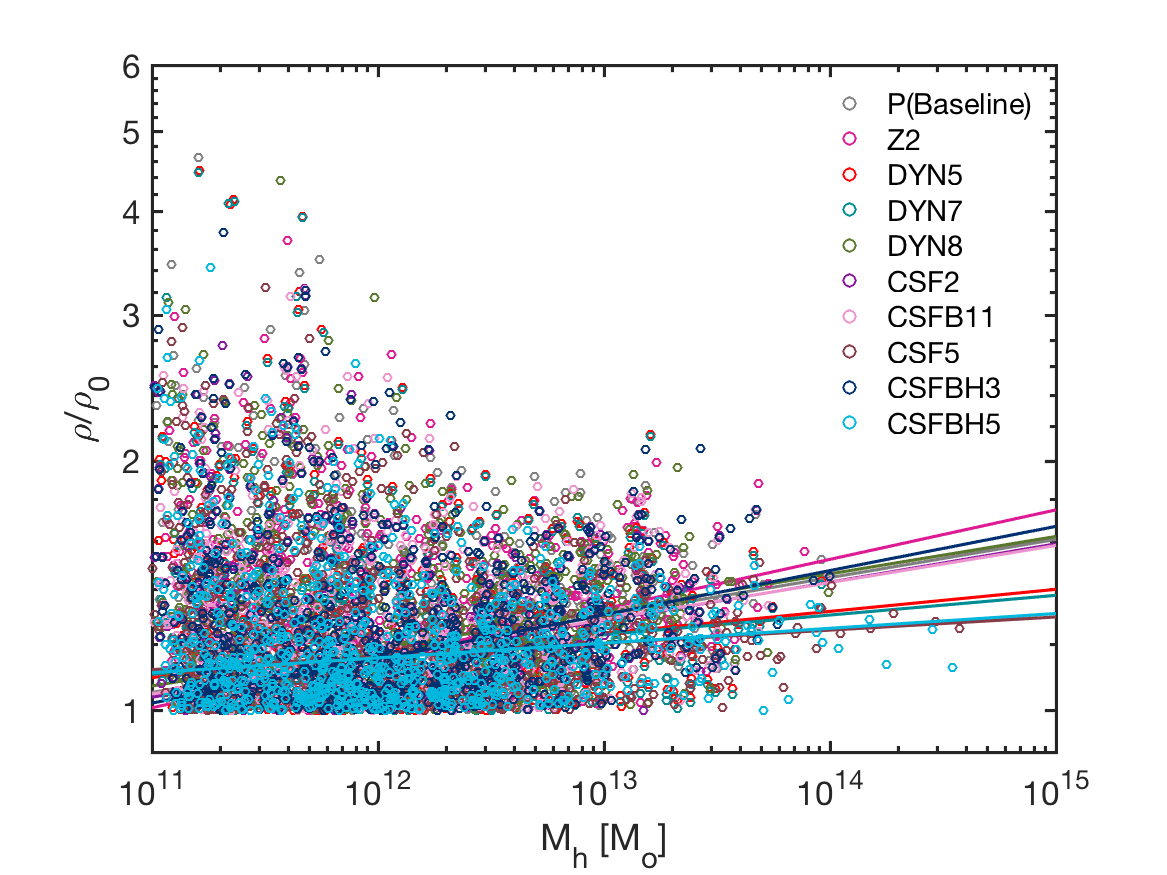} &
  \includegraphics[width=0.45\textwidth]{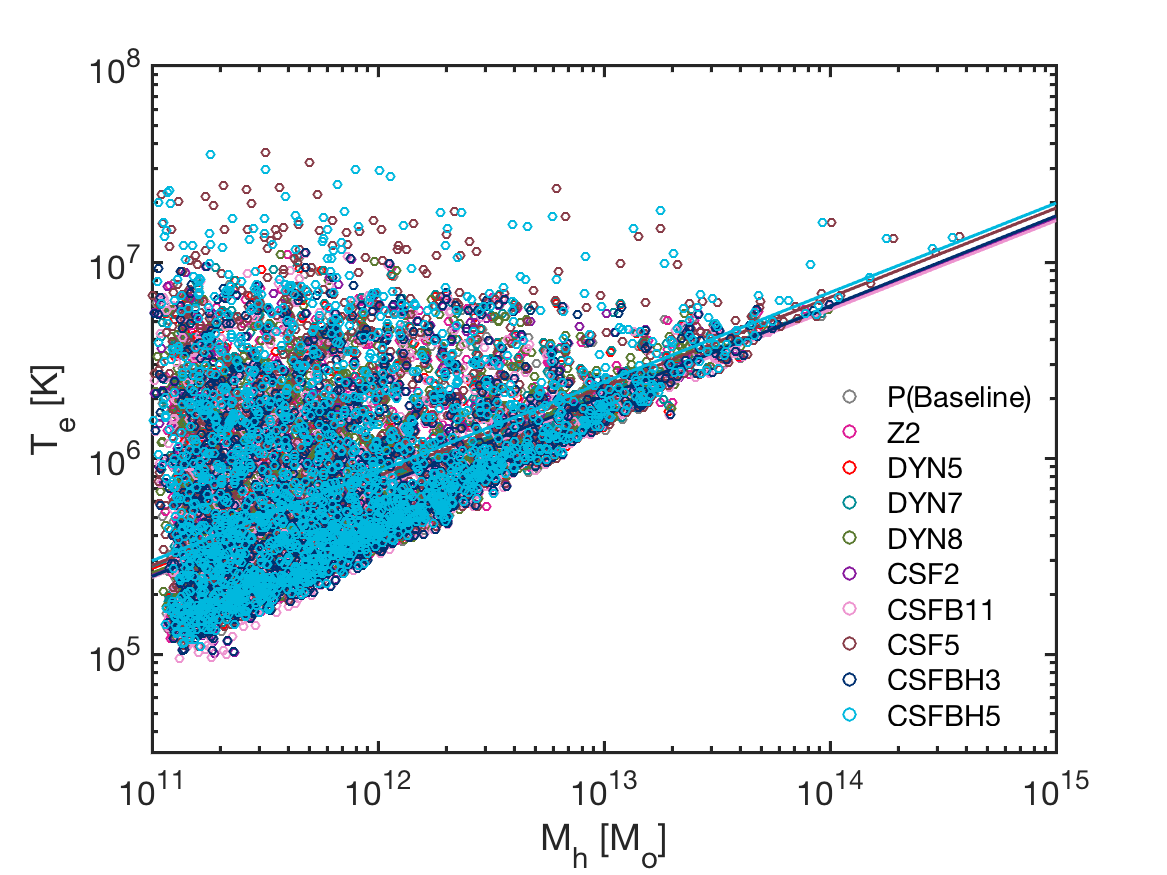} \\
  \includegraphics[width=0.45\textwidth]{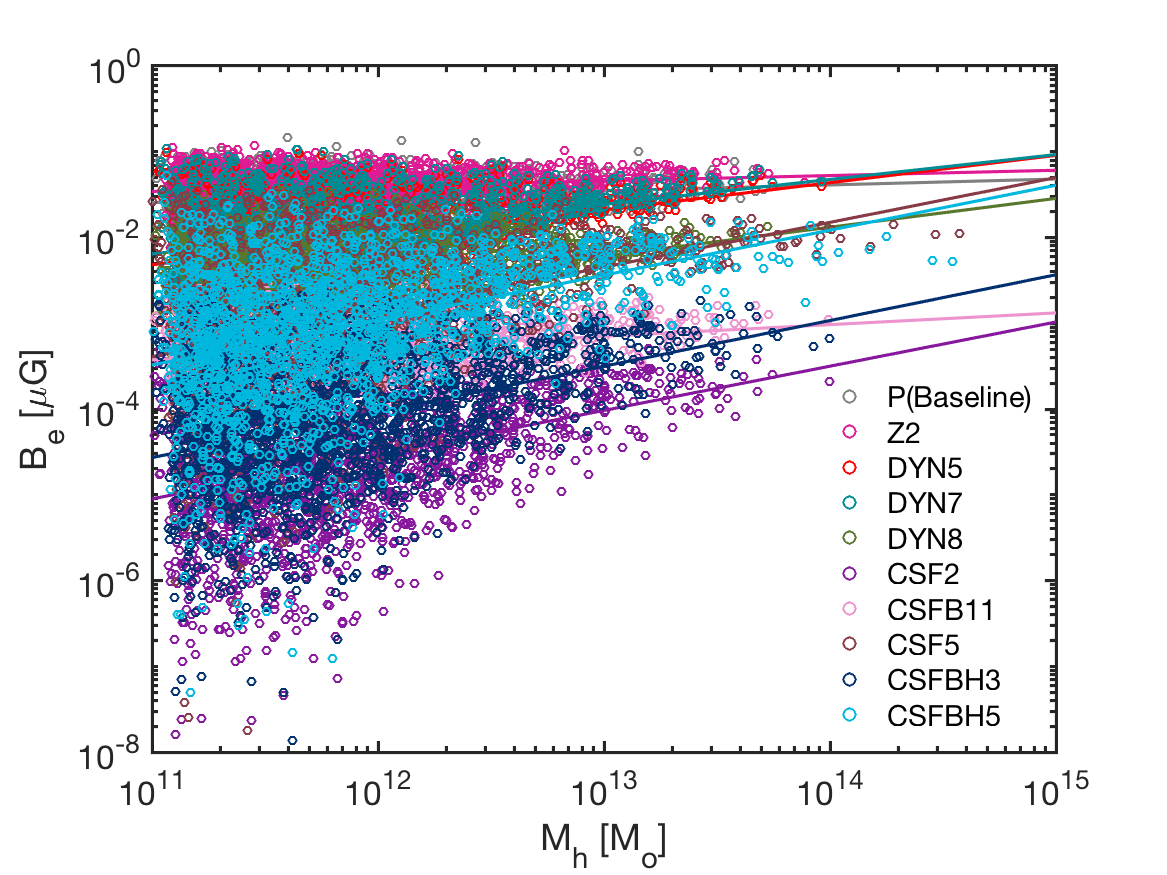}&
  \includegraphics[width=0.45\textwidth]{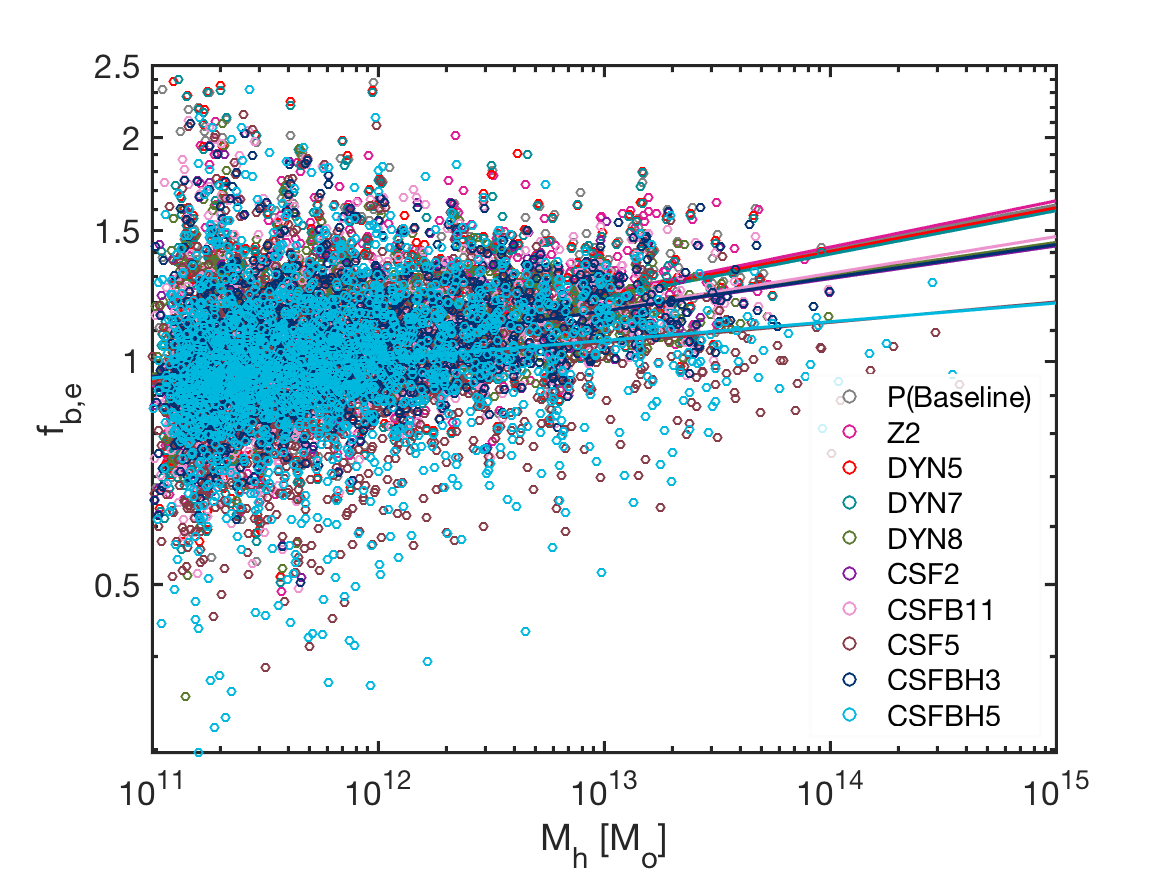}
  \end{tabular}
\caption{Correlation between halo mass and the environmental average quantities: baryonic mass density (in units of the critical mass $\rho_0$, top-left), temperature (top-right), magnetic field strength (bottom-left) and baryon fraction (bottom-right).}
\label{fig:mrt}
\end{figure*}

In Figure \ref{fig:mrt} we analyse the correlation between the halo masses and the average density (top left panel), temperature (top right panel), magnetic field (bottom left panel) and baryon fraction (bottom right panel) of the embedding environment. For all the models the distributions are highly scattered, especially at low masses. The dispersion tends to decrease for masses above $10^{13}M_{\odot}$, where the fitting curves are calculated. Once more, for all the models the temperature follows similar power laws. The average fitting parameters (calculated as above) result to be $\beta_{T} = 0.454 \pm 0.005$, $\alpha_{T} = 0.431\pm 0.077$. At the highest masses this relation provides a precise estimate of the temperature dependence from the halo mass. At lower masses, due to large scatter, it gives a lower limit of the environmental temperature for a given halo mass. The density does not show a well defined trend, and in this case the lower limit is imposed by our filament identification procedure. A slight trend of the environmental average density to increase with the halo mass is identifiable, but, due to the large dispersion a clear behaviour cannot be assessed. The magnetic field shows a clear increasing trend  with the halo mass, which is sharper in astrophysical seeding models compared to primordial models, which also present larger normalisation. The baryon fraction presents a similar behaviour, its value slightly growing with the hosted halo mass. Once more, model CSFBH5 has a much larger scatter of values, and in general features the lowest baryon fraction values in the external medium, at all given halo masses, due to the extreme feedback strength here (which also explains the large values of magnetic fields measured external to halos). In this case, the feedback is capable of expelling a significant fraction of baryons even outside of the volume of filaments, as already noticed in Figure \ref{fig:ml}. 

\section{Discussion}
\label{sec:discussion}

While we defer to future work the detailed analysis of the observational signatures in the X-ray and radio bands of filaments in different physical scenarios, some general trends can be anticipated based on these new results. 
Thanks to the wide spectrum of models available in the Chronos++ suite, in this work we could extensively compare the effect of different physical processes on the thermodynamics and the magnetic properties of the gas in cosmological filaments. Based on our extraction procedure, we could separate three different types of target objects: filaments as a whole, galactic halos within filaments and the local environment where halos lies. 

For all the presented correlations, well defined trends emerge for the most massive objects (i.e. $ M^*_{BM} \geq  5\cdot 10^{12} M_{\odot}$ for filaments and $M^*_{TOT,h} \geq 10^{13} M_{\odot}$ for the halos).

For the mass-temperature relation we found only mild differences among different models, indicating that the final overall thermodynamic status of the gas is comparable in all cases, even if the local effect of heating and cooling can be considerably different, as shown by the broad dispersion in the environmental temperature correlated with the mass of the corresponding halos.  This is true  considering, both, the dependency of the filament temperature from the total baryonic mass of the filament and from the total mass of hosted halos. The slope of the various models tends to decrease for increasing total filament's gas mass. It remains instead almost uniform at all scales when halo masses are considered. Small mass filaments show indeed a  large scatter in their thermodynamic properties, due to the feedback from galaxies and AGN around them. These objects will be challenging to observe both in the soft X-ray band, having $T \leq 10^{6} ~\rm K$, and in the radio band, having $\langle B \rangle \sim 10^{-3}\div 1 ~\rm nG$ (depending on the model). Due to the large scatter of scaling relations in these objects, robustly constraining the thermal and magnetic properties of the WHIM in such filaments may be possible only with statistics of hundreds of objects.

The thermal properties of the largest filament ($M_{\rm BM} \geq 5 \cdot 10^{12} M_{\odot}$) do not allow an easy discrimination between the different scenarios considered in this work. Such discrimination would be even more challenging for observational data, due to their inherent large uncertainties.  The only relevant difference between model are found in the high density range ($n/\langle n \rangle \sim 10^2$), i.e. close to the central spine of filaments and in the proximity of halos, where feedback effects are more effective. On the other hand,  no systematic differences among models are found for the majority of the volume filled by WHIM gas. \\

Significant differences among models are found when the magnetic field is considered. In primordial models, where the only mechanism of amplification of the magnetic field is compression, the normalisation of the initial seed field has to be set high enough to get the observed level of magnetisation in galaxy clusters. At current time, this results in objects with magnetic field of the order of $10^{-1}\mu$G at all scales. Halos have a slightly higher values of $|B|$, while for filaments such value tends to mildly grow with the baryonic mass. Filaments from dynamo models show a stronger dependency from the mass, with values of $|B| \approx 10^{-2}\div 10^{-3}\mu$G at low masses, reaching values around $10^{-1}\mu$G at gas masses above $10^{13} M_{\odot}$. This follows from the efficiency of the dynamo mechanisms, which increases with the overdensity. A similar dependency from the mass is shown from the magnetic field in astrophysical models. At masses below $M^*_{BM}$, $|B|$ is always smaller than $10^{-2}\mu$G, with values below $10^{-4}\mu$G reached by various models. Above $M^*_{BM}$, no more dependence from the mass is shown, and only the CSF5 models reaches values for $|B|$ close to $10^{-1}\mu$G. The distinction between dynamo and astrophysical runs is larger when the local environment is considered, only the CSFBH5 model showing some overlap with DYN7. For embedded halos with masses above $M^*_{TOT,h}$, in all models the dispersion in the values of $|B|$ is small. Therefore, the local volume hosting massive halos appears to be a  promising target environment to characterise the mechanisms driving the evolution of magnetic fields in the cosmic web.

Our results suggest that the magnetic properties of the WHIM can effectively shed light on the mechanisms behind the evolution of cosmic filaments. 
For instance, the combination of the trend of magnetic fields and volume with filaments' mass found in our sample, indicates that the difference in Faraday Rotation ($\rm RM \propto L_{\rm LOS} ~n_e ~B $, where $L_{\rm LOS}$ is the length of the line of sight crossing the filament volume)  across the filament population varies by orders of magnitude, if extragalactic magnetic fields are mostly contributed by galaxy-formation related processes (i.e. in the astrophysical scenario). Conversely, the scatter across the population is expected to the of order  $\sim 10$, if magnetic fields have a primordial origin. In this case, in fact, the effect of compression is rather similar across filaments of very different masses. This addresses a viable test of magnetogenesis with filaments, provided that a large number of objects can be detected in the future using the SKA \citep[][]{va18frb}. Several tens of detected RM are in fact required to remove the contamination by RM intrinsic to sources \citep[e.g.][]{2018arXiv181111198L}. 
While these techniques rely on the signature by magnetic fields to detect the WHIM, we expect that with the SKA it should be possible to attempt also a detection of the coldest gas ($\sim 10^4 ~\rm K$) in the cosmic web using the HI hyperfine transition, even if the required sensitivity and statistics represent an even bigger challenge than that for synchrotron and Faraday Rotation studies  \citep[e.g.][]{2015aska.confE.132P,2017PASJ...69...73H}. 

Depending on their  true  magnetisation level, a significant role of filaments in the production of anisotropies in the propagation of UHECRs in the local Universe can be expected \citep[][]{Sigl:2003ay,Dolag:2003ra,hack16,hack18} even if a level of $\geq 10 \rm~ nG$ appears necessary  \citep[e.g.][]{2019arXiv190100627K}. \\

The baryon fraction results to be higher than in galaxies for most of the filaments, but in general less than the cosmic value in all models at almost all scales, except for the largest masses, where $ f_{\rm b}/f_{\rm b,cosm}$ is of the order of the unity or even bigger. In low mass objects, we measure a $\sim 5-10 \%$ significantly larger baryon fraction in low-mass filaments ($M_{\rm BM}\leq 10^{12} M_{\odot}$) simulated with cooling and feedback compared to non-radiative simulations. This highlights the potential role of AGN feedback and star winds from surrounding galaxies in expelling gas from their halos onto filaments \citep[e.g.][]{2006ApJ...650..560C,2016MNRAS.457.3024H}. In this case, also metal-rich gas is expected to be deposited in filaments, enriching their composition and boosting their line emission and absorption signatures \citep[e.g.][]{2006MNRAS.373.1265O,2018arXiv181001883M}. Critical to further explore these features with numerical simulation is the access to even higher resolution and more complex chemical networks \citep[e.g.][]{2018arXiv181112410H,2019MNRAS.482.4906P}. The analysis of the local volumes hosting halos shows, despite a large scatter, the tendency of the gas surrounding galaxies to have a large baryon fraction, often significantly above the cosmic value. On the contrary, the vast majority of the halos have $0.1 \le f_{\rm b}/f_{\rm b,cosm}\le 0.5$. Differently from the whole filament, no meaningful differences are found between the different models. Irrespective to the details of the galaxy formation physics, the gas in the surrounding environments is fed by the matter ejected from the corresponding halos, confirming the expectation of metal-enrichment discussed above.


In agreement with \citet{gh16}, we have found several scaling laws that relates the mass, the length and the temperature of a filament with the properties of the embedded halos. These scaling laws have small dispersion with slightly different trends at low and high ($>M^*_{TOT,h}$) masses, the latter having a higher slope. These relations proved to be substantially the same in all models, providing a general tool to infer the overall properties of a filaments starting from those of galaxies, useful aid in their detection and identification. In addition, a sharp lower limit in the temperature of the WHIM surrounding halos of a given mass has been identified. Also this information can be exploited to constrain the properties and help in the detection of the filament's gas pointing to the neighbourhood of the resident halos. 

\section{Conclusions}

Cosmic filaments are a fundamental building block of the cosmic web \citep[e.g.][]{1996Natur.380..603B,2001ApJ...552..473D,2014MNRAS.441.2923C,2018arXiv181001883M} and potentially outstanding instruments to access the past evolution of cosmic gas, preserving traces of the fundamental processes which took place several $\rm Gyr$ in the past,  from the enrichment of complex chemical elements and metals \citep[e.g.][]{2006MNRAS.373.1265O,2011ApJ...731....6S,2016xnnd.confE..27N}, to the injection and evolution of non-thermal particles in large-scale structures\citep[e.g.][]{pf07,va14curie}, to the seeding and dynamo amplification of extragalactic magnetic fields \citep[e.g.][]{ry08,va14mhd,2015MNRAS.453.3999M}. \\

In this new work, thanks to a large statistics of cosmological simulation, we accomplished an extensive survey the thermal and non-thermal properties of filaments in the cosmic web. 

In the following summary we list those characteristics that can be considered robust predictions of our simulations. We start from those properties that show only a weak dependency on the investigated variations of baryonic physics or magnetic field seeding scenario: 

\begin{itemize}
\item  the global structural properties of filaments (in particular their length, mass and average temperature);
\item the global baryon fraction enclosed within the filaments volume;
\item the average profiles of gas and dark matter, temperature and baryon fraction with respect to the central axis of filaments in the $ \geq 10^{13} M_{\odot}$ mass range;
\item the mass, temperature and velocity distribution of galaxy halos in filaments, as well as the clustering properties, as shown by the 3D 2-points correlation function of halos within filaments; 
\item the temperature and velocity of the halos, which are consistent with the environment in which they live. In particular the velocity of the halos traces the velocity of the fluid;
\item the baryon fraction, for which different values, significantly smaller than the unity in halos, of the order of the cosmic value or even larger for the local halo's environment and slightly smaller than one in the rest of the filament, can be identified.
\end{itemize}

Our study highlighted also several properties which appear to  significantly depend on the assumed baryonic physics or magnetic physical models: 

\begin{itemize}
\item  the $T \propto M_{fil,BM}^{\rm \alpha}$ relation is steeper in radiative runs with feedback, compared to non-radiative runs (i.e. $\alpha \approx 0.35$ vs $\alpha \approx 0.3$), and the most massive filaments are on average a factor $\sim 2$ hotter if feedback from galaxies and AGN is active. Differences between models are smaller when the total halo mass (instead of the total baryonic mass) is considered. In this case the relation appears to be bi-modal with the slope becoming higher above $10^{13} M_{\odot}$;
\item  the relation between the mass and the mean magnetic fields of filaments hugely varies in slope and normalisation if different seeding scenarios are assumed: $B_M \propto M_{fil,BM}^{0.15}$ in primordial models, $B_M \propto M^{0.4}_{fil,BM}$ in dynamo models and $B_M \propto M^{1.8}_{fil,BM}$ in astrophysical seeding models. Similar differences are found considering the halo mass. Galactic halos present a larger scatter in their magnetic fields in astrophysical seeding models, as well as larger variations with respect to their environment;
\item  the velocity profile and the magnetic field profile with respect to the central axis of filaments in the $ \geq 10^{13} M_{\odot}$ mass range is different in the different scenarios. Feedback produces a flatter velocity profile out to $\sim 4 ~\rm Mpc$ from the axis of filaments due to outflows, while the magnetic field profile in astrophysical seeding models significantly drops for $\geq 2$ Mpc from the filaments' central axis, due to the scarcity of magnetisation sources.
\end{itemize}

While this study attempted an exhaustive study of the large-scale properties of filaments in the cosmic web, based on the most complete to date suite of physical variations of WHIM, a natural (yet challenging) follow-up is represented by the access to a higher spatial and mass resolution, besides the adoption of more sophisticated and accurate description of the physics underlying the evolution of cosmic filaments and galactic halos. This will allow us to include and probe galaxy formation processes, providing a direct link between galaxies and the large scale distribution of the gas. 

Furthermore, our current results will be extended in order to investigate in details how the properties of filaments maps to observables, in particular in radio, potentially detectable by the coming generations of radiotelescopes, focusing on synchrotron and HI emission and Faraday Rotation measurements. In addition other interesting phenomena, like the potential impact of filaments on UHECRs emission, can be subject of additional in-depth analysis. 


\section*{Acknowledgements}

The cosmological simulations were performed with the {\enzo} code (http://enzo-project.org), which is the product of a collaborative effort of scientists at many universities and national laboratories. We gratefully acknowledge the {\enzo} development group for providing extremely helpful and well-maintained on-line documentation and tutorials.
F.V. acknowledges financial support from the ERC  Starting Grant "MAGCOW", no. 714196.   
The simulations on which this work is based have been produced on Piz Daint supercomputer at CSCS-ETHZ (Lugano, Switzerland) under projects s701 and s805 and on the J\"ulich Supercomputing Centre (JFZ) under project HHH42, 
in numerical projects with F.V. as PI. 
We also acknowledge the usage of online storage tools kindly provided by the Inaf Astronomica Archive (IA2) initiave (http://www.ia2.inaf.it).  We thank Marcus Br\"{u}ggen for his support in the first production of the Chronos++ suite of simulations employed in this work, and to Jean Favre for his support in the first design of the Visit filament finder employed in this work.

\bibliographystyle{mnras}
\bibliography{main}

\begin{thebibliography}{116}
\expandafter\ifx\csname natexlab\endcsname\relax\def\natexlab#1{#1}\fi

\bibitem[{Ade {et~al}\mbox{.}(2015)Ade {et~al.}}]{PLANCK2015}
Ade P. A.~R., {et~al.}, 2015

\bibitem[{{Akahori} {et~al}\mbox{.}(2016){Akahori}, {Ryu}, \&
  {Gaensler}}]{2016ApJ...824..105A}
{Akahori} T., {Ryu} D., {Gaensler} B.~M., 2016, \apj, 824, 105

\bibitem[{{Alpaslan} {et~al}\mbox{.}(2014){Alpaslan}, {Robotham}, {Driver},
  {Norberg}, {Baldry}, {Bauer}, {Bland-Hawthorn}, {Brown}, {Cluver}, {Colless},
  {Foster}, {Hopkins}, {Van Kampen}, {Kelvin}, {Lara-Lopez}, {Liske},
  {Lopez-Sanchez}, {Loveday}, {McNaught-Roberts}, {Merson}, \&
  {Pimbblet}}]{alp14a}
{Alpaslan} M. {et~al.}, 2014, \mnras, 438, 177

\bibitem[{{Andreon}(2010)}]{2010MNRAS.407..263A}
{Andreon} S., 2010, \mnras, 407, 263

\bibitem[{{Arag{\'o}n-Calvo} {et~al}\mbox{.}(2007){Arag{\'o}n-Calvo}, {Jones},
  {van de Weygaert}, \& {van der Hulst}}]{2007A&A...474..315A}
{Arag{\'o}n-Calvo} M.~A., {Jones} B.~J.~T., {van de Weygaert} R., {van der
  Hulst} J.~M., 2007, \aap, 474, 315

\bibitem[{{Aragon-Calvo} {et~al}\mbox{.}(2010){Aragon-Calvo}, {Shandarin}, \&
  {Szalay}}]{2010arXiv1006.4178A}
{Aragon-Calvo} M.~A., {Shandarin} S.~F., {Szalay} A., 2010, arXiv e-prints

\bibitem[{{Arlen} {et~al}\mbox{.}(2014){Arlen}, {Vassilev}, {Weisgarber},
  {Wakely}, \& {Yusef Shafi}}]{2014ApJ...796...18A}
{Arlen} T.~C., {Vassilev} V.~V., {Weisgarber} T., {Wakely} S.~P., {Yusef Shafi}
  S., 2014, \apj, 796, 18

\bibitem[{{Beck} {et~al}\mbox{.}(2013){Beck}, {Hanasz}, {Lesch}, {Remus}, \&
  {Stasyszyn}}]{beck13}
{Beck} A.~M., {Hanasz} M., {Lesch} H., {Remus} R.-S., {Stasyszyn} F.~A., 2013,
  \mnras, 429, L60

\bibitem[{{Beresnyak} \& {Miniati}(2016)}]{bm16}
{Beresnyak} A., {Miniati} F., 2016, \apj, 817, 127

\bibitem[{{Blasi} {et~al}\mbox{.}(1999){Blasi}, {Burles}, \&
  {Olinto}}]{Blasi.Burles..1999}
{Blasi} P., {Burles} S., {Olinto} A.~V., 1999, \apjl, 514, L79

\bibitem[{{Bond} {et~al}\mbox{.}(1996){Bond}, {Kofman}, \&
  {Pogosyan}}]{1996Natur.380..603B}
{Bond} J.~R., {Kofman} L., {Pogosyan} D., 1996, \nat, 380, 603

\bibitem[{{Borgani} {et~al}\mbox{.}(2008){Borgani}, {Diaferio}, {Dolag}, \&
  {Schindler}}]{borgani08}
{Borgani} S., {Diaferio} A., {Dolag} K., {Schindler} S., 2008, \ssr, 134, 269

\bibitem[{{Broderick} {et~al}\mbox{.}(2012){Broderick}, {Chang}, \&
  {Pfrommer}}]{2012ApJ...752...22B}
{Broderick} A.~E., {Chang} P., {Pfrommer} C., 2012, \apj, 752, 22

\bibitem[{{Brown} {et~al}\mbox{.}(2017){Brown}, {Vernstrom}, {Carretti},
  {Dolag}, {Gaensler}, {Staveley-Smith}, {Bernardi}, {Haverkorn}, {Kesteven},
  \& {Poppi}}]{brown17}
{Brown} S. {et~al.}, 2017, \mnras, 468, 4246

\bibitem[{{Brown}(2011)}]{2011JApA...32..577B}
{Brown} S.~D., 2011, Journal of Astrophysics and Astronomy, 32, 577

\bibitem[{{Br{\"u}ggen} {et~al}\mbox{.}(2005){Br{\"u}ggen}, {Ruszkowski},
  {Simionescu}, {Hoeft}, \& {Dalla Vecchia}}]{br05}
{Br{\"u}ggen} M., {Ruszkowski} M., {Simionescu} A., {Hoeft} M., {Dalla Vecchia}
  C., 2005, \apjl, 631, L21

\bibitem[{{Bryan} {et~al}\mbox{.}(2014){Bryan}, {Norman}, {O'Shea}, {Abel},
  {Wise}, {Turk}, {Reynolds}, {Collins}, {Wang}, {Skillman}, {Smith},
  {Harkness}, {Bordner}, {Kim}, {Kuhlen}, {Xu}, {Goldbaum}, {Hummels},
  {Kritsuk}, {Tasker}, {Skory}, {Simpson}, {Hahn}, {Oishi}, {So}, {Zhao},
  {Cen}, {Li}, \& {Enzo Collaboration}}]{enzo14}
{Bryan} G.~L. {et~al.}, 2014, \apjs, 211, 19

\bibitem[{{Caprini} \& {Gabici}(2015)}]{2015PhRvD..91l3514C}
{Caprini} C., {Gabici} S., 2015, \prd, 91, 123514

\bibitem[{{Cautun} {et~al}\mbox{.}(2014){Cautun}, {van de Weygaert}, {Jones},
  \& {Frenk}}]{2014MNRAS.441.2923C}
{Cautun} M., {van de Weygaert} R., {Jones} B.~J.~T., {Frenk} C.~S., 2014,
  \mnras, 441, 2923

\bibitem[{{Cen} \& {Ostriker}(2006)}]{2006ApJ...650..560C}
{Cen} R., {Ostriker} J.~P., 2006, \apj, 650, 560

\bibitem[{{Chen} {et~al}\mbox{.}(2015){Chen}, {Ho}, {Tenneti}, {Mandelbaum},
  {Croft}, {DiMatteo}, {Freeman}, {Genovese}, \&
  {Wasserman}}]{2015MNRAS.454.3341C}
{Chen} Y.-C. {et~al.}, 2015, \mnras, 454, 3341

\bibitem[{Childs {et~al}\mbox{.}(2011)Childs, Brugger, Whitlock, Meredith,
  Ahern, Bonnell, Miller, Weber, Harrison, Fogal, Garth, S, Bethel, Durant,
  Camp, Favre, Rübel, Navrátil, Α, Α, \& Vivodtzev}]{Childs11visit:an}
Childs H. {et~al.}, 2011, in In Proceedings of SciDAC

\bibitem[{{Dav{\'e}} {et~al}\mbox{.}(2001){Dav{\'e}}, {Cen}, {Ostriker},
  {Bryan}, {Hernquist}, {Katz}, {Weinberg}, {Norman}, \&
  {O'Shea}}]{2001ApJ...552..473D}
{Dav{\'e}} R. {et~al.}, 2001, \apj, 552, 473

\bibitem[{{de Graaff} {et~al}\mbox{.}(2017){de Graaff}, {Cai}, {Heymans}, \&
  {Peacock}}]{2017arXiv170910378D}
{de Graaff} A., {Cai} Y.-C., {Heymans} C., {Peacock} J.~A., 2017, arXiv
  e-prints

\bibitem[{{Dedner} {et~al}\mbox{.}(2002){Dedner}, {Kemm}, {Kr{\"o}ner}, {Munz},
  {Schnitzer}, \& {Wesenberg}}]{ded02}
{Dedner} A., {Kemm} F., {Kr{\"o}ner} D., {Munz} C.-D., {Schnitzer} T.,
  {Wesenberg} M., 2002, Journal of Computational Physics, 175, 645

\bibitem[{Dolag {et~al}\mbox{.}(2003)Dolag, Grasso, Springel, \&
  Tkachev}]{Dolag:2003ra}
Dolag K., Grasso D., Springel V., Tkachev I., 2003

\bibitem[{{Donnert} {et~al}\mbox{.}(2009){Donnert}, {Dolag}, {Lesch}, \&
  {M{\"u}ller}}]{donn09}
{Donnert} J., {Dolag} K., {Lesch} H., {M{\"u}ller} E., 2009, \mnras, 392, 1008

\bibitem[{{Donnert} {et~al}\mbox{.}(2018){Donnert}, {Vazza}, {Br{\"u}ggen}, \&
  {ZuHone}}]{review_dynamo}
{Donnert} J., {Vazza} F., {Br{\"u}ggen} M., {ZuHone} J., 2018, ArXiv e-prints

\bibitem[{{Eckert} {et~al}\mbox{.}(2015){Eckert}, {Jauzac}, {Shan}, {Kneib},
  {Erben}, {Israel}, {Jullo}, {Klein}, {Massey}, {Richard}, \&
  {Tchernin}}]{2015Natur.528..105E}
{Eckert} D. {et~al.}, 2015, \nat, 528, 105

\bibitem[{{Eckmiller} {et~al}\mbox{.}(2011){Eckmiller}, {Hudson}, \&
  {Reiprich}}]{2011A&A...535A.105E}
{Eckmiller} H.~J., {Hudson} D.~S., {Reiprich} T.~H., 2011, \aap, 535, A105

\bibitem[{{Federrath} {et~al}\mbox{.}(2011){Federrath}, {Chabrier},
  {S2015aska.confE..95Bber}, {Banerjee}, {Klessen}, \&
  {Schleicher}}]{2011PhRvL.107k4504F}
{Federrath} C., {Chabrier} G., {S2015aska.confE..95Bber} J., {Banerjee} R.,
  {Klessen} R.~S., {Schleicher} D.~R.~G., 2011, Physical Review Letters, 107,
  114504

\bibitem[{{Federrath} {et~al}\mbox{.}(2014){Federrath}, {Schober}, {Bovino}, \&
  {Schleicher}}]{fed14}
{Federrath} C., {Schober} J., {Bovino} S., {Schleicher} D.~R.~G., 2014, \apjl,
  797, L19

\bibitem[{{Finoguenov} {et~al}\mbox{.}(2003){Finoguenov}, {Briel}, \&
  {Henry}}]{2003A&A...410..777F}
{Finoguenov} A., {Briel} U.~G., {Henry} J.~P., 2003, \aap, 410, 777

\bibitem[{{Genel} {et~al}\mbox{.}(2014){Genel}, {Vogelsberger}, {Springel},
  {Sijacki}, {Nelson}, {Snyder}, {Rodriguez-Gomez}, {Torrey}, \&
  {Hernquist}}]{2014MNRAS.445..175G}
{Genel} S. {et~al.}, 2014, \mnras, 445, 175

\bibitem[{{Gheller} {et~al}\mbox{.}(2018){Gheller}, {Vazza}, \&
  {Bonafede}}]{cosmodeep}
{Gheller} C., {Vazza} F., {Bonafede} A., 2018, \mnras, 480, 3749

\bibitem[{{Gheller} {et~al}\mbox{.}(2016){Gheller}, {Vazza}, {Br{\"u}ggen},
  {Alpaslan}, {Holwerda}, {Hopkins}, \& {Liske}}]{gh16}
{Gheller} C., {Vazza} F., {Br{\"u}ggen} M., {Alpaslan} M., {Holwerda} B.~W.,
  {Hopkins} A.~M., {Liske} J., 2016, \mnras, 462, 448

\bibitem[{{Gheller} {et~al}\mbox{.}(2015){Gheller}, {Vazza}, {Favre}, \&
  {Br{\"u}ggen}}]{gh15}
{Gheller} C., {Vazza} F., {Favre} J., {Br{\"u}ggen} M., 2015, \mnras, 453, 1164

\bibitem[{{Giodini} {et~al}\mbox{.}(2009){Giodini}, {Pierini}, {Finoguenov},
  {Pratt}, {Boehringer}, {Leauthaud}, {Guzzo}, {Aussel}, {Bolzonella}, {Capak},
  {Elvis}, {Hasinger}, {Ilbert}, {Kartaltepe}, {Koekemoer}, {Lilly}, {Massey},
  {McCracken}, {Rhodes}, {Salvato}, {Sanders}, {Scoville}, {Sasaki}, {Smolcic},
  {Taniguchi}, {Thompson}, \& {COSMOS Collaboration}}]{2009ApJ...703..982G}
{Giodini} S. {et~al.}, 2009, \apj, 703, 982

\bibitem[{{Gonz{\'a}lez} \& {Padilla}(2010)}]{2010MNRAS.407.1449G}
{Gonz{\'a}lez} R.~E., {Padilla} N.~D., 2010, \mnras, 407, 1449

\bibitem[{{Grete} {et~al}\mbox{.}(2016){Grete}, {Vlaykov}, {Schmidt}, \&
  {Schleicher}}]{gr16}
{Grete} P., {Vlaykov} D.~G., {Schmidt} W., {Schleicher} D.~R.~G., 2016, Physics
  of Plasmas, 23, 062317

\bibitem[{{Hackstein} {et~al}\mbox{.}(2016){Hackstein}, {Vazza}, {Br{\"u}ggen},
  {Sigl}, \& {Dundovic}}]{hack16}
{Hackstein} S., {Vazza} F., {Br{\"u}ggen} M., {Sigl} G., {Dundovic} A., 2016,
  \mnras, 462, 3660

\bibitem[{{Hackstein} {et~al}\mbox{.}(2018){Hackstein}, {Vazza}, {Br{\"u}ggen},
  {Sorce}, \& {Gottl{\"o}ber}}]{hack18}
{Hackstein} S., {Vazza} F., {Br{\"u}ggen} M., {Sorce} J.~G., {Gottl{\"o}ber}
  S., 2018, \mnras, 475, 2519

\bibitem[{{Hahn} {et~al}\mbox{.}(2007){Hahn}, {Porciani}, {Carollo}, \&
  {Dekel}}]{2007MNRAS.375..489H}
{Hahn} O., {Porciani} C., {Carollo} C.~M., {Dekel} A., 2007, \mnras, 375, 489

\bibitem[{{Haider} {et~al}\mbox{.}(2016){Haider}, {Steinhauser},
  {Vogelsberger}, {Genel}, {Springel}, {Torrey}, \&
  {Hernquist}}]{2016MNRAS.457.3024H}
{Haider} M., {Steinhauser} D., {Vogelsberger} M., {Genel} S., {Springel} V.,
  {Torrey} P., {Hernquist} L., 2016, \mnras, 457, 3024

\bibitem[{Hoekstra {et~al}\mbox{.}(2005)Hoekstra, Hsieh, Yee, Lin, \&
  Gladders}]{0004-637X-635-1-73}
Hoekstra H., Hsieh B.~C., Yee H. K.~C., Lin H., Gladders M.~D., 2005, The
  Astrophysical Journal, 635, 73

\bibitem[{{Hopkins} \& {Raives}(2016)}]{2016MNRAS.455...51H}
{Hopkins} P.~F., {Raives} M.~J., 2016, \mnras, 455, 51

\bibitem[{{Horii} {et~al}\mbox{.}(2017){Horii}, {Asaba}, {Hasegawa}, \&
  {Tashiro}}]{2017PASJ...69...73H}
{Horii} T., {Asaba} S., {Hasegawa} K., {Tashiro} H., 2017, \pasj, 69, 73

\bibitem[{{Hummels} {et~al}\mbox{.}(2018){Hummels}, {Smith}, {Hopkins},
  {O'Shea}, {Silvia}, {Werk}, {Lehner}, {Wise}, {Collins}, \&
  {Butsky}}]{2018arXiv181112410H}
{Hummels} C.~B. {et~al.}, 2018, arXiv e-prints

\bibitem[{{Jones} {et~al}\mbox{.}(2011){Jones}, {Porter}, {Ryu}, \&
  {Cho}}]{jones11}
{Jones} T.~W., {Porter} D.~H., {Ryu} D., {Cho} J., 2011, \memsai, 82, 588

\bibitem[{{Kang} {et~al}\mbox{.}(2007){Kang}, {Ryu}, {Cen}, \&
  {Ostriker}}]{ka07}
{Kang} H., {Ryu} D., {Cen} R., {Ostriker} J.~P., 2007, \apj, 669, 729

\bibitem[{{Kennicutt}(1998)}]{1998ApJ...498..541K}
{Kennicutt}, Jr. R.~C., 1998, \apj, 498, 541

\bibitem[{{Kim} {et~al}\mbox{.}(2019){Kim}, {Ryu}, {Kang}, {Kim}, \&
  {Rey}}]{2019arXiv190100627K}
{Kim} J., {Ryu} D., {Kang} H., {Kim} S., {Rey} S.-C., 2019, arXiv e-prints

\bibitem[{{Kim} {et~al}\mbox{.}(2011){Kim}, {Wise}, {Alvarez}, \&
  {Abel}}]{2011ApJ...738...54K}
{Kim} J.-h., {Wise} J.~H., {Alvarez} M.~A., {Abel} T., 2011, \apj, 738, 54

\bibitem[{{Kravtsov}(2003)}]{2003ApJ...590L...1K}
{Kravtsov} A.~V., 2003, \apjl, 590, L1

\bibitem[{{Kritsuk} {et~al}\mbox{.}(2011){Kritsuk}, {Nordlund}, {Collins},
  {Padoan}, {Norman}, {Abel}, {Banerjee}, {Federrath}, {Flock}, {Lee}, {Li},
  {M{\"u}ller}, {Teyssier}, {Ustyugov}, {Vogel}, \& {Xu}}]{kri11}
{Kritsuk} A.~G. {et~al.}, 2011, \apj, 737, 13

\bibitem[{{Locatelli} {et~al}\mbox{.}(2018){Locatelli}, {Vazza}, \&
  {Dom{\'{\i}}nguez-Fern{\'a}ndez}}]{2018arXiv181111198L}
{Locatelli} N., {Vazza} F., {Dom{\'{\i}}nguez-Fern{\'a}ndez} P., 2018, arXiv
  e-prints

\bibitem[{{Madau} \& {Dickinson}(2014)}]{2014ARA&A..52..415M}
{Madau} P., {Dickinson} M., 2014, \araa, 52, 415

\bibitem[{{Marinacci} \& {Vogelsberger}(2016)}]{2016MNRAS.456L..69M}
{Marinacci} F., {Vogelsberger} M., 2016, \mnras, 456, L69

\bibitem[{{Marinacci} {et~al}\mbox{.}(2015){Marinacci}, {Vogelsberger}, {Mocz},
  \& {Pakmor}}]{2015MNRAS.453.3999M}
{Marinacci} F., {Vogelsberger} M., {Mocz} P., {Pakmor} R., 2015, \mnras, 453,
  3999

\bibitem[{{Marinacci} {et~al}\mbox{.}(2017){Marinacci}, {Vogelsberger},
  {Pakmor}, {Torrey}, {Springel}, {Hernquist}, {Nelson}, {Weinberger},
  {Pillepich}, {Naiman}, \& {Genel}}]{ma17}
{Marinacci} F. {et~al.}, 2017, ArXiv e-prints

\bibitem[{{Martizzi} {et~al}\mbox{.}(2018){Martizzi}, {Vogelsberger}, {Artale},
  {Haider}, {Torrey}, {Marinacci}, {Nelson}, {Pillepich}, {Weinberger},
  {Hernquist}, {Naiman}, \& {Springel}}]{2018arXiv181001883M}
{Martizzi} D. {et~al.}, 2018, arXiv e-prints

\bibitem[{Meredith(2004)}]{Meredith2004}
Meredith J.~S., 2004, in Nuclear Explosives Code Developers Conference (NECDC)

\bibitem[{{Mogavero} \& {Schekochihin}(2014)}]{2014MNRAS.440.3226M}
{Mogavero} F., {Schekochihin} A.~A., 2014, \mnras, 440, 3226

\bibitem[{{Nagai} \& {Lau}(2011)}]{nala11}
{Nagai} D., {Lau} E.~T., 2011, \apjl, 731, L10

\bibitem[{{Neronov} \& {Vovk}(2010)}]{2010Sci...328...73N}
{Neronov} A., {Vovk} I., 2010, Science, 328, 73

\bibitem[{{Nicastro}(2016)}]{2016xnnd.confE..27N}
{Nicastro} F., 2016, in XMM-Newton: The Next Decade, p.~27

\bibitem[{{Nicastro} {et~al}\mbox{.}(2010){Nicastro}, {Krongold}, {Fields},
  {Conciatore}, {Zappacosta}, {Elvis}, {Mathur}, \&
  {Papadakis}}]{2010ApJ...715..854N}
{Nicastro} F., {Krongold} Y., {Fields} D., {Conciatore} M.~L., {Zappacosta} L.,
  {Elvis} M., {Mathur} S., {Papadakis} I., 2010, \apj, 715, 854

\bibitem[{{Oppenheimer} \& {Dav{\'e}}(2006)}]{2006MNRAS.373.1265O}
{Oppenheimer} B.~D., {Dav{\'e}} R., 2006, \mnras, 373, 1265

\bibitem[{{Oppermann} {et~al}\mbox{.}(2015){Oppermann}, {Junklewitz},
  {Greiner}, {En{\ss}lin}, {Akahori}, {Carretti}, {Gaensler}, {Goobar},
  {Harvey-Smith}, {Johnston-Hollitt}, {Pratley}, {Schnitzeler}, {Stil}, \&
  {Vacca}}]{2015A&A...575A.118O}
{Oppermann} N. {et~al.}, 2015, \aap, 575, A118

\bibitem[{{O'Sullivan} {et~al}\mbox{.}(2018){O'Sullivan}, {Machalski}, {Van
  Eck}, {Heald}, {Brueggen}, {Fynbo}, {Heintz}, {Lara-Lopez}, {Vacca},
  {Hardcastle}, {Shimwell}, {Tasse}, {Vazza}, {Andernach}, {Birkinshaw},
  {Haverkorn}, {Horellou}, {Williams}, {Harwood}, {Brunetti}, {Anderson},
  {Mao}, {Nikiel-Wroczynski}, {Takahashi}, {Carretti}, {Vernstrom}, {van
  Weeren}, {Orru}, {Morabito}, \& {Callingham}}]{2018arXiv181107934O}
{O'Sullivan} S.~P. {et~al.}, 2018, ArXiv e-prints

\bibitem[{{Papastergis} {et~al}\mbox{.}(2012){Papastergis}, {Cattaneo},
  {Huang}, {Giovanelli}, \& {Haynes}}]{2012ApJ...759..138P}
{Papastergis} E., {Cattaneo} A., {Huang} S., {Giovanelli} R., {Haynes} M.~P.,
  2012, \apj, 759, 138

\bibitem[{{Pfrommer} {et~al}\mbox{.}(2007){Pfrommer}, {En{\ss}lin}, {Springel},
  {Jubelgas}, \& {Dolag}}]{pf07}
{Pfrommer} C., {En{\ss}lin} T.~A., {Springel} V., {Jubelgas} M., {Dolag} K.,
  2007, \mnras, 378, 385

\bibitem[{{Pfrommer} {et~al}\mbox{.}(2006){Pfrommer}, {Springel}, {En{\ss}lin},
  \& {Jubelgas}}]{pf06}
{Pfrommer} C., {Springel} V., {En{\ss}lin} T.~A., {Jubelgas} M., 2006, \mnras,
  367, 113

\bibitem[{{Planck Collaboration} {et~al}\mbox{.}(2013){Planck Collaboration},
  {Ade}, {Aghanim}, {Arnaud}, {Ashdown}, {Atrio-Barandela}, {Aumont},
  {Baccigalupi}, {Balbi}, {Banday}, \& et~al.}]{2013A&A...550A.134P}
{Planck Collaboration} {et~al.}, 2013, \aap, 550, A134

\bibitem[{{Planck Collaboration} {et~al}\mbox{.}(2016){Planck Collaboration},
  {Ade}, {Aghanim}, {Arnaud}, {Ashdown}, {Aumont}, {Baccigalupi}, {Banday},
  {Barreiro}, {Bartlett}, \& et~al.}]{2016A&A...594A..13P}
{Planck Collaboration} {et~al.}, 2016, \aap, 594, A13

\bibitem[{{Popping} {et~al}\mbox{.}(2015){Popping}, {Meyer}, {Staveley-Smith},
  {Obreschkow}, {Jozsa}, \& {Pisano}}]{2015aska.confE.132P}
{Popping} A., {Meyer} M., {Staveley-Smith} L., {Obreschkow} D., {Jozsa} G.,
  {Pisano} D.~J., 2015, Advancing Astrophysics with the Square Kilometre Array
  (AASKA14), 132

\bibitem[{{Popping} {et~al}\mbox{.}(2019){Popping}, {Narayanan}, {Somerville},
  {Faisst}, \& {Krumholz}}]{2019MNRAS.482.4906P}
{Popping} G., {Narayanan} D., {Somerville} R.~S., {Faisst} A.~L., {Krumholz}
  M.~R., 2019, \mnras, 482, 4906

\bibitem[{{Porter} {et~al}\mbox{.}(2015){Porter}, {Jones}, \& {Ryu}}]{po15}
{Porter} D.~H., {Jones} T.~W., {Ryu} D., 2015, \apj, 810, 93

\bibitem[{{Pshirkov} {et~al}\mbox{.}(2016){Pshirkov}, {Tinyakov}, \&
  {Urban}}]{2016PhRvL.116s1302P}
{Pshirkov} M.~S., {Tinyakov} P.~G., {Urban} F.~R., 2016, Physical Review
  Letters, 116, 191302

\bibitem[{{Reichert} {et~al}\mbox{.}(2011){Reichert}, {B{\"o}hringer},
  {Fassbender}, \& {M{\"u}hlegger}}]{2011A&A...535A...4R}
{Reichert} A., {B{\"o}hringer} H., {Fassbender} R., {M{\"u}hlegger} M., 2011,
  \aap, 535, A4

\bibitem[{{Rieder} \& {Teyssier}(2016)}]{2016MNRAS.457.1722R}
{Rieder} M., {Teyssier} R., 2016, \mnras, 457, 1722

\bibitem[{{Ryu} {et~al}\mbox{.}(2008){Ryu}, {Kang}, {Cho}, \& {Das}}]{ry08}
{Ryu} D., {Kang} H., {Cho} J., {Das} S., 2008, Science, 320, 909

\bibitem[{{Ryu} {et~al}\mbox{.}(2003){Ryu}, {Kang}, {Hallman}, \&
  {Jones}}]{ry03}
{Ryu} D., {Kang} H., {Hallman} E., {Jones} T.~W., 2003, \apj, 593, 599

\bibitem[{{Samui} {et~al}\mbox{.}(2017){Samui}, {Subramanian}, \&
  {Srianand}}]{sam17}
{Samui} S., {Subramanian} K., {Srianand} R., 2017, ArXiv e-prints

\bibitem[{{Schleicher} {et~al}\mbox{.}(2013){Schleicher}, {Schober},
  {Federrath}, {Bovino}, \& {Schmidt}}]{2013NJPh...15b3017S}
{Schleicher} D.~R.~G., {Schober} J., {Federrath} C., {Bovino} S., {Schmidt} W.,
  2013, New Journal of Physics, 15, 023017

\bibitem[{{Schober} {et~al}\mbox{.}(2013){Schober}, {Schleicher}, \&
  {Klessen}}]{schober13}
{Schober} J., {Schleicher} D.~R.~G., {Klessen} R.~S., 2013, \aap, 560, A87

\bibitem[{{Shu} \& {Osher}(1988)}]{1988JCoPh..77..439S}
{Shu} C.-W., {Osher} S., 1988, Journal of Computational Physics, 77, 439

\bibitem[{Sigl {et~al}\mbox{.}(2003)Sigl, Miniati, \& Ensslin}]{Sigl:2003ay}
Sigl G., Miniati F., Ensslin T.~A., 2003, Phys. Rev., D68, 043002

\bibitem[{{Smith} {et~al}\mbox{.}(2011){Smith}, {Hallman}, {Shull}, \&
  {O'Shea}}]{2011ApJ...731....6S}
{Smith} B.~D., {Hallman} E.~J., {Shull} J.~M., {O'Shea} B.~W., 2011, \apj, 731,
  6

\bibitem[{{Sousbie} {et~al}\mbox{.}(2008){Sousbie}, {Pichon}, {Colombi},
  {Novikov}, \& {Pogosyan}}]{2008MNRAS.383.1655S}
{Sousbie} T., {Pichon} C., {Colombi} S., {Novikov} D., {Pogosyan} D., 2008,
  \mnras, 383, 1655

\bibitem[{{Stasyszyn} {et~al}\mbox{.}(2013){Stasyszyn}, {Dolag}, \&
  {Beck}}]{2013MNRAS.428...13S}
{Stasyszyn} F.~A., {Dolag} K., {Beck} A.~M., 2013, \mnras, 428, 13

\bibitem[{{Stoica} {et~al}\mbox{.}(2005){Stoica}, {Mart{\'{\i}}nez}, {Mateu},
  \& {Saar}}]{2005A&A...434..423S}
{Stoica} R.~S., {Mart{\'{\i}}nez} V.~J., {Mateu} J., {Saar} E., 2005, \aap,
  434, 423

\bibitem[{{Tanimura} {et~al}\mbox{.}(2017){Tanimura}, {Hinshaw}, {McCarthy},
  {Van Waerbeke}, {Aghanim}, {Ma}, {Mead}, {Hojjati}, \&
  {Tr{\"o}ster}}]{2017arXiv170905024T}
{Tanimura} H. {et~al.}, 2017, arXiv e-prints

\bibitem[{{Tricco} {et~al}\mbox{.}(2016){Tricco}, {Price}, \&
  {Federrath}}]{2016MNRAS.461.1260T}
{Tricco} T.~S., {Price} D.~J., {Federrath} C., 2016, \mnras, 461, 1260

\bibitem[{{Trivedi} {et~al}\mbox{.}(2014){Trivedi}, {Subramanian}, \&
  {Seshadri}}]{2014PhRvD..89d3523T}
{Trivedi} P., {Subramanian} K., {Seshadri} T.~R., 2014, \prd, 89, 043523

\bibitem[{{Vacca} {et~al}\mbox{.}(2018){Vacca}, {Murgia}, {Loi}, {Vazza},
  {Finoguenov}, {Carretti}, {Feretti}, {Giovannini}, {Concu}, {Melis},
  {Gheller}, {Paladino}, {Poppi}, {Valente}, {Bernardi}, {Boschin}, {Brienza},
  {Clarke}, {Colafrancesco}, {En{\ss}lin}, {Ferrari}, {de Gasperin},
  {Gastaldello}, {Girardi}, {Gregorini}, {Johnston-Hollitt}, {Junklewitz},
  {Orr{\`u}}, {Parma}, {Perley}, \& {Taylor}}]{2018MNRAS.tmp.1093V}
{Vacca} V. {et~al.}, 2018, \mnras

\bibitem[{Vazza {et~al}\mbox{.}(2017)Vazza, Brueggen, Gheller, Hackstein,
  Wittor, \& Hinz}]{va17cqg}
Vazza F., Brueggen M., Gheller C., Hackstein S., Wittor D., Hinz P.~M., 2017,
  Classical and Quantum Gravity

\bibitem[{{Vazza} {et~al}\mbox{.}(2013{\natexlab{a}}){Vazza}, {Br{\"u}ggen}, \&
  {Gheller}}]{va13feedback}
{Vazza} F., {Br{\"u}ggen} M., {Gheller} C., 2013{\natexlab{a}}, \mnras, 428,
  2366

\bibitem[{{Vazza} {et~al}\mbox{.}(2014{\natexlab{a}}){Vazza}, {Br{\"u}ggen},
  {Gheller}, \& {Wang}}]{va14mhd}
{Vazza} F., {Br{\"u}ggen} M., {Gheller} C., {Wang} P., 2014{\natexlab{a}},
  \mnras, 445, 3706

\bibitem[{{Vazza} {et~al}\mbox{.}(2018{\natexlab{a}}){Vazza}, {Br{\"u}ggen},
  {Hinz}, {Wittor}, {Locatelli}, \& {Gheller}}]{va18frb}
{Vazza} F., {Br{\"u}ggen} M., {Hinz} P.~M., {Wittor} D., {Locatelli} N.,
  {Gheller} C., 2018{\natexlab{a}}, \mnras, 480, 3907

\bibitem[{{Vazza} {et~al}\mbox{.}(2016){Vazza}, {Br{\"u}ggen}, {Wittor},
  {Gheller}, {Eckert}, \& {Stubbe}}]{va16scienzo}
{Vazza} F., {Br{\"u}ggen} M., {Wittor} D., {Gheller} C., {Eckert} D., {Stubbe}
  M., 2016, \mnras, 459, 70

\bibitem[{{Vazza} {et~al}\mbox{.}(2018{\natexlab{b}}){Vazza}, {Brunetti},
  {Br{\"u}ggen}, \& {Bonafede}}]{va18mhd}
{Vazza} F., {Brunetti} G., {Br{\"u}ggen} M., {Bonafede} A., 2018{\natexlab{b}},
  \mnras, 474, 1672

\bibitem[{{Vazza} {et~al}\mbox{.}(2013{\natexlab{b}}){Vazza}, {Eckert},
  {Simionescu}, {Br{\"u}ggen}, \& {Ettori}}]{va13clump}
{Vazza} F., {Eckert} D., {Simionescu} A., {Br{\"u}ggen} M., {Ettori} S.,
  2013{\natexlab{b}}, \mnras, 429, 799

\bibitem[{{Vazza} {et~al}\mbox{.}(2015{\natexlab{a}}){Vazza}, {Ferrari},
  {Bonafede}, {Br{\"u}ggen}, {Gheller}, {Braun}, \& {Brown}}]{va15ska}
{Vazza} F., {Ferrari} C., {Bonafede} A., {Br{\"u}ggen} M., {Gheller} C.,
  {Braun} R., {Brown} S., 2015{\natexlab{a}}, ArXiv e-prints

\bibitem[{{Vazza} {et~al}\mbox{.}(2015{\natexlab{b}}){Vazza}, {Ferrari},
  {Br{\"u}ggen}, {Bonafede}, {Gheller}, \& {Wang}}]{va15radio}
{Vazza} F., {Ferrari} C., {Br{\"u}ggen} M., {Bonafede} A., {Gheller} C., {Wang}
  P., 2015{\natexlab{b}}, \aap, 580, A119

\bibitem[{{Vazza} {et~al}\mbox{.}(2014{\natexlab{b}}){Vazza}, {Gheller}, \&
  {Br{\"u}ggen}}]{va14curie}
{Vazza} F., {Gheller} C., {Br{\"u}ggen} M., 2014{\natexlab{b}}, ArXiv e-prints

\bibitem[{{Vazza} {et~al}\mbox{.}(2017){Vazza}, {Jones}, {Br{\"u}ggen},
  {Brunetti}, {Gheller}, {Porter}, \& {Ryu}}]{va17turb}
{Vazza} F., {Jones} T.~W., {Br{\"u}ggen} M., {Brunetti} G., {Gheller} C.,
  {Porter} D., {Ryu} D., 2017, \mnras, 464, 210

\bibitem[{{Vernstrom} {et~al}\mbox{.}(2017){Vernstrom}, {Gaensler}, {Brown},
  {Lenc}, \& {Norris}}]{vern17}
{Vernstrom} T., {Gaensler} B.~M., {Brown} S., {Lenc} E., {Norris} R.~P., 2017,
  \mnras, 467, 4914

\bibitem[{{Wang} \& {Abel}(2009)}]{wa09}
{Wang} P., {Abel} T., 2009, \apj, 696, 96

\bibitem[{{Wang} {et~al}\mbox{.}(2010){Wang}, {Abel}, \& {Kaehler}}]{wang10}
{Wang} P., {Abel} T., {Kaehler} R., 2010, \na, 15, 581

\bibitem[{{Werner} {et~al}\mbox{.}(2008){Werner}, {Finoguenov}, {Kaastra},
  {Simionescu}, {Dietrich}, {Vink}, \& {B{\"o}hringer}}]{2008A&A...482L..29W}
{Werner} N., {Finoguenov} A., {Kaastra} J.~S., {Simionescu} A., {Dietrich}
  J.~P., {Vink} J., {B{\"o}hringer} H., 2008, \aap, 482, L29

\bibitem[{{Wittor} {et~al}\mbox{.}(2017{\natexlab{a}}){Wittor}, {Jones},
  {Vazza}, \& {Br{\"u}ggen}}]{wi17b}
{Wittor} D., {Jones} T., {Vazza} F., {Br{\"u}ggen} M., 2017{\natexlab{a}},
  \mnras, 471, 3212

\bibitem[{{Wittor} {et~al}\mbox{.}(2017{\natexlab{b}}){Wittor}, {Vazza}, \&
  {Br{\"u}ggen}}]{wi17}
{Wittor} D., {Vazza} F., {Br{\"u}ggen} M., 2017{\natexlab{b}}, \mnras, 464,
  4448

\bibitem[{{Xu} {et~al}\mbox{.}(2009){Xu}, {Li}, {Collins}, {Li}, \&
  {Norman}}]{xu09}
{Xu} H., {Li} H., {Collins} D.~C., {Li} S., {Norman} M.~L., 2009, \apjl, 698,
  L14

\bibitem[{{Xu} \& {Han}(2015)}]{2015RAA....15.1629X}
{Xu} J., {Han} J.~L., 2015, Research in Astronomy and Astrophysics, 15, 1629

\bibitem[{{Zaritsky} {et~al}\mbox{.}(2014){Zaritsky}, {Courtois},
  {Mu{\~n}oz-Mateos}, {Sorce}, {Erroz-Ferrer}, {Comer{\'o}n}, {Gadotti}, {Gil
  de Paz}, {Hinz}, {Laurikainen}, {Kim}, {Laine}, {Men{\'e}ndez-Delmestre},
  {Mizusawa}, {Regan}, {Salo}, {Seibert}, {Sheth}, {Athanassoula}, {Bosma},
  {Cisternas}, {Ho}, \& {Holwerda}}]{2014AJ....147..134Z}
{Zaritsky} D. {et~al.}, 2014, \aj, 147, 134

\end{thebibliography}

\appendix

\section{Sub-grid models for galaxy formation physics}
\label{feedback}
Our runs adopts the \citet[][]{2003ApJ...590L...1K} star formation model implemented in {\enzo}, with a few ad-hoc modifications to better perform in large unigrid simulations. Star forming particles (actually {\it stellar populations}) in {\enzo} are formed on the fly whenever specific gas conditions are met in cells \citep[][]{enzo14}.  Following \citet[][]{2003ApJ...590L...1K},  four criteria must be met to form a star particle: a) the gas in the cell must exceed a threshold density, $n \geq n_{*}$; b) the cell is part of a converging flow ($\nabla \cdot \vec{v} < 0$); c) the local cooling time is smaller than the dynamical timescale ($t_{\rm cool} \leq t_*$); d) the baryonic mass is larger than the minimum star mass ($m_b=\rho \Delta x^3 \geq m_*$ ). When all these conditions are met in one cell, {\enzo} forms at run-time a stellar particle with a mass $m_* =m_b \Delta t/t_{\rm *}$, where $\Delta t$ is the timestep. 

Stars can also return thermal energy, gas mass and metals back into the surrounding gas. 
The feedback from each star particle (coming from supernovae explosions) depends on the assumed fractions of energy/momentum/mass ejected per each formed star particles, $E_{SN}= \epsilon_{SF} m_* c^2$. In our runs and considering the resolution at which we work, the feedback energy is entirely released as thermal (i.e. hot supernovae-driven winds), which generally promotes  pressure-driven winds around stellar particles ($v_{\rm wind} \sim 10-10^2$ km/s). Given the relatively large time steps in our simulations ($\Delta t \approx 5-10 ~ \rm Myr$ at low redshift) we decided to model the feedback effect in the same timestep in which each stellar particle form, in an instantaneous recycling assumption, which bypasses unnecessary particle-to grid loops that kills the performances for large unigrid simulations.
The \citet[][]{2003ApJ...590L...1K}  model is suitable for uniform grid simulations and has been
designed to reproduce the observed Kennicutt's law \citep[][]{1998ApJ...498..541K}. In \citet{va17cqg} we have further shown how in our runs it can well reproduce the observed cosmic star formation history,  compiled from infrared and ultraviolet observations by  \cite{2014ARA&A..52..415M}, if an appropriate set of model parameters is set (see Table 1). 

In the CHRONOS++ suite we also modified {\enzo} to support the injection of the magnetic fields by stars, introducing magnetic dipoles at each feedback episode. The dipoles are randomly oriented along one of the simulation axis. The total injected energy scales with the feedback thermal energy, i.e. $E_{\rm b,SN}=\epsilon_{\rm SF,b} \cdot E_{\rm SN}$. For the subset of simulations presented in this work, we  used  $\epsilon_{\rm SF,b}~\sim 1-10 \%$ (see Table 1).  We notice that the combination of spatial and mass resolution ($\Delta x=83.3$ kpc and $m_{\rm DM}=6.19 \cdot 10^{7}M_{\odot}$, respectively) should quench the formation of galaxies at the low mass end $ \leq 10^8 M_{\odot}$, and therefore our runs still lacks a contributor to the cosmic star formation rate \citep[e.g.][]{2014MNRAS.445..175G}. However,  the contribution to both the chemical, thermal and magnetic enrichment of the intergalactic medium by the implemented feedback from $\sim 10^{10}-10^{11} M_{\odot}$ galaxies is well resolved and is expected to dominate in the full cosmic volume \citep[e.g.][]{sam17}.

\bigskip

The injection, growth and feedback from  supermassive black holes (SMBH) on the surrounding cosmic gas has been included as well, implemented with few modifications on top of existing {\enzo} routines, based on \cite{2011ApJ...738...54K}.
Unlike star forming particles, SMBH particles in {\enzo} must be seeded at a given redshift (we chose $z=4$ in all runs) with an initial mass ($M_{\rm BH}=10^4 M_{\odot}$ in our case), and from that moment on, they are advected at run time (similar to dark matter particles as well as star particles, they are  assumed to be collisionless 
and only subject to the local gravity acceleration)  grow in mass and produce feedback events. 
We assumed for each SMBH  a Bondi-Hoyle accretion rate, with the additional model assumptions (motivated by the low resolution), as in Table 1:  a) the gas accretes onto SMBH assuming a fixed temperature of $3 \cdot 10^5 \rm K$ ; b) the Bondi accretion rate gets boosted by a factor $\alpha_{Bondi}=10^2-10^3$ if the gas density which can be resolved around each SMBH is too low. 

SMBH particles release thermal energy to the surrounding gas with an efficiency $\epsilon_{\rm BH} = \Delta M c^2 \Delta t/E_{\rm BH}$, $\Delta M$ being the accreted gas mass and $E_{\rm BH}$ being the feedback energy.  We have implemented the release of magnetic feedback energy from SMBH in {\enzo} in the form of magnetic dipoles, assuming $E_{b,AGN}= \epsilon_{b,AGN} E_{\rm BH}$. In this paper we explored values in the range $\epsilon_{\rm b,AGN}=1-10 \%$. Our tests in \citet{va17cqg} show that such treatment of SMBH and associated feedback (combined with star formation) is reasonably capable of modifying the mass-temperature scaling relations of galaxy groups and clusters, in the $10^{13} M_{\odot} \leq M \leq 10^{15} M_{\odot}$  mass range, towards X-ray observations  \citep[][]{2011A&A...535A...4R,2011A&A...535A.105E}, by increasing the typical temperature of $M_{\rm 500} \leq 10^{14}M_{\odot}$ systems in a realistic way.

\section{Sub-grid models for magnetic dynamo amplification}

Our  run-time implementation of a sub-grid (SG) model for small-scale magnetic dynamo  in {\enzo} attempts to compute the maximal 
energy that can be channelled into magnetic fields, based on the dissipation of solenoidal turbulent motions  that we can measure
at run-time in the simulation. This is done in order to bracket, at least in as set of ad-hoc models, the maximum effect of turbulent dynamo amplification on scales that our simulation cannot resolve, and that will probably remain challenging to resolve for a long time \citep[e.g.][]{review_dynamo}. 
Our extended resolution studies on magnetic amplification in filaments \citep[][]{va14mhd} tend to exclude the possibility of a significant development of a volume filling dynamo in filaments, induced by the decay of fast and supersonic turbulent motions measured there. However, our simple MHD view cannot exclude that in real plasmas ``microscopic" small-scale instabilities arising from kinetic plasma  effects \citep[e.g.][for a review]{2014MNRAS.440.3226M}
can further amplify the field,  starting from sub-kpc scales, even if there are presently no observational reason to think this is the case.  
The maximum energy that can be channelled into magnetic fields during a root-grid timestep, $\Delta t$ ($\sim 10 ~\rm Myr$) is of the order of $E_{\rm B,dyn} = \epsilon_{\rm dyn}(\mathcal{M})F_{\rm turb} \Delta t$, where 
$F_{\rm turb} \simeq \eta_t \rho \epsilon_{\omega}^3/L$ is the turbulent kinetic energy flux, with $ \epsilon_{\omega}=(\nabla \times \vec{v})^2$ is the flow enstrophy which we measure at run-time on a $L=3$ cells stencil, while  $\eta_t$ is the estimated dissipation rate onto magnetic fields, which must be guessed or calibrated based on other simulations  \citep[][]{jones11,po15,bm16,va17turb,wi17b}. \\

In our runs we adopted a general approach in which $\eta_t$ can scale with the local plasma parameters, as its value can critically change going from the sub-sonic to the supersonic regime \citep{ry08}, based on \citet{fed14}, who simulated small-scale dynamo in a variety of conditions. We could estimate the saturation level and the typical growth time of magnetic fields as a function of the local turbulent Mach number of the flow, $\mathcal{M_{\rm turb}}$. In particular, $\epsilon_{\rm dyn}=\epsilon_{\rm d}(\mathcal{M_{\rm turb}})$ can be approximated  as  $\epsilon_d(\mathcal{M}_{\rm turb}) \approx (E_B/E_k) \Gamma \Delta t$, where $E_B/E_k$ is the estimated ratio between magnetic and kinetic energy at saturation, and $\Gamma$ is the typical growth rate, which we both take from \citet{fed14} as a function of the flow Mach number. Once we have computed the amplified magnetic energy, $E_{\rm B,dyn}$, we can generate at run-time a magnetic field vector, $\vec{\delta B}$, to be added to the already existing field, imposing it to be parallel to the local direction of the gas vorticity, so that the new generated field is also solenoidal by construction. A corresponding budget in kinetic energy is removed from cells. Momentum is subtracted assuming an isotropic dissipation of the small-scale velocity vectors.  Our procedure is manifestly simpler than more sophisticated SG models on the market \citep{gr16} and the resulting topology of magnetic fields is not consistently attached to the underlying turbulent flow. However, the overall energy budget in amplified magnetic fields well matches the prediction derived in other works \citep[e.g.][]{ry08} and is overall suitable to bracket the (possible) amplification of magnetic fields in the WHIM under extreme conditions. In our CHRONOS++ suite \citep[][]{va17cqg} and in the simulation presented here (see  Table 1) we explored a few variations in the amplitude of seed fields before the dynamo acts, as well as in the normalisation of the $\epsilon_d(\mathcal{M}_{\rm turb})$ dynamo efficiency, finding only mild differences. 

\end{document}